\documentclass[12pt,english]{article}
\usepackage{lmodern}
\usepackage[T1]{fontenc}
\usepackage[latin9]{inputenc}
\usepackage{amsmath}
\usepackage{amssymb}
\usepackage{esint}

\usepackage{geometry}
\makeatletter
\usepackage{float} 
\usepackage[colorlinks=true, linkcolor=ForestGreen, citecolor=blue]{hyperref}
\usepackage{cite}
\usepackage{babel}
\usepackage{mathdots}
\usepackage{amsfonts}
\usepackage{mathrsfs}
\usepackage{amsmath}
\usepackage{esint}
\usepackage{graphicx}
\usepackage{youngtab}
\usepackage{multicol}
\usepackage{multirow}
\usepackage{slashed}
\usepackage{simplewick}
\usepackage{braket}
\usepackage{bm}
\usepackage{relsize}
\usepackage[dvipsnames,table,xcdraw]{xcolor}
\usepackage{subfigure}
\usepackage{feynmp}
\usepackage{pifont}
\usepackage{cancel}
\numberwithin{equation}{section}
\allowdisplaybreaks

\makeatother

\usepackage[dvipsnames]{xcolor}

\begin{document}
\author{Wan-Zhe Feng\footnote{Email: vicf@tju.edu.cn}~~and~Zi-Hui Zhang\footnote{Email: zhangzh\_@tju.edu.cn}\\
\textit{\small{Center for Joint Quantum Studies and Department of Physics,}}\\
\textit{\small{School of Science, Tianjin University, Tianjin 300350, PR. China}}}
\title{Gauge-independent gravitational waves
from a minimal dark $U(1)$ sector with viable dark matter candidates}
\date{Happy Year of the Horse!}

\maketitle

\begin{abstract}

Searches for stochastic gravitational wave backgrounds generated by first-order phase transitions offer a powerful probe of hidden sectors, but quantitative predictions in gauge theories are obstructed by the gauge dependence of the finite-temperature effective potential and the associated tunneling action. We study a minimal gauged $U(1)$ dark sector containing a dark Higgs and a dark photon, optionally supplemented by a vectorlike dark fermion, coupled to the Standard Model through the Higgs portal or kinetic mixing. Using the Nielsen identity together with a controlled derivative expansion and power counting, we construct a gauge-independent effective action in the high- and low-temperature limits, enabling model-intrinsic nucleation dynamics and robust gravitational wave predictions. We perform dedicated Monte Carlo scans in both limits and map viable microscopic parameters to detector-facing peak frequencies and amplitudes, spanning bands relevant to pulsar timing arrays and planned space-based interferometers. In our scans, supercooled phase transitions typically produce much stronger signals and are more likely to fall within the sensitivity range of current and future gravitational wave detectors, whereas parametrically high-temperature phase transitions generally yield weaker signals and are of less phenomenological interest. We further connect the phase transition phenomenology to viable dark matter candidates within the same minimal field content, providing benchmark targets for dark photon dark matter and dark fermion dark matter, and highlighting their complementarity with gravitational wave observables. Overall, our results provide an end-to-end, gauge-independent pipeline from a minimal hidden sector Lagrangian to gravitational wave spectra and cosmologically viable dark matter benchmarks, yielding the most reliable and concrete predictions to date for a minimal gauged $U(1)$ dark sector.

\end{abstract}

\newpage{}

{  \hrule height 0.4mm \hypersetup{colorlinks=black,linktocpage=true}
\tableofcontents
\vspace{0.5cm}
 \hrule height 0.4mm}
\newpage{}

\section{Introduction}

The discovery of gravitational waves provides a novel probe of hidden sectors in which dark matter may reside.
Searches for stochastic gravitational wave backgrounds produced by first-order phase transitions,
either from the Standard Model (SM)
augmented by new physics~\cite{Anderson:1991zb,Carrington:1991hz,Arnold:1992rz,Kajantie:1996mn,Grojean:2004xa,Morrissey:2012db,Athron:2023xlk},
or in a dark  sector~\cite{Schwaller:2015tja,Jaeckel:2016jlh,Jinno:2016knw,Addazi:2016fbj,Chao:2017vrq,Addazi:2017gpt,Marzola:2017jzl,Baldes:2018emh,Marzo:2018nov,Breitbach:2018ddu,Fairbairn:2019xog,Nakai:2020oit,Addazi:2020zcj,Kierkla:2022odc,Fujikura:2023lkn,Ertas:2021xeh,Wang:2022akn,Chen:2023rrl,Bringmann:2023opz,Li:2023bxy,Kanemura:2023jiw,Bringmann:2023iuz,Banik:2024zwj,Feng:2024pab,Balan:2025uke,Baules:2025pww,Goncalves:2024lrk,Costa:2025csj,Li:2025nja,Wang:2026wwp,Feng:2025wvc,Houtz:2025ogg,Mahapatra:2026fyv},
offer a powerful and complementary pathway to uncovering new physics,
especially in light of the continuing non-detection of dark matter in conventional experiments.
Gravitational wave observations thus constitute a key component of a multi-messenger program
to test particle physics in the early Universe.
Accordingly, the interplay between gravitational wave phenomenology and physics beyond the SM,
particularly in scenarios aimed at explaining the nature of dark matter,
has become one of the most active frontiers of current research.

In well-motivated dark sector models, gravitational waves from first-order phase transitions are particularly compelling:
New gauge symmetries and scalar fields are ubiquitous in BSM frameworks,
and their symmetry breaking can naturally occur at scales both below and above the electroweak scale,
leading to gravitational wave signals that span a wide range of frequencies.
Consequently, it is essential to establish robust and concrete predictions for gravitational wave signals
from dark sectors that can be confronted with data across multiple frequency bands.
If detected, such a signal can reveal information about the underlying dark sector parameters
including the gauge coupling, the scalar quartic coupling, and the scalar vacuum expectation value (vev),
that may be difficult or impossible to access directly through collider probes or conventional dark matter searches.

However, a central obstacle to making reliable gravitational wave predictions in gauge theories
is the well-known gauge dependence of the finite-temperature effective potential and the associated tunneling action.
The Higgs effective potential as a function of an arbitrary background field is not a physical observable.
In $R_\xi$ gauges, the result depends explicitly on the gauge-fixing parameter $\xi$,
and thus a conventional computation based directly on the gauge-dependent effective potential
leads to gauge-dependent gravitational wave spectra that can hardly be regarded as genuine ``predictions''.
This gauge ambiguity obscures the interpretation of parameter scans
and hampers the extraction of robust, model-intrinsic conclusions.

To address this issue, we adopt a gauge-independent formulation of the effective action based on the Nielsen identity~\cite{Metaxas:1995ab,Patel:2011th,Garny:2012cg,Andreassen:2014eha,Andreassen:2014gha,Arunasalam:2021zrs,Lofgren:2021ogg,Hirvonen:2021zej,Zhu:2025pht,Liu:2025ipj,Liu:2026ask},
with particular emphasis on the low-temperature treatment developed in~\cite{Feng:2025wvc}.
The key payoff is a well-defined, gauge-independent tunneling action
and thus gauge-independent predictions for the phase transition
and the resulting gravitational wave signal within a controlled power counting scheme.
We provide a detailed discussion of the validity of the derivative expansion of the effective action,
along with the associated gauge coupling power counting rules,
and the scaling relations between the gauge coupling and the quartic coupling
at zero, high, and low temperatures, thereby clarifying several common sources of confusion in the literature.
Building on the gauge-independent effective action in the relevant high- and low-temperature limits,
we perform dedicated Monte Carlo scans in both regimes.
This yields concrete, detector-facing predictions:
each viable parameter point maps to a peak frequency and amplitude of the stochastic background,
enabling a direct comparison with the sensitivity targets of current and planned experiments across various frequency bands.

We focus in particular on supercooled first-order phase transitions,
which are among the strongest and most distinctive possible sources of a stochastic gravitational wave
background~\cite{Leitao:2015fmj,Megevand:2016lpr,Kobakhidze:2017mru,Iso:2017uuu,Ellis:2019oqb,Wang:2020jrd,Ellis:2020nnr,Nakai:2020oit,Addazi:2020zcj,Kawana:2022fum,Freese:2022qrl,Lewicki:2022pdb,Kierkla:2022odc,Sagunski:2023ynd,Madge:2023dxc,Fujikura:2023lkn,Athron:2023mer,Ghosh:2023aum,Athron:2023rfq,Goncalves:2024lrk,Athron:2025pog,Costa:2025csj,Li:2025nja,Feng:2025wvc,Wang:2026wwp,Pascoli:2026tuu}.
In such transitions, bubble nucleation and percolation occur well below the critical temperature, $T_{n,p}\ll T_c$.
The false vacuum can dominate the energy density of the Universe for an extended period before the transition completes,
leading to a large release of vacuum energy and typically enhancing the resulting gravitational wave amplitude.
At the same time, the lower transition temperature and larger characteristic length scale shift the peak frequency to lower values,
in some cases into the nanohertz band, thereby attracting substantial recent attention.
For scenarios with small portal couplings to the SM, which may be difficult to probe through collider searches or conventional dark matter experiments,
gravitational waves from supercooled phase transitions can provide a leading discovery channel for this otherwise elusive region of parameter space.

In this work, we focus on the simplest representative setup:
a minimal $U(1)_x$ dark sector featuring a dark Higgs field and a dark photon,
and optionally a dark fermion, together with portal interactions
connecting the hidden sector to the SM.
This minimality is not merely aesthetic.
It makes the parameter dependence transparent,
allows for a systematic exploration of distinct thermal regimes,
and admits dark matter candidates whose cosmological evolution
can be computed consistently alongside the phase transition dynamics.
In particular, we highlight two viable dark matter candidates already present in the minimal field content:
(1) dark photon dark matter, which requires vanishing or negligibly small kinetic mixing; and
(2) dark fermion dark matter, which relies on a non-negligible dark photon portal.

This paper therefore provides robust and predictive results
for gravitational waves generated by the symmetry breaking of a minimal gauged $U(1)$ dark sector
with a viable single dark matter candidate.
Such results are significant in the current high energy physics research,
where no compelling hints of new physics have emerged from any experiments,
and precision tests as well as multi-messenger searches are becoming increasingly essential for new physics discovery.

The remainder of this paper is organized as follows.
In Section~\ref{Sec:U1ex} we introduce the minimal dark $U(1)_x$ sector and portal interactions that connect it to SM states.
In Section~\ref{Sec:Gauge} we review the gauge dependence of the finite-temperature effective potential
and the construction of gauge-independent effective actions based on the Nielsen identity,
with particular emphasis on the low-temperature regime relevant for supercooled phase transitions.
In Section~\ref{Sec:GW} we present the computation of gravitational wave signatures from first-order phase transitions,
including the tunneling action and transition rate, characteristic temperatures, hydrodynamic parameters, and the resulting gravitational wave power spectra, and we compare gauge-dependent results and gauge-independent predictions.
In Section~\ref{Sec:Ph} we analyze the cosmological evolution of dark sector particles,
derive the coupled Boltzmann equations for the dark matter scenarios of interest,
and present benchmark models and phenomenological implications.
Finally, we summarize our conclusions and present an outlook in Section~\ref{Sec:Con}.

\section{The $U(1)_{x}$ extension and portals to the SM}\label{Sec:U1ex}

In this section, we review the minimal $U(1)_x$ extension of the SM,
featuring a dark Higgs, a dark photon, and optionally dark fermions,
with their portal interactions to SM particles.

\subsection{The $U(1)_{x}$ extension of the SM}

We consider a $U(1)_{x}$ gauge extension of the SM. The
full Lagrangian can be written as
\begin{equation}
\mathcal{L}=\mathcal{L}_{{\rm SM}}+\mathcal{L}_{{\rm hid}}+\mathcal{L}_{{\rm mix}}\,,
\end{equation}
where $\mathcal{L}_{{\rm SM}}$ and $\mathcal{L}_{{\rm hid}}$ are
given by
\begin{align}
\mathcal{L}_{{\rm SM}}\supset & -\frac{1}{4}F_{\mu\nu}F^{\mu\nu}-\mu_{{\rm SM}}^{2}H^{\dagger}H+\lambda_{{\rm SM}}(H^{\dagger}H)^{2}\,,\\
\mathcal{L}_{\mathrm{hid}}= & \;-\frac{1}{4}F_{x\,\mu\nu}F_{x}^{\mu\nu}+\bigl|(\partial_{\mu}-ig_{x}A_{\mu}^{\prime})\Phi\bigr|^{2}+\mu_{x}^{2}\Phi^{\ast}\Phi-\lambda_{x}(\Phi^{\ast}\Phi)^{2}\nonumber \\
 & +g_{x}\overline{\chi}\gamma^{\mu}\chi A_{\mu}^{\prime}+\overline{\chi}(i\gamma^{\mu}\partial_{\mu}-m_{\chi})\chi\,,
\end{align}
where $F_{\mu\nu},F_{x\,\mu\nu}$ are the field strengths of the hypercharge
and $U(1)_{x}$ gauge fields respectively, $H$ is the SM
Higgs doublet and $\Phi$ is the $U(1)_{x}$ Higgs field with $U(1)_{x}$
charge $+1$. $A_{\mu}^{\prime}$ is the $U(1)_{x}$ gauge field,
and $\chi$ is a dark fermion with vector mass $m_{\chi}$ and $U(1)_{x}$
charge $Q_{x}=+1$. The most important parameters relevant for the
study of both $U(1)_{x}$ gravitational wave physics and dark matter
physics are the $U(1)_{x}$ gauge coupling $g_{x}$, the $U(1)_{x}$
Higgs vev $v_{x}$ and its quartic coupling $\lambda_{x}$.

The mixing terms contain kinetic mixing and Higgs mixing contributions,
written as
\begin{equation}
\mathcal{L}_{{\rm mix}}=\;-\frac{\delta}{2}F_{\mu\nu}F_{x}^{\mu\nu}-\lambda_{{\rm mix}}(H^{\dagger}H)(\Phi^{\ast}\Phi)\,,\label{eq:KM}
\end{equation}
where $\delta$ and $\lambda_{{\rm mix}}$ are mixing parameters,
both receiving stringent constraints from experiments. The experimental
constraint of $\lambda_{{\rm mix}}$ will be discussed in Section~\ref{sec:BM}.
The kinetic mixing can arise from either integrating out bifundamental
fields charged under both the $U(1)_{Y}$ and $U(1)_{x}$, which can
be absent; or from graviton loop effects, which are in general suppressed.
We focus on two classes of dark matter candidates in this paper:
\begin{itemize}
\item \textbf{Dark photon dark matter}, which requires vanishing or extremely small kinetic mixing, $\delta \lesssim 10^{-16}$.
\item \textbf{Dark fermion dark matter}, which requires a non-negligible kinetic mixing.
\end{itemize}

Expanding $\Phi$ around the background field $\phi_{c}$ as
\begin{equation}
\Phi=\frac{1}{\sqrt{2}}\bigl(\phi_{c}+h_{x}+iG\bigr)\,,
\end{equation}
where $h_{x}$ and $G$ are the dark Higgs and Goldstone boson respectively,
we obtain
\begin{align}
\bigl|(\partial_{\mu}-ig_{x}A_{\mu}^{\prime})\Phi\bigr|^{2} & =\frac{1}{2}(\partial_{\mu}h_{x}+g_{x}A_{\mu}^{\prime}G)^{2}+\frac{1}{2}\bigl[\partial_{\mu}G-g_{x}A_{\mu}^{\prime}(\phi_{c}+h_{x})\bigr]^{2}\nonumber \\
 & =\frac{1}{2}(\partial_{\mu}h_{x})^{2}+\frac{1}{2}(\partial_{\mu}G)^{2}-g_{x}\phi_{c}(\partial_{\mu}G)A^{\prime\mu}+\cdots\,.\label{eq:Dphi}
\end{align}
Together with the gauge-fixing term and the ghost terms,
\begin{align}
\mathcal{L}_{{\rm gf}} & =-\frac{1}{2\xi}\bigl(\partial_{\mu}A^{\prime}_\mu+\xi g_{x}\phi_{c}G\bigr)^{2}\,,\\
\mathcal{L}_{{\rm ghost}} & =\overline{c}\,\bigl[-\partial^{2}-\xi g_{x}^{2}\phi_{c}(\phi_{c}+h_{x})\bigl]\,c\,,
\end{align}
where $\xi$ is the gauge parameter, $c  (\overline{c})$ is the
ghost (anti-ghost) field, the mixing between $A^{\prime}$ and $G$
in Eq.~(\ref{eq:Dphi}) is canceled by the cross term
in $\mathcal{L}_{{\rm gf}}$. The field-dependent masses in the $R_\xi$ gauge are given by
\begin{align}
m_{A^{\prime}}^{2} & =g_{x}^{2}\phi_{c}^{2}\,, \qquad
m_{A^{\prime}_0}^{2} =\xi m_{A^{\prime}}^{2} = \xi g_{x}^{2}\phi_{c}^{2}\,,\label{eq:mA}\\
m_{h_{x}}^{2} & =-\mu_{x}^{2}+3\lambda_{x}\phi_{c}^{2}\,,\label{eq:mh}\\
m_{c}^{2} & = \xi m_{A^{\prime}}^{2} = \xi g_{x}^{2}\phi_{c}^{2} \,,\label{eq:mc}\\
m_{G}^{2} & = -\mu_{x}^{2}+\lambda_{x}\phi_{c}^{2} +\xi m_{A^{\prime}}^{2}\,,\label{eq:mG}
\end{align}
where the masses of the unphysical time-like gauge component $A^{\prime}_0$,
the ghost and the Goldstone boson depend on the gauge parameter $\xi$.

\subsection{Higgs portal through scalar mixing}

We now discuss the Higgs portal interactions
from the quartic mixing term between $H$ and $\Phi$.
The full scalar potential reads
\begin{equation}
V(H,\phi_{x})=-\mu_{{\rm SM}}^{2}H^{\dagger}H-\mu_{x}^{2}\Phi^{\ast}\Phi+\lambda_{{\rm SM}}\bigl(H^{\dagger}H\bigr)^{2}+\lambda_{x}\bigl(\Phi^{\ast}\Phi\bigr)^{2}+\lambda_{{\rm mix}}\bigl(H^{\dagger}H\bigr)\bigl(\Phi^{\ast}\Phi\bigr)\,.
\end{equation}
Working in the unitary gauge, $H$ and $\Phi$ can be expanded around
their minima as
\begin{equation}
H=\frac{1}{\sqrt{2}}\left(\begin{array}{c}
0\\
v_{{\rm SM}}+h^{0}
\end{array}\right)\,,\qquad\Phi=\frac{1}{\sqrt{2}}\bigl(v_{x}+h_{x}^{0}\bigr)\,,
\end{equation}
where $h^{0}$, $h_{x}^{0}$ are real scalars in the flavor eigenbasis.
The vacuum expectation values $v_{{\rm SM}}$, $v_{x}$ are given
by
\begin{align}
v_{{\rm SM}}^{2} & =\frac{4\mu_{{\rm SM}}^{2}\lambda_{x}-2\mu_{x}^{2}\lambda_{{\rm mix}}}{4\lambda_{x}\lambda_{{\rm SM}}-\lambda_{{\rm mix}}^{2}}\,,\\
v_{x}^{2} & =\frac{4\mu_{x}^{2}\lambda_{{\rm SM}}-2\mu_{{\rm SM}}^{2}\lambda_{{\rm mix}}}{4\lambda_{x}\lambda_{{\rm SM}}-\lambda_{{\rm mix}}^{2}}\,.
\end{align}
The kinetic and mass terms are then
\begin{align}
\bigl|(\partial_{\mu}-ig_{x}A_{\mu}^{\prime})\Phi\bigr|^{2} & =\frac{1}{2}\bigl(\partial_{\mu}h_{x}^{0}\bigr)^{2}+\frac{1}{2}g_{x}^{2}\Bigl[A^{\prime2}\bigl(h_{x}^{0}\bigr)^{2}+2v_{x}A^{\prime2}h_{x}^{0}+v_{x}^{2}A^{\prime2}\Bigr]\,,\\
-\mathcal{L}_{{\rm mass}} & =\frac{1}{2}\bigl(\begin{array}{cc}
h^{0} & h_{x}^{0}\end{array}\bigr)\left(\begin{array}{cc}
2\lambda_{{\rm SM}}v_{{\rm SM}}^{2} & \lambda_{{\rm mix}}v_{{\rm SM}}v_{x}\\
\lambda_{{\rm mix}}v_{{\rm SM}}v_{x} & 2\lambda_{x}v_{x}^{2}
\end{array}\right)\left(\begin{array}{c}
h^{0}\\
h_{x}^{0}
\end{array}\right)\,.
\end{align}
An orthogonal transformation is performed to diagonalize the mass
matrix, which transforms the original eigenbasis $(h^{0}\;\;h_{x}^{0})^{T}$
into mass eigenbasis $(h\;\;h_{x})^{T}$ as
\begin{equation}
\left(\begin{array}{c}
h\\
h_{x}
\end{array}\right)=\left(\begin{array}{cc}
\cos\theta & -\sin\theta\\
\sin\theta & \cos\theta
\end{array}\right)\left(\begin{array}{c}
h^{0}\\
h_{x}^{0}
\end{array}\right)\,,
\end{equation}
where the mixing angle $\theta$ is calculated to be
\begin{equation}
\theta=\frac{1}{2}\arctan \frac{\lambda_{{\rm mix}}v_{{\rm SM}}v_{x}}{\lambda_{x}v_{x}^{2}-\lambda_{{\rm SM}}v_{{\rm SM}}^{2}}\,. \label{eq: HMang}
\end{equation}
The mass eigenvalues are
\begin{align}
m_{h}^{2} & =\lambda_{{\rm SM}}v_{{\rm SM}}^{2}+\lambda_{x}v_{x}^{2}+(\lambda_{{\rm SM}}v_{{\rm SM}}^{2}-\lambda_{x}v_{x}^{2})\sqrt{1+\tan^{2}2\theta}\,,\\
m_{h_{x}}^{2} & =\lambda_{{\rm SM}}v_{{\rm SM}}^{2}+\lambda_{x}v_{x}^{2}-(\lambda_{{\rm SM}}v_{{\rm SM}}^{2}-\lambda_{x}v_{x}^{2})\sqrt{1+\tan^{2}2\theta}\,.
\end{align}
Define $\Delta m^{2}=m_{h_{x}}^{2}-m_{h}^{2}$ and we obtain
\begin{align}
v_{x} & =\frac{\Delta m^{2}\sin2\theta}{2v_{{\rm SM}}\lambda_{{\rm mix}}}\,,\\
\lambda_{{\rm SM}} & =\frac{m_{h}^{2}}{2v_{{\rm SM}}^{2}}+\frac{\Delta m^{2}\sin^{2}\theta}{2v_{{\rm SM}}^{2}}\,,\\
\lambda_{x} & =\frac{2v_{{\rm SM}}^{2}\lambda_{{\rm mix}}^{2}}{\sin^{2}2\theta\,\Delta m^{2}}\biggl(\frac{m_{h_{x}}^{2}}{\Delta m^{2}}-\sin^{2}\theta\biggr)\,,
\end{align}
where the quartic coupling $\lambda_{{\rm mix}}$ must satisfy all
current experimental constraints. In the mass eigenbasis, the scalar
potential is now written as
\begin{align}
V(h,h_{x})= & \;C_{h^{3}}h^{3}+C_{h^{2}h_{x}}h^{2}h_{x}+C_{hh_{x}^{2}}hh_{x}^{2}+C_{h_{x}^{3}}h_{x}^{3}\nonumber \\
 & +C_{h^{4}}h^{4}+C_{h^{3}h_{x}}h^{3}h_{x}+C_{h^{2}h_{x}^{2}}h^{2}h_{x}^{2}+C_{hh_{x}^{3}}hh_{x}^{3}+C_{h_{x}^{4}}h_{x}^{4}\nonumber \\
 & +\frac{1}{2}\bigl(\begin{array}{cc}
h & h_{x}\end{array}\bigr)\left(\begin{array}{cc}
m_{h}^{2} & 0\\
0 & m_{h_{x}}^{2}
\end{array}\right)\left(\begin{array}{c}
h\\
h_{x}
\end{array}\right)\,,
\end{align}
where the trilinear couplings are given by
\begin{gather}
C_{h^{3}} =\frac{m_{h}^{2}}{2v_{{\rm SM}}c_{\theta}}\Bigl(c_{\theta}^{4}-\frac{\lambda_{{\rm mix}}v_{{\rm SM}}^{2}}{\Delta m^{2}}s_{\theta}^{2}\Bigr)\,,\quad C_{h^{2}h_{x}}=\frac{2m_{h}^{2}+m_{h_{x}}^{2}}{2v_{{\rm SM}}}s_{\theta}\Bigl(c_{\theta}^{2}+\frac{\lambda_{{\rm mix}}v_{{\rm SM}}^{2}}{\Delta m^{2}}\Bigr)\,,\\
C_{h_{x}^{3}} =\frac{m_{h_{x}}^{2}}{2v_{{\rm SM}}s_{\theta}}\Bigl(s_{\theta}^{4}+\frac{\lambda_{{\rm mix}}v_{{\rm SM}}^{2}}{\Delta m^{2}}c_{\theta}^{2}\Bigr)\,,\quad C_{hh_{x}^{2}}=\frac{m_{h}^{2}+2m_{h_{x}}^{2}}{2v_{{\rm SM}}}c_{\theta}\Bigl(s_{\theta}^{2}-\frac{\lambda_{{\rm mix}}v_{{\rm SM}}^{2}}{\Delta m^{2}}\Bigr)\,,
\end{gather}
and the quartic couplings are
\begin{gather}
C_{h^{4}} =\frac{1}{4}\Bigl(\frac{s_{2\theta}^{2}}{4}\lambda_{{\rm mix}}+c_{\theta}^{4}\lambda_{{\rm SM}}+s_{\theta}^{4}\lambda_{x}\Bigr)\,,\quad C_{h^{3}h_{x}}=\frac{1}{4}s_{2\theta}\bigl(-c_{2\theta}\lambda_{{\rm mix}}+2c_{\theta}^{2}\lambda_{{\rm SM}}-2s_{\theta}^{2}\lambda_{x}\bigr)\,,\\
C_{h_{x}^{4}} =\frac{1}{4}\Bigl(\frac{s_{2\theta}^{2}}{4}\lambda_{{\rm mix}}+s_{\theta}^{4}\lambda_{{\rm SM}}+c_{\theta}^{4}\lambda_{x}\Bigr)\,,\quad C_{hh_{x}^{3}}=\frac{1}{4}s_{2\theta}\bigl(c_{2\theta}\lambda_{{\rm mix}}+2s_{\theta}^{2}\lambda_{{\rm SM}}-2c_{\theta}^{2}\lambda_{x}\bigr)\,,\\
C_{h^{2}h_{x}^{2}} =\frac{1}{4}\Bigl[\bigl(c_{2\theta}^{2}+s_{2\theta}^{2}\bigr)\lambda_{{\rm mix}}+\frac{3}{2}s_{2\theta}^{2}\bigl(-\lambda_{{\rm mix}}+\lambda_{{\rm SM}}+\lambda_{x}\bigr)\Bigr]\,,
\end{gather}
with $c_{\theta}\equiv\cos\theta$ and $s_{\theta}\equiv\sin\theta$.
The $U(1)_{x}$ gauge boson $A^{\prime}$ acquires a mass $M_{A^{\prime}}=g_{x}v_{x}$,
referred to as a dark photon $\gamma^{\prime}$. The couplings of
$\gamma^{\prime}$ to two Higgs fields can be obtained from
\begin{align}
\bigl|(\partial_{\mu}-ig_{x}A_{\mu}^{\prime})\Phi\bigr|^{2} & =\frac{1}{2}\Bigl[s_{\theta}^{2}\bigl(\partial_{\mu}h\bigr)^{2}+c_{\theta}^{2}\bigl(\partial_{\mu}h_{x}\bigr)^{2}-2s_{\theta}c_{\theta}\bigl(\partial_{\mu}h\bigr)\bigl(\partial^{\mu}h_{x}\bigr)\Bigr]\nonumber \\
 & \quad +\frac{1}{2}g_{x}^{2}A^{\prime2}\Bigl[s_{\theta}^{2}h^{2}+c_{\theta}^{2}h_{x}^{2}-2s_{\theta}c_{\theta}h\,h_{x}+2v_{x}(s_{\theta}h+c_{\theta}h_{x})+v_{x}^{2}\Bigr]\,.
\end{align}
The couplings of $h$ and $h_{x}$ to SM particles are then given by
\begin{align}
\mathcal{L}= & \;\Bigl(M_{W}^{2}W^{\mu+}W_{\mu}^{-}+\frac{1}{2}M_{Z}^{2}Z^{\mu}Z_{\mu}\Bigr)\Bigl(1+\frac{h^{0}}{v_{{\rm SM}}}\Bigr)^{2}-\sum_{i}m_{i}\overline{f_{i}}f_{i}\frac{h^{0}}{v_{{\rm SM}}}\nonumber \\
& \to \;\Bigl(\frac{c_{\theta}^{2}}{v_{{\rm SM}}^{2}}h^{2}+\frac{2c_{\theta}}{v_{{\rm SM}}}h\Bigr)\Bigl(M_{W}^{2}W^{\mu+}W_{\mu}^{-}+\frac{1}{2}M_{Z}^{2}Z^{\mu}Z_{\mu}\Bigr)\nonumber \\
 & \ \  +\Bigl(\frac{s_{\theta}^{2}}{v_{{\rm SM}}^{2}}h_{x}^{2}+\frac{2s_{\theta}}{v_{{\rm SM}}}h_{x}\Bigr)\Bigl(M_{W}^{2}W^{\mu+}W_{\mu}^{-}+\frac{1}{2}M_{Z}^{2}Z^{\mu}Z_{\mu}\Bigr)\nonumber \\
 & \ \ +\frac{2c_{\theta}s_{\theta}}{v_{{\rm SM}}^{2}}hh_{x}\Bigl(M_{W}^{2}W^{\mu+}W_{\mu}^{-}+\frac{1}{2}M_{Z}^{2}Z^{\mu}Z_{\mu}\Bigr)-\frac{c_{\theta}h+s_{\theta}h_{x}}{v_{{\rm SM}}}\sum_{i}m_{i}\overline{f_{i}}f_{i}\,.
\end{align}

\subsection{$U(1)$ portal through kinetic mixing}

We now discuss $U(1)_{x}$ portal interactions from the kinetic mixing.
In the gauge eigenbasis $V^{T}=(C,B,A^{3})$ where $C,B$ and $A^{3}$
denote the dark $U(1)_{x}$ gauge field, SM hypercharge field, and
the neutral $SU(2)_{L}$ gauge field respectively, the mixing matrices
take the form~\cite{Feldman:2007wj,Feng:2023ubl}
\begin{equation}
\mathcal{K}=\left(\begin{array}{ccc}
1 & \delta & 0\\
\delta & 1 & 0\\
0 & 0 & 1
\end{array}\right)\,,\qquad M_{0}^{2}=\left(\begin{array}{ccc}
g_{x}^{2}v_{x}^{2} & 0 & 0\\
0 & \frac{1}{4}g_{Y}^{2}v_{{\rm SM}}^{2} & -\frac{1}{4}g_{2}g_{Y}v_{{\rm SM}}^{2}\\
0 & -\frac{1}{4}g_{2}g_{Y}v_{{\rm SM}}^{2} & \frac{1}{4}g_{2}^{2}v_{{\rm SM}}^{2}
\end{array}\right)\,,
\end{equation}
where $g_{2}$ and $g_{Y}$ denote the SM couplings. To diagonalize
these two matrices simultaneously, we first perform a non-unitary
transformation $K$ that diagonalizes the kinetic terms with $K^{T}\mathcal{K}K=\boldsymbol{1}$.
Under the transformation $V^{\prime}=K^{-1}V$, the mass matrix becomes
$M_{0}^{\prime2}=K^{T}M_{0}^{2}K$. We then apply an orthogonal transformation
$O$ such that $O^{T}K^{T}M_{0}^{2}KO$ is diagonal. The transformation
matrices $K$ and $O$ are found to be
\begin{equation}
K=\left(\begin{array}{ccc}c_{\delta} & 0 & 0\\-s_{\delta} & 1 & 0\\0 & 0 & 1\end{array}\right)\,,\qquad O=\left(\begin{array}{ccc}1 & 0 & 0\\0 & \cos\theta_{W} & -\sin\theta_{W}\\0 & \sin\theta_{W} & \cos\theta_{W}\end{array}\right)\left(\begin{array}{ccc}\cos\psi & 0 & \sin\psi\\0 & 1 & 0\\-\sin\psi & 0 & \cos\psi\end{array}\right)\,,
\end{equation}
where $c_{\delta}=1/\sqrt{1-\delta^{2}}$, $s_{\delta}=\delta/\sqrt{1-\delta^{2}}$,
$\cos\theta_{W}=g_{2}/\sqrt{g_{2}^{2}+g_{Y}^{2}}$, $\sin\theta_{W}=g_{Y}/\sqrt{g_{2}^{2}+g_{Y}^{2}}$.
The angle $\psi$ is
\begin{equation}
\psi=\frac{1}{2}\arctan\frac{2\delta\sqrt{1-\delta^{2}}\,\sin\theta_{W}}{1-\varepsilon^{2}-\delta^{2}(1+\sin^{2}\theta_{W})}\,,
\end{equation}
where we define $\varepsilon^{2}\equiv M_{1}^{2}/M_{Z}^{2}$ with
$M_{1}=g_{x}v_{x}$ and $M_{Z}=\tfrac{1}{2}v_{{\rm SM}}\sqrt{g_{2}^{2}+g_{Y}^{2}}$.

The full rotation matrix is thus
\begin{equation}
\mathcal{R}\equiv KO=\left(\begin{array}{ccc}
c_{\delta}\cos\psi & 0 & c_{\delta}\sin\psi\\
-s_{\delta}\cos\psi+\sin\psi\sin\theta_{W} & \cos\theta_{W} & -s_{\delta}\sin\psi-\cos\psi\sin\theta_{W}\\
-\sin\psi\cos\theta_{W} & \sin\theta_{W} & \cos\psi\cos\theta_{W}
\end{array}\right)\,,
\end{equation}
which diagonalizes the kinetic and mass matrices simultaneously and
transforms the original eigenbasis $V^{T}=(C,B,A^{3})$ to the mass
eigenbasis $E^{T}=(A^{\prime},A_{\gamma},Z)$ with $E=\mathcal{R}^{T}V$.
The resulting mass matrix is
\begin{equation}
M^{2}=\mathcal{R}^{T}M_{0}^{2}\mathcal{R}=\left(\begin{array}{ccc}
M_{A^{\prime}}^{2} & 0 & 0\\
0 & 0 & 0\\
0 & 0 & M_{Z}^{\prime2}
\end{array}\right)\,,
\end{equation}
and the interactions of gauge bosons with fermions can be obtained
from
\begin{equation}
\mathcal{L}_{{\rm int}}=(g_{x}J_{x},\;g_{Y}J_{Y},\;g_{2}J_{3})\,V=(g_{x}J_{x},\;g_{Y}J_{Y},\;g_{2}J_{3})\,\mathcal{R}E\,.
\end{equation}

The neutral current couplings are summarized as follows
\begin{align}
\mathcal{L}_{\chi} & =g_{x}Q_{x}(\mathcal{R}_{11}A_{\mu}^{\prime}+\mathcal{R}_{13}Z_{\mu})\overline{\chi}\gamma^{\mu}\chi\,,\\
\mathcal{L}_{A^{\prime},Z,A_{\gamma}} & =\frac{1}{2}\overline{f_{i}}\gamma^{\mu}\left[(v_{i}^{\prime}-a_{i}^{\prime}\gamma^{5})f_{i}A_{\mu}^{\prime}+(v_{i}-a_{i}\gamma^{5})f_{i}Z_{\mu}\right]+eQ_{i}\overline{f_{i}}\gamma^{\mu}f_{i}A_{\gamma\,\mu}\,,
\end{align}
where $eQ_{i}$ are the electric charges of the SM fermions, and
\begin{align}
v_{i} & =\left(g_{2}\mathcal{R}_{33}-g_{Y}\mathcal{R}_{23}\right)T_{i}^{3}+2g_{Y}\mathcal{R}_{23}Q_{i}\,,\\
a_{i} & =\left(g_{2}\mathcal{R}_{33}-g_{Y}\mathcal{R}_{23}\right)T_{i}^{3}\,,\\
v_{i}^{\prime} & =\left(g_{2}\mathcal{R}_{31}-g_{Y}\mathcal{R}_{21}\right)T_{i}^{3}+2g_{Y}\mathcal{R}_{21}Q_{i}\,,\\
a_{i}^{\prime} & =\left(g_{2}\mathcal{R}_{31}-g_{Y}\mathcal{R}_{21}\right)T_{i}^{3}\,.
\end{align}

\section{Gauge dependence of the scalar effective action}\label{Sec:Gauge}

In gauge theories, the Higgs effective potential as a function of
arbitrary background field is not a physical observable and is in general
gauge-dependent. Consequently, the effective potential depends
not only on the $U(1)_x$ parameters $g_{x}$, $\lambda_{x}$, and $v_{x}$,
but also on the background field value $\phi_{c}$ and on
the gauge-fixing parameter $\xi$.
In the conventional approach, one evaluates the gauge-dependent effective
potential and the resulting effective action, which leads to
gauge-dependent gravitational wave results.

In this section, we review the gauge-independent formulation of the effective
action based on the Nielsen identity.
This framework is essential for obtaining well-defined
gravitational wave predictions from cosmological phase transitions.

\subsection{The gauge-dependent scalar effective potential}\label{sec:GDEP}

The full (gauge-dependent) effective potential is given by
\begin{equation}
V_{{\rm eff}}(\phi_{c},T,\xi)=V_{0}(\phi_{c})+V_{0}^{{\rm \text{1-loop}}}(\phi_{c},\xi)+V_{T}^{{\rm {\rm \text{1-loop}}}}(\phi_{c},T,\xi)+V_{{\rm daisy}}(\phi_{c},T,\xi)\,, \label{eq: VFull}
\end{equation}
where $V_{0}$ is the tree-level potential, $V_{0}^{{\rm {\rm \text{1-loop}}}}$
and $V_{T}^{{\rm {\rm \text{1-loop}}}}$ represent the one-loop
corrections at zero and finite temperatures respectively, and $V_{{\rm daisy}}$
corresponds to the daisy contribution arising from higher-loop resummation.
The tree-level potential $V_{0}$ is written as
\begin{equation}
V_{0}(\phi_{c})=-\frac{\mu_{x}^{2}}{2}\phi_{c}^{2}+\frac{\lambda_{x}}{4}\phi_{c}^{4}\,.
\end{equation}
At zero temperature, the one-loop effective potential after dimensional
regularization is expressed as
\begin{equation}
V_{0}^{{\rm \text{1-loop}}}(\phi_{c})=\sum_{i}\pm g_{i}\frac{m_{i}^{4}(\phi_{c})}{64\pi^{2}}\Bigl[\log\Bigl(\frac{m_{i}^{2}(\phi_{c})}{\Lambda^{2}}\Bigr)-C_{i}-C_{{\rm UV}}\Bigr]\,,
\end{equation}
where the prefactor $\pm1$ corresponds to bosons and fermions respectively,
$g_{i}$ denotes the degree of freedom of particle $i$, and $\Lambda$
is the renormalization scale.
In this work, we set the scale $\Lambda$ to be of order the $U(1)_x$ gauge boson mass, $\Lambda = g_x v_x$.

The renormalization constants are $C_{i}=5/6$
for gauge bosons and $C_{i}=3/2$ for scalars and fermions, while
$C_{{\rm UV}}=\frac{2}{\epsilon}-\gamma+\log4\pi$ with $\epsilon=4-D$.
It is convenient to adopt the $\overline{\text{MS}}$ renormalization scheme\footnote{
Alternatively, one may adopt an on-shell renormalization scheme, which is
commonly used in supercooled phase transition analyses, by introducing an
additional counterterm potential $V_{\rm ct}$,
written as
\begin{equation}
V_{{\rm ct}}(\phi_{c})=-\frac{\delta\mu_{x}^{2}}{2}\phi_{c}^{2}+\frac{\delta\lambda_{x}}{4}\phi_{c}^{4}\,,
\end{equation}
which cancels the ultraviolet divergence of the one-loop correction at zero temperature.
$V_{{\rm ct}}$ is determined by imposing the tadpole and on-shell mass conditions at zero temperature
\begin{equation}
\frac{\partial(V_{{\rm CW}}+V_{{\rm ct}})}{\partial\phi_{c}}\biggr|_{\phi_{c}=v_{x}}=0\,,\qquad\;\frac{\partial^{2}(V_{{\rm CW}}+V_{{\rm ct}})}{\partial\phi_{c}^{2}}\biggr|_{\phi_{c}=v_{x}}=0\,,
\end{equation}
ensuring that the potential minimum and the field mass coincide with their tree-level values.

However, in this scheme some of the counterterms are $\xi$-dependent and may
spoil the power-counting procedure required for the gauge-independent analysis
presented in the next subsection. We therefore adopt the $\overline{\rm MS}$
scheme throughout this paper.}
to remove the ultraviolet divergence, under which the one-loop
correction at zero temperature reduces to the Coleman-Weinberg
potential, written as
\begin{align}
V_{{\rm CW}}(\phi_{c})= & \;\sum_{i}\pm g_{i}\frac{m_{i}^{4}(\phi_{c})}{64\pi^{2}}\Bigl[\log\Bigl(\frac{m_{i}^{2}(\phi_{c})}{\Lambda^{2}}\Bigr)-C_{i}\Bigr]\nonumber \\
= & \;\frac{1}{64\pi^{2}}\biggl[m_{h_{x}}^{4}\Bigl(\log\frac{m_{h_{x}}^{2}}{\Lambda^{2}}-\frac{3}{2}\Bigr)
 +3m_{A^{\prime}}^{4}\Bigl(\log\frac{m_{A^{\prime}}^{2}}{\Lambda^{2}}-\frac{5}{6}\Bigr)
 +m_{A_0^{\prime}}^{4}\Bigl(\log\frac{m_{A_0^{\prime}}^{2}}{\Lambda^{2}}-\frac{3}{2}\Bigr) \nonumber \\
& \qquad\, +m_{G}^{4}\Bigl(\log\frac{m_{G}^{2}}{\Lambda^{2}}-\frac{3}{2}\Bigr)
 -2m_{c}^{4}\Bigl(\log\frac{m_{c}^{2}}{\Lambda^{2}}-\frac{3}{2}\Bigr)\biggr]\,.\label{eq:Vcw}
\end{align}
where the contributions from the dark Higgs,
the physical dark photon $A^\prime$ with three degrees of freedom,
the unphysical scalar gauge mode $A_0^\prime$ with one degree of freedom, c.f., Eq.~(\ref{eq:mA}),
the Goldstone boson, and the ghost are all included.
The $m_{A_0^\prime}^4$ term combines with the $m_c^4$ term,
and eventually the overall coefficient of the $m_c^4$ contribution becomes $-1$.

The thermal contribution to the one-loop effective potential is given by
\begin{equation}
V_{T}^{{\rm 1-loop}}(\phi_{c},T)=\sum_{i}\pm \, g_{i}\,
\frac{T^{4}}{2\pi^{2}}\,J_{B/F}\biggl(\frac{m_{i}^{2}(\phi_{c})}{T^{2}}\biggr)\,,\label{eq:VT-1loop}
\end{equation}
where the upper (lower) sign corresponds to bosons (fermions),
and the thermal function $J$ is defined as
\begin{equation}
J_{B/F}(x^{2})\equiv\int_{0}^{\infty}{\rm d}y\,y^{2}\log\Bigl[1\mp\exp\bigl(-\sqrt{y^{2}+x^{2}}\bigr)\Bigr]\,.
\end{equation}
In the high-temperature limit ($x=m/T\ll1$), the thermal function
can be expanded as~\cite{Dolan:1973qd}
\begin{align}
J_{B}(x^{2}) & \approx-\frac{\pi^{4}}{45}+\frac{\pi^{2}}{12}x^{2}-\frac{\pi}{6}x^{3}-\frac{1}{32}x^{4}\log x^{2}+\cdots\,,\label{eq:JBB}\\
J_{F}(x^{2}) & \approx\frac{7\pi^{4}}{360}-\frac{\pi^{2}}{24}x^{2}-\frac{1}{32}x^{4}\log x^{2}+\cdots\,.\label{eq:JBF}
\end{align}
In the low-temperature regime ($x\gg1$), the expansion takes the
form~\cite{Curtin:2016urg,Laine:2016hma}
\begin{equation}
J_{B/F}(x^{2})\approx\mp\sum_{l=1}^{k}\frac{(\pm1)^{l}}{l^{2}}x^{2}K_{2}(xl)\,,\label{eq:JBl}
\end{equation}
where $K_{2}$ is the modified Bessel function of the second kind,
and $k$ denotes the order at which the series is truncated. It has
been shown that truncating at $k=2,3$ provides a good approximation~\cite{Curtin:2016urg}.

Daisy (or ring) resummation~\cite{Carrington:1991hz}, which sums the leading
infrared-divergent bosonic zero-mode contributions to all orders\footnote{Another approach is to integrate out the temperature dependence by matching to a three-dimensional EFT~\cite{Ginsparg:1980ef,Appelquist:1981vg,Nadkarni:1982kb,Farakos:1994kx,Kajantie:1995dw,Braaten:1995cm}.}
and thereby renders the finite-temperature effective potential infrared finite in regimes
where naive perturbation theory would otherwise break down, must be included
in the full thermal Higgs effective potential.
Two approaches are commonly used for daisy resummation: the
Parwani method~\cite{Parwani:1991gq} and the Arnold-Espinosa method~\cite{Arnold:1992rz}. In the Parwani method, the daisy correction
is implemented by replacing the tree-level masses in $J_{B}$ functions
with thermal masses,
\begin{equation}
m_{i}^{2}(\phi_{c})\to m_{i}^{2}(\phi_{c})+\Pi_{i}(T)\,.
\end{equation}
In the Arnold-Espinosa method that we follow in this work, an additional
term is added to the effective potential,
\begin{equation}
V_{{\rm daisy}}(\phi_{c},T)=-\frac{T}{12\pi}\sum_{i}g_{i}\Bigl\{\bigl[m_{i}^{2}(\phi_{c})+\Pi_{i}(T)\bigr]^{3/2}-\bigl[m_{i}^{2}(\phi_{c})\bigr]^{3/2}\Bigr\}\,,
\end{equation}
where the sum runs over the longitudinal gauge boson and scalar modes.
The one-loop thermal masses for the $U(1)_x$ dark sector particles are given by~\cite{Feng:2024pab}
\begin{equation}
\Pi_{A^{\prime}}=\frac{1}{3}g_{x}^{2}T^{2}+\sum_{i=1}^{n}\frac{1}{6}g_{x}^{2}Q_{i}^{2}T^{2}\,,
\qquad\Pi_{h_{x}}=\frac{1}{3}\lambda_{x}T^{2}+\frac{1}{4}g_{x}^{2}T^{2}\,,\qquad\Pi_{G}=\frac{1}{4}\lambda_{x}T^{2}\,,\label{eq: thermal mass}
\end{equation}
where the second term in $\Pi_{A^{\prime}}$ arises from chiral fermion
loops. For a Dirac fermion $\chi$ with $U(1)_{x}$ charge $Q_{x}$,
it can be treated as two chiral fermions with the same charge, one
obtains
\begin{equation}
\Pi_{A^{\prime}}=\frac{1}{3}g_{x}^{2}(1+Q_{x}^{2})T^{2}\,.\label{eq: PiZp}
\end{equation}
The two prescriptions agree at leading infrared order but can differ by
higher-order terms. Although the Parwani approach is often numerically
smoother, the Arnold--Espinosa prescription is generally preferred for a
controlled perturbative expansion and is therefore more suitable for the
gauge-independent analysis.

In the conventional gravitational wave analysis,
one starts from the full gauge-dependent effective potential in Eq.~(\ref{eq: VFull}),
and solves the bounce equation Eq.~(\ref{eq: S3EoM}) in Landau gauge $\xi=0$.
This leads to a gauge-dependent bounce solution, and consequently, gauge-dependent gravitational wave results,
preventing definitive predictions for the minimal $U(1)$ dark sector.
The results are nevertheless presented in Section~\ref{sec:GDGW},
to enable a direct comparison with the gauge-independent analysis.

In the high-temperature regime, using the expansions in Eqs.~(\ref{eq:JBB}) and~(\ref{eq:JBF}),
the finite-temperature effective potential can be conveniently approximated by the simple form
\begin{equation}
	V(\phi, T) = D\,(T^2-T_0^2)\,\phi^2 - E\,T\,\phi^3 + \frac{\lambda_x}{4}\,\phi^4\,, \label{eq: EffPo}
\end{equation}
where $T_0$ is the temperature at which $\phi=0$ ceases to be a local minimum.
A caveat is that the parametrization in Eq.~(\ref{eq: EffPo}) is valid only within the high-temperature regime
and should not be applied at intermediate or low temperatures,
as will be discussed in Section~\ref{sec:GDGW}.

\subsection{Gauge-independent effective action}\label{sec:GIEA}

As discussed above, the effective potential evaluated at a fixed background
field value is in general gauge-dependent, and therefore cannot by itself
be used to define an unambiguous gravitational wave signal.
A gauge-independent treatment can be formulated using the Nielsen identity,
which states that an infinitesimal change in the gauge parameter $\xi$ can be compensated by
a corresponding field redefinition of the background configuration.
In this sense, the $\xi$ dependence of $V_{\rm eff}$ at fixed $\phi$ is a coordinate effect in field space.
The induced shift of the background field from the field redefinition
is of higher order in the $\hbar$ expansion and therefore does not appear at the leading order.

Gauge independence of physical quantities, such as the bounce action,
the nucleation rate, and related observables,
follows when they are evaluated
on-shell, i.e., on configurations satisfying $\delta S/\delta \phi = 0$,
within a consistent loop ($\hbar$) expansion.
Under this procedure, the relevant effective action is gauge-independent,
leading to well-defined gravitational wave predictions for a given model.

In a weakly coupled $U(1)_{x}$ theory, the $\hbar$ order tracks
powers of $g_{x}^2$ and $\lambda_{x}$. Every gauge interaction
brings at least two powers of $g_{x}$ per closed loop in the
Abelian $U(1)_{x}$ theory, and scalar self-interactions bring one
power of $\lambda_x$ at one loop. Hence, each extra loop corresponds
to one extra factor of $g_{x}^{2}$ or $\lambda_x$.
With an appropriate choice of power counting, $\lambda_x \sim g_x^{\,n}$,
which depends on the temperature regime and will be specified in this subsection,
one can organize the effective potential as an expansion in powers of the $U(1)_x$
gauge coupling $g_x$.
This ordering separates leading and subleading contributions in a controlled
manner and ensures that the resulting effective action is gauge-independent at the order considered,
guaranteed by the Nielsen identity.

In this subsection, we review the gauge-independent analysis of the scalar
effective action for phase transitions at zero temperature, in the
high-temperature regime, and in the low-temperature (supercooled) regime.

\subsubsection{The field-dependent effective action}

In quantum field theory, the tree-level potential is modified by higher-order
radiative corrections, which can be calculated using the functional
method. Consider a scalar field $\phi$ coupled to a classical source
$J$, the partition function is given by
\begin{equation}
Z[J]={\rm e}^{{\rm i}W[J]}=\int\mathcal{D}\phi\,\exp\Bigl\{{\rm i}\bigl[S[\phi]+\int{\rm d}^{4}x\,J\phi\bigr]\Bigr\}\,,
\end{equation}
where $S[\phi]=\int{\rm d}^{4}x\,\mathcal{L}[\phi]$ denotes the action.
The functional derivative of $W[J]$ with respect to $J(x)$ is computed
as
\begin{equation}
\frac{\delta W[J]}{\delta J(x)}=-{\rm i}\frac{\delta\log Z[J]}{\delta J(x)}=\frac{\int\mathcal{D}\phi\,{\rm e}^{{\rm i}\int(\mathcal{L}+J\phi)}\phi(x)}{\int\mathcal{D}\phi\,{\rm e}^{{\rm i}\int(\mathcal{L}+J\phi)}}\,,
\end{equation}
corresponding to the vacuum expectation value of $\phi(x)$ in the
presence of a non-zero external source $J(x)$. This quantity defines
the classical field (mean field) $\phi_{c}(x)$ as
\begin{equation}
\frac{\delta W[J]}{\delta J(x)}=\bigl\langle\Omega\bigr|\phi(x)\bigr|\Omega\bigr\rangle_{J}\equiv\phi_{c}(x)\,.
\end{equation}
The effective action $\Gamma[\phi]$ is then defined through the Legendre
transformation of $W[J]$ as
\begin{equation}
\Gamma[\phi_{c}]\equiv W[J]-\int{\rm d}^{4}x\,J(x)\phi_{c}(x)\,,
\end{equation}
satisfying
\begin{equation}
\frac{\delta}{\delta\phi_{c}(x)}\Gamma[\phi_{c}]=-J(x)\,.
\end{equation}
When the external source $J(x)$ is set to zero, one obtains
\begin{equation}
\frac{\delta}{\delta\phi_{c}(x)}\Gamma[\phi_{c}]\Bigr|_{J=0}=0\,.
\end{equation}

For a translation-invariant vacuum state, one may obtain an $x$-independent
solution $\phi_{c}$ and express the effective action in form of
\begin{equation}
\Gamma[\phi_{c}]=-\int{\rm d}^{4}x\,V_{{\rm eff}}(\phi_{c})\,,
\end{equation}
where $V_{{\rm eff}}$ denotes the effective potential. Applying a
Fourier transformation from position to momentum space, $\tilde{\phi}(p)=\int{\rm d}^{4}x\,{\rm e}^{-{\rm i}px}\phi(x)$,
the effective potential can be recast as
\begin{equation}
V_{{\rm eff}}(\phi_{c})=-\sum_{n=0}^{\infty}\,\frac{\phi_{c}^{n}}{n!}\,\Gamma^{(n)}(p=0)\,,
\end{equation}
where $\Gamma^{(n)}$ represents the $n$-point effective vertex in
momentum space with all external momenta set to zero.

In what follows, we use $\phi_c$ to denote the background (mean) field. For a homogeneous configuration, the true vacuum at temperature $T$ is given by the constant field value $\phi_c$ that minimizes the finite-temperature effective potential $V_{\rm eff}(\phi_c,T)$ globally.

Bubble nucleation is governed by the Euclidean effective action evaluated on the bounce configuration.
In practice, $\Gamma_E$ is approximated by a derivative expansion truncated at $\mathcal{O}(\partial^2)$:
at zero temperature this yields Eq.~(\ref{eq: SE0T});
while at finite temperature it leads to Eqs.~(\ref{eq: SET}) and~(\ref{eq: S3}).
Accordingly,
the Euclidean actions $S_E$ employed later in the paper are
controlled approximations to the underlying $1$PI effective action $\Gamma[\phi]$,
evaluated on the relevant bounce configuration $\phi_b$.

\subsubsection{Phase transition at zero temperature}

The gauge-independent bubble nucleation rate at
zero temperature is derived in~\cite{Metaxas:1995ab}
by employing the power counting of the gauge coupling
and a derivative expansion of the Nielsen identity~\cite{Nielsen:1975fs}. In this subsection, we review the computation in~\cite{Metaxas:1995ab} demonstrating the gauge independence of the effective action at zero
temperature, and the extension to finite temperature will be discussed
in Section~\ref{sec:HT} and~\ref{sec:LT}.

The gauge dependence of the effective action in $R_{\xi}$ gauges governed by the Nielsen identity is expressed as
\begin{equation}
\xi\frac{\partial S_{{\rm eff}}}{\partial\xi}=-\int{\rm d}^{4}x\,\frac{\delta S_{{\rm eff}}}{\delta\phi_{c}(x)}\,C(x)\,,\label{eq:Nielsen Identity}
\end{equation}
where $\phi_{c}$ denotes the background field, $C(x)$ is the Nielsen
function, given by
\begin{equation}
C(x)=\frac{ig_{x}}{2}\int{\rm d}^{4}y\,\bigl\langle\bar{c}(x)G(x)c(y)\times\bigl(\partial_{\mu}A^{\prime\mu}(y)+\sqrt{2}\xi g_{x}\phi_{c}G(y)\bigr)\bigr\rangle\,,
\end{equation}
where $g_{x}$ is a gauge coupling, $c,\bar{c}$ and $G$ represent
the ghost, anti-ghost and Goldstone boson, respectively.
A derivative expansion of $S_{{\rm eff}}$, $C(x)$ and $\delta S_{{\rm eff}}/\delta\phi_{c}$
can then be performed in powers of the derivatives of $\phi_{c}$,
yielding
\begin{align}
S_{{\rm eff}} & =\int{\rm d}^{4}x\,\bigl[V_{{\rm eff}}(\phi_{c})+\tfrac{1}{2}Z(\phi_{c})(\partial_{\mu}\phi_{c})^{2}+\mathcal{O}(\partial^{4})\bigr]\,,\label{eq:GradSeff}\\
C(x) & =C_{0}(\phi_{c})+D(\phi)(\partial_{\mu}\phi_{c})^{2}-\partial^{\mu}\bigl[\tilde{D}(\phi_{c})\partial_{\mu}\phi_{c}\bigr]+\mathcal{O}(\partial^{4})\,,\label{eq:GradCx}\\
\frac{\delta S_{{\rm eff}}}{\delta\phi_{c}} & =\frac{\partial V_{{\rm eff}}(\phi_{c})}{\partial\phi_{c}}+\frac{1}{2}\frac{\partial Z}{\partial\phi_{c}}(\partial_{\mu}\phi_{c})^{2}-\partial^{\mu}\bigl[Z(\phi_{c})\partial_{\mu}\phi_{c}\bigr]+\mathcal{O}(\partial^{4})\,,\label{eq:GradDelSeff}
\end{align}
where $Z(\phi_{c})$ denotes the field renormalization
factor and $C_{0},D,\tilde{D}$ are Nielsen coefficients; the total derivative term containing $\tilde{D}$ is discussed in~\cite{Garny:2012cg}. Substituting
the derivative expansions given in Eqs.~(\ref{eq:GradSeff})--(\ref{eq:GradDelSeff}) into Eq.~(\ref{eq:Nielsen Identity}), the Nielsen
identity at $\mathcal{O}(\partial^{0})$ and $\mathcal{O}(\partial^{2})$ is derived as
\begin{align}
\xi\frac{\partial V_{{\rm eff}}}{\partial\xi} & =-C_{0}\frac{\partial V_{{\rm eff}}}{\partial\phi_{c}}\,,\label{eq:Nexp1}\\
\xi\frac{\partial Z}{\partial\xi} & =-C_{0}\frac{\partial Z}{\partial\phi_{c}}-2Z\frac{\partial C_{0}}{\partial\phi_{c}}+2D\frac{\partial V_{{\rm eff}}}{\partial\phi_{c}}+2\tilde{D}\frac{\partial^{2}V_{{\rm eff}}}{\partial\phi_{c}^{2}}\,,\label{eq:Nexp2}
\end{align}
where integration by parts is applied and the surface term vanishes
as follows:
\begin{align}
\int{\rm d}^{4}x\,C_{0}(\phi_{c})\partial_{\mu}\bigl[Z(\phi_{c})\partial_{\mu}\phi_{c}\bigr] & =\int{\rm d}^{4}x\,\Bigl\{\partial_{\mu}\bigl[C_{0}(\phi_{c})Z(\phi_{c})\partial_{\mu}\phi_{c}\bigr]-\bigl[\partial_{\mu}C_{0}(\phi_{c})\bigr]Z(\phi_{c})\partial_{\mu}\phi_{c}\Bigr\}\nonumber \\
 & =\int{\rm d}^{4}x\,\bigl[\partial_{\mu}C_{0}(\phi_{c})\bigr]Z(\phi_{c})\partial_{\mu}\phi_{c}\nonumber \\
 & =\int{\rm d}^{4}x\,Z(\phi_{c})\,\frac{\partial C_{0}}{\partial\phi_{c}}\,(\partial_{\mu}\phi_{c})^{2}\,.
\end{align}
Now take the loop ($\hbar$) expansion to the potential and the field strength renormalization as
\begin{equation}
V=V_{0}+\hbar V_{1}+\hbar^{2}V_{2}+\cdots\,,\qquad Z=1+\hbar Z_{1}+\hbar^{2}Z_{2}+\cdots\,.
\end{equation}
In a weakly coupled $U(1)_x$ theory, the order in $\hbar$ tracks powers of
$g_x^{2}$ and $\lambda_x$. The analysis below will proceed by expanding
$V$ and $Z$, as well as the Nielsen coefficients, in powers of the $U(1)_x$ gauge coupling $g_x$.

For the zero-temperature analysis, a Higgs-like scalar potential without
thermal effects does not generically produce a phase transition.
A first-order transition requires two minima separated by a barrier,
with the true vacuum lying below the false vacuum initially located at $\phi=0$.
One simple way to realize this in a toy setup is to choose a \emph{positive} mass term for the scalar,
\begin{equation}
V(\phi)\supset \frac{1}{2}\,\mu_x^2\phi^{2}\,,\qquad \mu_x^2>0
\end{equation}
in contrast to the usual Higgs potential with a negative mass-squared term.
Additionally, since the barrier is generated by the Coleman-Weinberg contribution,
the positive mass term $\sim \mu^{2}$ cannot be parametrically large,
otherwise the barrier will not develop.
This setup can thus support a zero-temperature phase transition and is adopted for pioneering studies~\cite{Metaxas:1995ab}.

For a gauge theory in which the gauge symmetry is broken by a Higgs field,
a crucial prerequisite for a controlled expansion is to choose an appropriate
power counting for the Higgs quartic coupling $\lambda_x$ in terms of the gauge coupling $g_x$.
In the zero-temperature regime, one commonly adopts the scaling
\begin{equation}
\lambda_x \sim g_x^{4}\,. \label{eq:counting_0T}
\end{equation}
This scaling has two key insights:
\begin{itemize}
  \item \textbf{The generation of the barrier at zero temperature:}
  In a gauged $U(1)_x$ theory, the dominant Coleman-Weinberg contribution scales as
  \begin{equation}
  V_{\rm CW}(\phi) \sim \frac{g_x^4}{16 \pi^2} \phi^4 \log \frac{g_x^2 \phi^2}{\Lambda^2}
  \sim \frac{g_x^4}{16 \pi^2} \phi^4 \,.
  \end{equation}
  The extremum condition $V^\prime(v_x) = 0$ implies
  \begin{equation}
  \mu_x^2 \sim v_x^2 \left(\lambda_x + \frac{g_x^4}{16 \pi^2}\right)\,.
  \end{equation}
  For the Coleman-Weinberg contribution to compete with the tree-level quartic
  term and thereby generate a barrier, one requires $\lambda_x \sim g_x^{4}$.
  It then follows that $\mu_x^{2} \sim g_x^{4}\phi_c^{2}$.
  \item \textbf{The validity of the derivative expansion:}
  The derivative expansion is a low-momentum expansion obtained after integrating
  out the heavy modes of the theory,
  and it is expanded in powers of the ratio of
  the characteristic momentum scale associated with spatial variations of the
  background field, $k_{\rm field}$, to the mass scale of heavy modes that have
  been integrated out, $\Lambda(\phi_{\rm wall})$. Here $\phi_{\rm wall}$
  denotes the typical field value within the bubble wall.
  The derivative expansion is thus valid when $k_{\rm field} / \Lambda(\phi_{\rm wall}) \ll 1$,
  i.e., $k_{\rm field} \ll \Lambda(\phi_{\rm wall})$.

  The characteristic momentum is estimated as
  \begin{equation}
  k_{\rm field} \sim \frac{1}{L_{\rm wall}} \sim \sqrt{V^{\prime \prime}(\phi_{\rm wall})}\,, 
  \end{equation}
  where $L_{\rm wall}$ is the characteristic thickness of the bubble wall.
  For the zero-temperature phase transition,
  \begin{equation}
  k_{\rm field} 
  \sim \frac{g_x^2}{4\pi} \, \phi_{\rm wall}\,,
  \end{equation}
  which is parametrically much smaller than the scale at which the dark photon has
  been integrated out, $\Lambda = g_x\,\phi_{\rm wall}$,
  and thus the derivative expansion is valid at zero temperature.
\end{itemize}

At zero temperature, the effective potential $V_{{\rm eff}}$ is given by
\begin{equation}
V^{0{\rm T}}_{{\rm eff}} =V^{0{\rm T}}_{g_{x}^{4}}+V^{0{\rm T}}_{g_{x}^{6}}+\mathcal{O}(g_{x}^{8})\,,
\end{equation}
with the leading and subleading effective potentials written as
\begin{align}
V^{0{\rm T}}_{g_{x}^{4}} & =\frac{\mu_{x}^{2}}{2}\phi_{c}^{2}+\frac{\lambda_{x}}{4}\phi_{c}^{4}+3\,\frac{m_{A^{\prime}}^{4}}{64\pi^{2}}\Bigl(\log\frac{m_{A^{\prime}}^{2}}{\Lambda^{2}}-\frac{5}{6}\Bigr)\,,\label{eq:Vg4_0T}\\
V^{0{\rm T}}_{g_{x}^{6}} & =\frac{m_{G}^{4}}{64\pi^{2}}\Bigl(\log\frac{m_{G}^{2}}{\Lambda^{2}}-\frac{3}{2}\Bigr)-\frac{m_{c}^{4}}{64\pi^{2}}\Bigl(\log\frac{m_{c}^{2}}{\Lambda^{2}}-\frac{3}{2}\Bigr)\,,\label{eq:Vg6_0T}
\end{align}

The field strength renormalization $Z$ is expanded in powers of the gauge coupling $g_{x}$ as
\begin{align}
Z &=1+Z_{g_{x}^{2}}+\mathcal{O}(g_{x}^{4})\,.\nonumber\\
&= 1+ \frac{g_{x}^{2}}{16\pi^{2}}\Bigl(\xi\log\frac{m_{c}^{2}}{\Lambda^{2}}+3\log\frac{m_{A^{\prime}}^{2}}{\Lambda^{2}}+\xi\Bigr)
+\mathcal{O}(g_{x}^{4})\,,
\label{eq:Z_0T}
\end{align}
and perturbation analysis shows that the Nielsen coefficient $C_{0}(\phi_{c})$ starts at order
$g_{x}^{2}$ and $D,\tilde{D}$ starts at unity, expressed as
\begin{align}
C_{0} & =C_{g_{x}^{2}}+\mathcal{O}(g_{x}^{4}) \nonumber\\
 &= -\frac{\xi g_{x}^{2}\phi_{c}}{32\pi^{2}}\,\log\frac{m_{c}^{2}}{\Lambda^{2}}+\mathcal{O}(g_{x}^{4})\,,\label{eq:C_0T}\\
D,\tilde{D} & =1+\mathcal{O}(g_{x}^{2})\,.
\end{align}
The leading and subleading contributions to the Nielsen identity are thus obtained as follows
\begin{align}
\xi\frac{\partial V^{0{\rm T}}_{g_{x}^{4}}}{\partial\xi} & =0\,,\label{eq:NI-0T-Vg4}\\
\xi\frac{\partial V^{0{\rm T}}_{g_{x}^{6}}}{\partial\xi} & =-C_{g_{x}^{2}}\frac{\partial V^{0{\rm T}}_{g_{x}^{4}}}{\partial\phi_{c}}\,,\label{eq:NI-0T-Vg6}\\
\xi\frac{\partial Z_{g_{x}^{2}}}{\partial\xi} & =-2\frac{\partial C_{g_{x}^{2}}}{\partial\phi_{c}}\,,\label{eq:NI-0T-Zg2}
\end{align}
which hold analytically.

At zero temperature, the tunneling rate is governed by the four-dimensional Euclidean action
\begin{equation}
S_E(\phi) = \int {\rm d}^4 x_E \left[\frac{1}{2}\,(\partial_\mu \phi)^2 + V(\phi)\right]\,,
\qquad
x_E=(\tau,\vec{x})\,. \label{eq: SE0T}
\end{equation}
Varying $S_E$ with respect to $\phi$ leads to a partial differential equation for the bounce configuration.
The solution that extremizes $S_E$ is invariant under rotations in Euclidean space and hence possesses an $O(4)$ symmetry.
It thus depends only on the radial coordinate $\rho = \sqrt{\tau^2 + x^2 + y^2 +z^2}$, i.e., $\phi=\phi(\rho)$.
The field equation then reduces to the ordinary differential equation
\begin{equation}
\phi^{\prime \prime}(\rho) + \frac{3}{\rho} \, \phi^{\prime}(\rho) = \frac{{\rm d}V}{{\rm d}\phi}\,,
\end{equation}
subject to boundary conditions
\begin{equation}
\phi\,(\rho\to\infty)=0,
\qquad
\left.\frac{{\rm d}\phi}{{\rm d}\rho}\right|_{\rho=0}=0\,.
\end{equation}

The gauge-independent procedure for computing the zero-temperature bubble nucleation rate proceeds
by determining the bounce solution $\phi_{b}$~\cite{Coleman:1977py,Callan:1977pt}
from the leading-order effective potential, which is automatically independent of the gauge-fixing parameter $\xi$
\begin{equation}
\Box\phi_{b}=\frac{\partial V^{0{\rm T}}_{g_{x}^{4}}}{\partial\phi_{c}}\biggr|_{\phi_{c}=\phi_{b}}\,,\qquad\phi_{b}(\infty)=0\,,\qquad\phi_{b}^{\prime}(0)=0\,.
\end{equation}

The nucleation rate is then obtained as
\begin{equation}
\Gamma=\mathcal{A}\,{\rm e}^{-(S^{0{\rm T}}_{0}+S^{0{\rm T}}_{1})}\,,
\end{equation}
where the coefficient $\mathcal{A}$ is expressed in terms of functional
determinants associated with the characteristic mass scales of the
theory, $S^{0{\rm T}}_{0}$ and $S^{0{\rm T}}_{1}$ denote the Euclidean action of the bounce
solution, given by
\begin{align}
S^{0{\rm T}}_{0} & =\int{\rm d}^{4}x\,\Bigl[V^{0{\rm T}}_{g_{x}^{4}}(\phi_{b})+\frac{1}{2}(\partial_{\mu}\phi_{b})^{2}\Bigr]\,,\\
S^{0{\rm T}}_{1} & =\int{\rm d}^{4}x\,\Bigl[V^{0{\rm T}}_{g_{x}^{6}}(\phi_{b},\xi)+\frac{1}{2}Z_{g_{x}^{2}}(\phi_{b},\xi)\,(\partial_{\mu}\phi_{b})^{2}\Bigr]\,.
\end{align}
$S_0$ is gauge-independent since $V^{0{\rm T}}_{g_{x}^{4}}$ is $\xi$-independent.
Using Eqs.~(\ref{eq:Vg6_0T}) and (\ref{eq:Z_0T}), one finds
\begin{align}
\xi\frac{\partial S^{0{\rm T}}_{1}}{\partial\xi} & =\int{\rm d}^{4}x\,\Bigl[\xi\frac{\partial V^{0{\rm T}}_{g_{x}^{6}}}{\partial\xi}+\frac{1}{2}\xi\frac{\partial Z_{g_{x}^{2}}}{\partial\xi}\,(\partial_{\mu}\phi_{b})^{2}\Bigr]\nonumber \\
 & =-\int{\rm d}^{4}x\,\Bigl[C_{g_{x}^{2}}\frac{\partial V^{0{\rm T}}_{g_{x}^{4}}}{\partial\phi_{c}}+\frac{\partial C_{g_{x}^{2}}}{\partial\phi_{c}}\,(\partial_{\mu}\phi_{b})^{2}\Bigr]\biggr|_{\phi_{c}=\phi_{b}}\nonumber \\
 & =-\int{\rm d}^{4}x\Bigl[C_{g_{x}^{2}}\frac{\partial V^{0{\rm T}}_{g_{x}^{4}}}{\partial\phi_{c}}+(\partial_{\mu}C_{g_{x}^{2}})(\partial_{\mu}\phi_{b})\Bigr]\biggr|_{\phi_{c}=\phi_{b}}\nonumber \\
 & =-\int{\rm d}^{4}x\,C_{g_{x}^{2}}\Bigl[\frac{\partial V^{0{\rm T}}_{g_{x}^{4}}}{\partial\phi_{c}}-\Box\phi_{b}\Bigr]\biggr|_{\phi_{c}=\phi_{b}}\nonumber \\
 & =0\,,
\end{align}
the bounce solution $\phi_{b}$ ensures the gauge independence of the subleading effective action $S^{0{\rm T}}_{1}$.

\subsubsection{Phase transition at high temperatures ($T\gg m_{A^{\prime}}$)}\label{sec:HT}

High-temperature phase transitions typically occur at temperatures only
slightly below the critical temperature $T_c$, which is generally much larger
than the $U(1)_x$ gauge boson mass.
In this situation, the
percolation temperature that characterizes the onset of the phase transition is
usually close to the nucleation temperature, $T_p \approx T_n$, as will be discussed in Section~\ref{sec:CT}.
At $T_n$, the background field value satisfies $\phi_c \gtrsim v_x$,
and the transition temperature is much larger than the particle masses in the $U(1)_x$ sector, i.e.\ $T_n \gg m_{A^\prime}$.
In this regime, the calculation can be simplified by expanding in $m/T$ and
neglecting higher-order terms.

The bubble nucleation rate
at finite temperature is given by
\begin{align}
\Gamma(T) & =\mathcal{A}(T)\,{\rm e}^{-S_{3}(T)/T}\,,\\
S_{3}(T) & =\int{\rm d}^{3}x\,\bigl[V_{{\rm eff}}(\phi_{c},T)+\tfrac{1}{2}Z(\phi_{c},T)(\partial_{i}\phi_{c})^{2}+\mathcal{O}(\partial^{4})\bigr]\,,
\end{align}
where $\phi_{c}$ denotes the background field, $Z(\phi_{c},T)$ is
the field renormalization factor, and the prefactor $\mathcal{A}(T)$
will be given in Eq.~(\ref{eq:AT}).

We now implement  the power counting procedure
for the high-temperature phase transition,
following~\cite{Garny:2012cg,Lofgren:2021ogg,Hirvonen:2021zej}.
The nucleation temperature $T_{n}$ marks the
onset of the phase transition and lies slightly below the critical
temperature $T_{c}$, at which the two minima of the effective potential
become degenerate. Therefore $T_{c}$ can be taken as a representative
characteristic temperature of the transition. Including the one-loop
thermal corrections to $V_{{\rm eff}}$ and employing the high-temperature
expansion in Eq.~(\ref{eq:JBB}), the leading contribution to the effective potential becomes
\begin{align}
V_{{\rm LO}}= & -\frac{\mu_{x}^{2}}{2}\phi_{c}^{2}+\frac{\lambda_{x}}{4}\phi_{c}^{4}+\frac{T^{2}}{24}(m_{h_{x}}^{2}+3m_{A^{\prime}}^{2}+m_{G}^{2}-m_{c}^{2})+\cdots\nonumber \\
= & -\frac{1}{2}\mu_{{\rm eff}}^{2}\phi_{c}^{2}+\frac{\lambda_{x}}{4}\phi_{c}^{4}\cdots\,,
\end{align}
where particle masses are given in Eqs.~(\ref{eq:mA})--(\ref{eq:mG}), and
\begin{equation}
-\mu_{{\rm eff}}^{2}=-\mu_{x}^{2}+\bigl(\tfrac{1}{3}\lambda_{x}+\tfrac{1}{4}g_{x}^{2}\bigr)T^{2}\,.
\end{equation}
The negative mass term $-\mu_{x}^{2}$ competes with the positive
$T^{2}$ corrections, leading to a critical crossover temperature
$T_{o}$ where the Higgs effective mass is zero, i.e., $\mu_{{\rm eff}}^{2}=0$.
This temperature is just slightly below $T_c$ and a little bit above $T_n$, i.e.,
$T_c \gtrsim T_o \gtrsim T_n$.
Near the high-temperature phase transition, all terms in the potential are expected
to be of the same order of magnitude~\cite{Arnold:1992rz,Arunasalam:2021zrs},\footnote{
Using the high-temperature effective thermal potential given in Eq.~(\ref{eq: EffPo}),
the critical temperature $T_c$ is defined by
\begin{equation}
V^\prime(\phi,T_c) = 0\,,\qquad V(\phi_c, T_c) = V(0,T_c)\,,
\end{equation}
where $\phi_c \neq 0$ denotes the broken-phase minimum at $T_c$.
Solving these conditions yields
\begin{equation}
\phi_c = \frac{2 E}{\lambda_x}T_c\,,\qquad D(T^2-T_0^2) = \frac{E^2}{\lambda_x}T_c^2\,,\qquad{\rm at}\ T_c\,. \label{eq: SolTc}
\end{equation}
Plugging back into the potential Eq.~(\ref{eq: EffPo}) one finds
all three terms are the same parametric size $\sim E^4 T_c^4 / \lambda_x^3$.}
At $T_n$ which lies just slightly below $T_c$,
relation Eq.~(\ref{eq: SolTc}) approximately holds and thus
\begin{equation}
\frac{\phi_c}{T_n} \approx \frac{2E}{\lambda_x} \sim \mathcal{O}(1)\,.
\end{equation}
In a gauge theory one typically has $E\sim g_x^3$,
leading to the following power counting relation for high-temperature phase transitions:
\begin{equation}
\lambda_x \sim g_{x}^{3}\,,\qquad\mu_{{\rm eff}}^{2}\sim g_{x}^{3}T^{2}\,,\qquad T\sim \phi_{c}\,.\label{eq:counting_HT}
\end{equation}
For the scaling adopted above, the derivative expansion expanded in powers of $k_{\rm field}/\Lambda$,
where the characteristic momentum scale associated with field variations is
$k_{\rm field} \sim \sqrt{V^{\prime \prime}} \sim g^{3/2} \phi_c$, and the cutoff scale
$\Lambda$ is set by the non-zero Matsubara modes, $\Lambda \sim 2\pi T$.
Hence,
\begin{equation}
\frac{k_{\rm field}}{\Lambda}\sim \frac{g_x^{3/2}}{2\pi} \ll 1\,,
\end{equation}
ensuring that the derivative expansion is well controlled in the high-temperature regime.

In practice, our parameter scan shows that a high-temperature phase transition
can also occur for $\lambda_x \ll g_x^{3}$. In this case, the nucleation
temperature $T_n$ can lie well below $T_c$,
while the high-temperature expansion with $m_{A^\prime}/T_n \ll 1$ remains reasonably applicable.

With the thermal corrections, the corrected field-dependent masses become
\begin{align}
m_{A^{\prime}}^{2} & =g_{x}^{2}\phi_{c}^{2}\,, \quad
m_{A^{\prime}_0}^{2} = \xi g_{x}^{2}\phi_{c}^{2}\,, \quad
m_{c}^{2}=\xi g_{x}^{2}\phi_{c}^{2}\,,\\
m_{h_{x}}^{2} & =-\mu_{{\rm eff}}^{2}+3\lambda_{x}\phi_{c}^{2}\,,\\
m_{G}^{2} & =-\mu_{{\rm eff}}^{2}+\lambda_{x}\phi_{c}^{2}+\xi g_{x}^{2}\phi_{c}^{2}\,.
\end{align}

After expanding the one-loop thermal contribution as in Eq.~(\ref{eq:JBB})
and applying the power counting introduced in Eq.~(\ref{eq:counting_HT}),
the leading-order effective potential is obtained as
\begin{align}
V^{\rm HT}_{g_{x}^{3}}= & -\frac{\mu_{x}^{2}}{2}\phi_{c}^{2}+\frac{\lambda_{x}}{4}\phi_{c}^{4}+\frac{T^{2}}{24}(m_{h_{x}}^{2}+3m_{A^{\prime}}^{2}+m_{G}^{2}-m_{c}^{2})\nonumber \\
 & -\frac{T}{12}\bigl[2m_{A^{\prime}}^{3}+(m_{A^{\prime}}^{2}+\Pi_{A^{\prime}})^{3/2}\bigr]\nonumber \\
= & -\frac{1}{2}\mu_{{\rm eff}}^{2}\phi_{c}^{2}+\frac{\lambda_{x}}{4}\phi_{c}^{4}-\frac{g_{x}^{3}T}{12\pi}\bigl[2\phi_{c}^{3}+\bigl(\tfrac{1}{3}T^{2}+\phi_{c}^{2}\bigr)^{3/2}\bigr]\,.\label{eq:Vg3_highT}
\end{align}
To evaluate the next-to-leading order contribution, we first examine
the Coleman-Weinberg potential, c.f., Eq.~(\ref{eq:Vcw}),
\begin{equation}
V_{{\rm CW}}=F(m_{h_{x}}^{2})+3F(m_{A^{\prime}}^{2})+F(m_{G}^{2})-F(m_{c}^{2})\,,
\end{equation}
where
\begin{equation}
F(m_{i}^{2}) \equiv \frac{m_{i}^{4}}{64\pi^{2}}\Bigl(\log\frac{m_{i}^{2}}{\Lambda^{2}}-C_{i}\Bigr)\,.
\end{equation}
Among these terms, the gauge-independent contributions are $F(m_{h_{x}}^{2})\sim\mathcal{O}(g_{x}^{6})$
and $F(m_{A^{\prime}}^{2})\sim\mathcal{O}(g_{x}^{4})$. We then perform
a Taylor expansion $f(x)=f(a)+f^{\prime}(a)\,(x-a)+\cdots$ around
$m_{c}^{2}$, and we obtain
\begin{equation}
F(m_{G}^{2})-F(m_{c}^{2})\simeq F^{\prime}(m_{c}^{2})\,(m_{G}^{2}-m_{c}^{2})\sim\xi\,\mathcal{O}(g_{x}^{5})\,,
\end{equation}
where $F^{\prime}(m_{c}^{2})\sim\mathcal{O}(m_{c}^{2})$.
The remaining terms of the effective potential are
\begin{equation}
-\frac{T}{12\pi}\bigl[(m_{h_{x}}^{2}+\Pi_{h_{x}})^{3/2}
+(m_{G}^{2}+\Pi_{G})^{3/2}-m_{G}^{3}
+\underline{m_{G}^{3}-m_{c}^{3}}
\bigr]\,,
\end{equation}
where the first two $3/2$-power terms arise from daisy resummation,
while the two underlined terms originate from the high-temperature expansion of the
bosonic thermal function $J_B$, c.f., Eq.~(\ref{eq:JBB}).
The term $(m_{h_{x}}^{2}+\Pi_{h_{x}})^{3/2}\sim\mathcal{O}(g_{x}^{9/2})$
is of higher order in the power counting than $\mathcal{O}(g_x^{4})$ and is
therefore neglected. The gauge-dependent part is evaluated as
\begin{align}
(m_{G}^{2}+\Pi_{G})^{3/2}-m_{c}^{3} & =
\bigl(-\mu_{{\rm eff}}^{2}+\lambda_{x}\phi_{c}^{2}+\xi g_{x}^{2}\phi_{c}^{2} + \tfrac{1}{4}\lambda_{x}T^{2}\bigr)^{3/2}
-\bigl( \xi g_{x}^{2}\phi_{c}^{2} \bigr)^{3/2}\nonumber \\
& \approx \bigl(-\mu_{{\rm eff}}^{2}+\lambda_{x}\phi_{c}^{2}+\xi g_{x}^{2}\phi_{c}^{2}\bigr)^{3/2}
-\bigl(\xi g_{x}^{2}\phi_{c}^{2} \bigr)^{3/2} \sim\sqrt{\xi}\,\mathcal{O}(g_{x}^{4})\,,
\end{align}
where $m_{c}^{2}=\xi g_{x}^{2}\phi_{c}^{2}$ is much larger than
$-\mu_{{\rm eff}}^{2}+\lambda_{x}\phi_{c}^{2}+ \frac{1}{4}\lambda_{x}T^{2}$,
for $\xi \gtrsim g_{x}$.
For the limit $\xi\ll g_{x}$,
the gauge-dependent contribution becomes negligible.
In either case, this contribution is counted as a second-order effect and is
therefore included in the $\mathcal{O}(g_x^{4})$ potential, $V_{g_x^{4}}$.

Therefore, the next-to-leading order potential is written as
\begin{equation}
V^0_{g_{x}^{4}}
=3\,\frac{m_{A^{\prime}}^{4}}{64\pi^{2}}\Bigl(\log\frac{m_{A^{\prime}}^{2}}{\Lambda^{2}}-\frac{5}{6}\Bigr)-\frac{T}{12\pi}\left(m_{G}^{3}-m_{c}^{3}\right)\,.
\end{equation}

It was pointed out in~\cite{Metaxas:1995ab} that, under a specific choice of
power counting, certain higher-loop diagrams can yield sizable contributions.
In particular, diagrams containing transverse photon loops along a
scalar propagator do not increase the overall parametric order of the amplitude,
and therefore may contribute as the same order as lower-loop terms.
These contributions can be summed by replacing the tree-level Goldstone
mass with a \textit{dressed} mass~\cite{Metaxas:1995ab},
\begin{equation}
m_{G}^{2}\to\widetilde{m}_{G}^{2}\equiv\frac{1}{\phi_{c}}\frac{\partial V^{\rm HT}_{g_{x}^{3}}}{\partial\phi_{c}}+\xi g_{x}^{2}\phi_{c}^{2}\,.
\end{equation}
After the substitution, the subleading potential is written as
\begin{align}
\widetilde{V}_{g_{x}^{4}}^{0}= & \ 3\frac{m_{A^{\prime}}^{4}}{64\pi^{2}}\Bigl(\log\frac{m_{A^{\prime}}^{2}}{\Lambda^{2}}-\frac{5}{6}\Bigr)-\frac{T}{12\pi}\left(\widetilde{m}_{G}^{3}-m_{c}^{3}\right)\nonumber \\
\simeq & \ 3\frac{m_{A^{\prime}}^{4}}{64\pi^{2}}\Bigl(\log\frac{m_{A^{\prime}}^{2}}{\Lambda^{2}}-\frac{5}{6}\Bigr)-\frac{Tg_{x}\phi_{c}\sqrt{\xi}}{8\pi}\Bigl[-\mu_{{\rm eff}}^{2}+\lambda_{x}\phi_{c}^{2}-\frac{g_{x}^{3}T}{4\pi}\Bigl(2\phi_{c}+\sqrt{\tfrac{1}{3}T^{2}+\phi_{c}^{2}}\Bigr)\Bigr]\,.\label{eq:Vg4_highT}
\end{align}

For nucleation-related observables, one typically requires higher accuracy
than that provided by the leading order barrier alone. The next-to-leading order (NLO) effective
potential is derived within a three-dimensional finite-temperature effective field theory (EFT) framework.
The three-dimensional effective potential is computed in a constant background field $\phi_3$ using three-dimensional perturbation theory:
the leading-order piece is the tree-level potential plus the leading three-dimensional one-loop contributions,
while the NLO piece consists of the relevant two-loop vacuum diagrams in the three-dimensional EFT
together with the corrections induced by NLO-matched parameters.
After renormalization and consistent power counting, one arrives at~\cite{Lofgren:2021ogg,Hirvonen:2021zej}
\begin{align}
V^{\rm HT}_{g_{x}^{4}} & =-\frac{Tg_{x}\phi_{c}\sqrt{\xi}}{8\pi}\Bigl[-\mu_{{\rm eff}}^{2}+\lambda_{x}\phi_{c}^{2}-\frac{g_{x}^{3}T}{4\pi}\Bigl(2\phi_{c}+\sqrt{\tfrac{1}{3}T^{2}+\phi_{c}^{2}}\Bigr)\Bigr]\nonumber \\
 & +\frac{1}{(4\pi)^{2}}g_{x}^{4}T^{2}\phi_{c}^{2}\Bigl[-\frac{3}{2}+\ln\Bigl(\frac{4g_{x}^{2}\phi_{c}^{2}}{\Lambda^{2}}\Bigr)+\frac{1}{2}\ln\Bigl(\frac{4g_{x}^{2}(\tfrac{1}{3}T^{2}+\phi_{c}^{2})}{\Lambda^{2}}\Bigr)\Bigr]\,,
\end{align}
which will be used for the high-temperature phase transition analysis in the following sections.

With the above discussion, the gauge-independent effective action can be thus constructed.
The effective potential and the field renormalization
factor are expanded in powers of the coupling $g_{x}$ as
\begin{align}
V^{\rm HT}_{{\rm eff}} & =V^{\rm HT}_{g_{x}^{3}}+V^{\rm HT}_{g_{x}^{4}}+\mathcal{O}(g_{x}^{9/2}T^{4})\,,\\
Z & =1+Z_{g_{x}}+\mathcal{O}(g_{x}^{3/2})\:,
\end{align}
where $Z_{g_{x}}$ is given in~\cite{Garny:2012cg} as
\begin{equation}
Z_{g_{x}}=\frac{g_{x}T}{48\pi}\biggl[-\frac{22}{\phi_{c}}+\frac{\phi_{c}^{2}}{\bigl(\frac{1}{3}T^{2}+\phi_{c}^{2}\bigr)^{3/2}}\biggr]\,.\label{eq:Zg}
\end{equation}
The leading-order potential $V_{g_{x}^{3}}^{\rm HT}$, c.f., Eq.~(\ref{eq:Vg3_highT}), is manifestly gauge-independent, and the corresponding bounce solution
$\phi_{b}$ satisfies
\begin{equation}
\nabla^{2}\phi_{b}=\frac{\partial V^{\rm HT}_{g_{x}^{3}}}{\partial\phi_{c}}\biggr|_{\phi_{c}=\phi_{b}}\,,\qquad\phi_{b}(\infty)=0\,,\qquad\phi_{b}^{\prime}(0)=0\,.
\end{equation}
The nucleation rate then takes the form
\begin{align}
\Gamma & =\mathcal{A}\,{\rm e}^{-(S^{\rm HT}_{0}+S^{\rm HT}_{1})/T}\,,\\
S^{\rm HT}_{0} & =\int{\rm d}^{3}x\,\Bigl[V^{\rm HT}_{g_{x}^{3}}(\phi_{b})+\frac{1}{2}(\partial_{i}\phi_{b})^{2}\Bigr]\,,\\
S^{\rm HT}_{1} & =\int{\rm d}^{3}x\,\Bigl[V^{\rm HT}_{g_{x}^{4}}(\phi_{b})+\frac{1}{2}Z_{g_{x}}(\partial_{i}\phi_{b})^{2}\Bigr]\,.
\end{align}
The leading-order effective action is gauge-independent, since
the leading-order effective potential $V_{g_{x}^{3}}$ is constructed to be $\xi$-independent.

To analyze the gauge dependence of $V_{g_{x}^{4}}$ and $S_{1}$,
we employ the finite-temperature Nielsen identity and expand its coefficients
\begin{equation}
C=C_{g_{x}}+\mathcal{O}(g_{x}^{3/2})\,,\quad D,\tilde{D}=\mathcal{O}(g_{x}^{-1})\,,
\end{equation}
which yields
\begin{align}
\xi\frac{\partial V^{\rm HT}_{g_{x}^{4}}}{\partial\xi} & =-C_{g_{x}}\frac{\partial V^{\rm HT}_{g_{x}^{3}}}{\partial\phi_{c}}\,,\label{eq:Nielsen_T1}\\
\xi\frac{\partial Z_{g_{x}}}{\partial\xi} & =-2\frac{\partial C_{g_{x}}}{\partial\phi_{c}}\,,\label{eq:Nielsen_T2}
\end{align}
where the leading coefficient $C_{g_{x}}=\sqrt{\xi}g_{x}T/(16\pi)$~\cite{Garny:2012cg} is independent of the background field $\phi_{c}$,
i.e., $\partial C_{g_{x}}/\partial\phi_{c}=0$. Together with the
gauge-independent renormalization factor $Z_{g_{x}}$ in Eq.~(\ref{eq:Zg}), such that $\partial Z_{g_{x}}/\partial\xi=0$,
they ensure consistency with the Nielsen relation in Eq.~(\ref{eq:Nielsen_T2}).
Finally, applying Gauss's theorem and the asymptotic behavior of $\phi_{b}$,
one arrives
\begin{align}
\xi\frac{\partial S^{\rm HT}_{1}}{\partial\xi} & =\int{\rm d}^{3}x\,\Bigl[\xi\frac{\partial V^{\rm HT}_{g_{x}^{4}}}{\partial\xi}+\frac{1}{2}\xi\frac{\partial Z_{g_{x}}}{\partial\xi}(\partial_{i}\phi_{b})^{2}\Bigr]\nonumber \\
 & =\int{\rm d}^{3}x\,\Bigl[-C_{g_{x}}\frac{\partial V^{\rm HT}_{g_{x}^{3}}}{\partial\phi_{c}}+0\Bigr]\biggr|_{\phi_{c}=\phi_{b}}\nonumber \\
 & =-C_{g_{x}}\int{\rm d}^{3}x\,\nabla^{2}\phi_{b}\nonumber \\
 & =-C_{g_{x}}\int{\rm d}^{2}S\,\partial\phi_{b}\nonumber \\
 & =0\,.\label{eq:GI_T}
\end{align}
In the high-temperature regime, Eqs.~(\ref{eq:Nielsen_T1}) and~(\ref{eq:Nielsen_T2})
are satisfied explicitly by the analytic expressions given above, demonstrating
that the resulting high-temperature effective action is analytically gauge-independent.

\subsubsection{Phase transition at low temperatures ($T\ll m_{A^{\prime}}$)}\label{sec:LT}

The supercooled phase transition typically takes place at $T_p \ll m_{A^\prime}$.
In this regime, the nontrivial temperature-dependent contributions relevant to
the transition are generally subdominant to the pure field-dependent terms,
since $T_p \ll \phi_c(T_p) \sim v_x$.
In this case, however, unlike in the high-temperature regime,
the low-temperature thermal contribution does not admit a well-defined analytic expansion
in the small parameter $T/m$.
Moreover, the thermal effects are significant only when the field value approaches zero.
The basic strategy is therefore to treat the thermal corrections
as small perturbations added to the zero-temperature effective potential,
as was done in~\cite{Feng:2025wvc}.

Unlike the zero-temperature toy setup with a positive Higgs mass term, a
gauged $U(1)$ Higgs sector contains the standard tachyonic mass-squared term.
In the supercooled regime, thermal effects, although being small since $T_p \ll v_x$,
can lift the effective mass term positive and thereby create and maintain a false vacuum at  $\phi=0$,
as long as the tachyonic mass-squared parameter $\mu_x^2$ is parametrically small.

In addition to turning the effective mass term positive, a successful
supercooled phase transition requires thermal effects to generate a local barrier near $\phi=0$.
This barrier arises from the one-loop thermal contribution together with daisy resummation,
through the thermal cubic term $\sim T\,\phi^{3}$.
Although this term is parametrically suppressed in the low-temperature regime
since $T_p \ll v_x$, it can still induce a barrier in the vicinity of $\phi=0$, where $T_p \sim g_x \phi$.
Consequently, the thermally generated barrier peaks at field values of order $\phi_{\rm barrier} \sim {T_p}/{g_x}$
in the supercooled phase transition.

The thermal cubic term $\sim T\,\phi^{3}$ responsible for generating the
barrier is loop induced and is parametrically suppressed, especially in the low-temperature regime.
For this term to compete with the quartic contribution,
the quartic coupling $\lambda_x$ must itself be loop sized.
In particular, $\lambda_x$ should not exceed the size implied by the Coleman-Weinberg contribution:
\begin{equation}
\frac{\lambda_x}{4} \phi^4 \lesssim \frac{m_{A^\prime}^4}{64 \pi^2}\log \frac{m_{A^\prime}^2}{\Lambda^2}\,,
\end{equation}
which implies
\begin{equation}
\lambda_x \lesssim \mathcal{O}\left(\frac{g_x^4}{16\pi^2}\right)\,.
\end{equation}

For this scaling, the characteristic momentum scale associated with spatial
variations of the background field can be estimated as
\begin{equation}
k_{\rm field} \sim \sqrt{\lambda_x}\,\phi_{\rm wall} \sim \frac{g_x^{2}}{4\pi}\,
\phi_{\rm wall}\,,
\end{equation}
whereas the heavy scale that has been integrated out is set by the mass of
the $U(1)_x$ gauge boson, $\Lambda \sim g_x \phi_{\rm wall}$.
This implies
\begin{equation}
\frac{k_{\rm field}}{\Lambda}\sim \frac{g_x}{4\pi} \ll 1\,,
\end{equation}
and thus the derivative expansion remains parametrically controlled in the
low-temperature regime, even in the $\phi \to 0$ region.

Thus for the supercooled phase transition, we use the scaling~\cite{Feng:2025wvc}
to organize the field-dependent terms in the effective potential
\begin{equation}
\lambda_x \sim g_{x}^{4}\,,\qquad \mu_{x}^{2}\sim g_{x}^{4}\phi_{c}^{2}\,,
\label{eq:counting_LT}
\end{equation}

For the thermal contributions,
as $T\to 0$, the daisy contribution scales as $V_{\rm daisy}\sim T\,\phi_c^{3}$,
while the one-loop thermal contribution is further suppressed exponentially,
\begin{equation}
\frac{T^{4}}{2\pi^{2}} \,J_{B}\!\left(\frac{m^{2}}{T^{2}}\right) \sim T^{4}\, {\rm e}^{-m/T}\,.
\end{equation}
Therefore, away from the barrier region, thermal effects are generally
negligible since $T \ll \phi_c(T)$, and all thermal contributions can be
safely neglected relative to the purely field-dependent terms.

Nevertheless,
thermal effects are essential for generating the barrier around
$\phi \sim T_p/g_x$, enabling the supercooled phase transition. In the
vicinity of $\phi \sim T_p/g_x$, the temperature-dependent contributions can
be comparable to the field-dependent terms. We therefore retain all thermal
contributions in both the leading and subleading orders of effective potentials.
Consequently,
the leading and subleading-order effective potentials are thus written as
\begin{align}
V^{\rm LT}_{g_{x}^{4}} & = -\frac{\mu_{x}^{2}}{2}\phi_{c}^{2}+\frac{\lambda_{x}}{4}\phi_{c}^{4}+3\frac{m_{A^{\prime}}^{4}}{64\pi^{2}}\Bigl(\log\frac{m_{A^{\prime}}^{2}}{\Lambda^{2}}-\frac{5}{6}\Bigr) \nonumber\\
&\quad + V_{{\rm daisy}}^{A^{\prime}}(\phi,T) 
+\frac{T^{4}}{2\pi^{2}}\left[J_{B}\Bigl(\frac{m_{h_{x}}^{2}}{T^{2}}\Bigl)+3J_{B}\Bigl(\frac{m_{A^{\prime}}^{2}}{T^{2}}\Bigl)\right]\,,\\
V^{\rm LT}_{g_{x}^{6}} & = \frac{m_{G}^{4}}{64\pi^{2}}\Bigl(\log\frac{m_{G}^{2}}{\Lambda^{2}}-\frac{3}{2}\Bigr)-\frac{m_{c}^{4}}{64\pi^{2}}\Bigl(\log\frac{m_{c}^{2}}{\Lambda^{2}}-\frac{3}{2}\Bigr) \nonumber\\
& \quad +	V_{{\rm daisy}}^{G}(\phi,T)
+\frac{T^{4}}{2\pi^{2}}\left[J_{B}\Bigl(\frac{m_{G}^{2}}{T^{2}}\Bigr)-J_{B}\Bigl(\frac{m_{c}^{2}}{T^{2}}\Bigr)\right]\,,
\\
Z\left(\phi_{c}\right) & = 1+Z_{g_{x}^{2}}\left(\phi_{c},T=0\right)+\delta Z_{g_{x}^{2}}\left(\phi_{c},T\right)+\mathcal{O}(g_{x}^{4})\,,
\end{align}
where we incorporate the gauge-independent thermal contributions into the leading-order effective potential,
while the $\xi$-dependent thermal terms are assigned to the subleading effective potential.\footnote{
In the regime $\phi \to 0$ and $T_p \gg \phi$, which is most relevant for barrier formation,
the high-temperature expansions of the $\xi$-independent terms
$V_{{\rm daisy}}^{A^{\prime}}+J_B(m_{h_x}^2/T^2)+J_B(m_{A^{\prime}}^2/T^2)$
appear at lower order in the $g_x$ expansion than the $\xi$-dependent terms
$V_{{\rm daisy}}^{G}+J_B(m_G^2/T^2)+J_B(m_c^2/T^2)$,
as discussed in Section~\ref{sec:HT}.
The daisy contribution $V_{{\rm daisy}}^{h_x}$ is in general
more suppressed than the preceding terms and will therefore be neglected in the following discussion.}

The bounce solution is obtained by solving
\begin{equation}
	\nabla^{2}\phi_{b}=\frac{\partial V^{\rm LT}_{g_{x}^{4}}}{\partial\phi_{c}}\biggr|_{\phi_{c}=\phi_{b}}\,,\qquad\phi_{b}(\infty)=0\,,\qquad\phi_{b}^{\prime}(0)=0\,,
\end{equation}
and the nucleation rate is
\begin{align}
\Gamma & =\mathcal{A}\,{\rm e}^{-(S^{\rm LT}_{0}+S^{\rm LT}_{1})/T}\,,\\
S^{\rm LT}_{0} & =\int{\rm d}^{3}x\,\Bigl[V^{\rm LT}_{g_{x}^{4}}(\phi_{b})+\frac{1}{2}(\partial_{i}\phi_{b})^{2}\Bigr]\,,\\
S^{\rm LT}_{1} & =\int{\rm d}^{3}x\,\Bigl[V^{\rm LT}_{g_{x}^{6}}(\phi_{b})+\frac{1}{2}Z_{g_{x}^{2}}(\partial_{i}\phi_{b})^{2}\Bigr]\,,
\end{align}
where all terms in $S_{0}$ are $\xi$-independent. The Nielsen identities are analogous to the zero-temperature case as
\begin{align}
	\xi\frac{\partial V^{\rm LT}_{g_{x}^{4}}}{\partial\xi} & =0\,,\label{eq:NI-LT-Vg4}\\
	\xi\frac{\partial V^{\rm LT}_{g_{x}^{6}}}{\partial\xi} & =-C_{g_{x}^{2}}\frac{\partial V^{\rm LT}_{g_{x}^{4}}}{\partial\phi_{c}}\,,\label{eq:NI-LT-Vg6}\\
	\xi\frac{\partial Z_{g_{x}^{2}}}{\partial\xi} & =-2\frac{\partial C_{g_{x}^{2}}}{\partial\phi_{c}}\,.\label{eq:NI-LT-Zg2}
\end{align}
where $Z_{g_{x}^{2}}$ and $C_{g_{x}^{2}}$ are given in Eqs.~(\ref{eq:Z_0T})--(\ref{eq:C_0T}),
as in the zero-temperature analysis.
Using the identities above, we obtain
\begin{align}
	\xi\frac{\partial S^{\rm LT}_{1}}{\partial\xi} & =\int{\rm d}^{3}x\,\Bigl[\xi\frac{\partial V^{\rm LT}_{g_{x}^{6}}}{\partial\xi}+\frac{1}{2}\xi\frac{\partial Z_{g_{x}^{2}}}{\partial\xi}\,(\partial_{i}\phi_{b})^{2}\Bigr]\nonumber \\
	& =-\int{\rm d}^{3}x\,\Bigl[C_{g_{x}^{2}}\frac{\partial V^{\rm LT}_{g_{x}^{4}}}{\partial\phi_{c}}+\frac{\partial C_{g_{x}^{2}}}{\partial\phi_{c}}\,(\partial_{i}\phi_{b})^{2}\Bigr]\biggr|_{\phi_{c}=\phi_{b}}\nonumber \\
	& =-\int{\rm d}^{3}x\,\Bigl[C_{g_{x}^{2}}\frac{\partial V^{\rm LT}_{g_{x}^{4}}}{\partial\phi_{c}}+(\partial_{i}C_{g_{x}^{2}})\,(\partial_{i}\phi_{b})\Bigr]\biggr|_{\phi_{c}=\phi_{b}}\nonumber \\
	& =-\int{\rm d}^{3}x\,C_{g_{x}^{2}}\Bigl[\frac{\partial V^{\rm LT}_{g_{x}^{4}}}{\partial\phi_{c}}-\nabla^{2}\phi_{b}\Bigr]\biggr|_{\phi_{c}=\phi_{b}}\nonumber \\
	& =0\,,
\end{align}
demonstrating that the subleading effective action $S_1$ is gauge invariant.

In practice, once the thermal contributions are included,
Eq.~(\ref{eq:NI-LT-Vg6}) is still satisfied to a good approximation.
The subleading effective action $S^{\rm LT}_1$ is shown numerically
to be approximately gauge-independent.

\subsubsection{Comments and discussions}

Regarding gauge-independent treatments of the finite-temperature effective potential and their use in gravitational wave calculations,
a few practical points are worth recording:
\begin{itemize}
\item \textbf{The power counting rule:}
  For the high-temperature phase transition analysis in~\cite{Garny:2012cg,Lofgren:2021ogg,Hirvonen:2021zej},
  the Higgs quartic coupling is taken to scale as $\lambda_x \sim g_x^{3}$.
  More precisely, it is assumed to lie in the region $g_x^{4} \ll \lambda_x \ll g_x^{3}$~\cite{Arnold:1992rz}.\footnote{
  For high-temperature phase transitions, one expects
  \begin{equation}
  \frac{g_x \phi_c}{T_n} \sim \frac{g_x \phi_c}{T_c} \ll 1\,.
  \end{equation}
  Using the thermal effective potential in Eq.~(\ref{eq: EffPo}),
  one finds at $T_c$,
  \begin{equation}
  \frac{\phi_c}{T_c} \sim \frac{2 E}{\lambda_x} \sim \frac{g_x^3}{\lambda_x}\,.
  \end{equation}
  Substituting this into the previous condition gives
  \begin{equation}
  g_x \, \frac{\phi_c}{T_c} \sim \frac{g_x^4}{\lambda_x}  \ll 1\,. \label{eq: HTreSc}
  \end{equation}
  Hence, a consistent high-temperature phase transition typically requires
  $\lambda_x \gg g_x^{4}$, also implying that, for fixed $\lambda_x$,
  moving deeper into the high-temperature regime favors smaller values of $g_x$.}
  In our parameter scan, we find that all viable high-temperature phase
  transitions satisfy $\lambda_x \lesssim g_x^{3}$, while in many cases
  $\lambda_x \ll g_x^{4}$. This typically occurs for benchmark points in which
  $T_n$ lies somewhat below $T_c$,
  while the high-temperature approximation $m_{A^\prime}/T_n \ll 1$ remains reasonably valid.

  Nevertheless, adopting the scaling $\lambda_x \sim g_x^{3}$ to organize the
  leading and subleading terms in the potential does yield a gauge-independent
  effective action, as can be verified analytically using the Nielsen identity.
  However, alternative scalings such as $\lambda_x \sim g_x^{4}$ do not lead to consistent results,
  as the Nielsen identity is not satisfied at the corresponding order,
  and thus the resulting subleading effective action is not gauge-independent.

  \textit{What, then, is the role of the power counting?}
  In our view, it provides a systematic organization of the calculation:
  it singles out the leading contributions that determine the bounce solution,
  treats subleading terms as controlled corrections, and consistently discards higher-order terms
  in the gauge coupling $g_x$.
  \item \textbf{Gauge parameter dependence:}
  In the gauge-independent treatment, one must ensure that the terms neglected from
  the full effective potential are indeed parametrically small.
  Since some of the omitted contributions scale as $\sim \xi^n  g^{\text{higher powers}}$,
  the gauge parameter $\xi$ cannot be taken arbitrarily large.
  Otherwise, the power counting scheme is no longer reliable and the expansion breaks down,
  even if the Nielsen identity is satisfied for the effective potential up to subleading order.
%
\item \textbf{Solving for the bounce:}
  A central step in our gravitational wave analysis is to determine the bounce
  solution associated with the finite-temperature effective potential.
  In the usual (gauge-dependent) approach, one typically works in the Landau gauge
  and solves for the bounce using the full effective potential evaluated at $\xi=0$.
  By contrast, in the gauge-independent treatment one computes the bounce
  using the leading-order potential $V_{0}$, which is explicitly $\xi$-independent.
  For this procedure to be self-consistent, the theory must remain in the
  weak-coupling regime, $\alpha_{x}\ll 1$, such that the subleading contribution
  $V_{1}$ does not significantly distort the bounce solution determined from $V_{0}$.
  In practice, we enforce this by restricting the Monte Carlo scan to $g_x < 0.7$. 
  A more accurate treatment would therefore require solving for the bounce
  using the full thermal effective potential, which we leave for future work.
\item \textbf{Comments on the low-temperature analysis:}
  Our scan indicates that a supercooled phase transition typically
  favors a moderately large gauge coupling,
  e.g., $g_x \gtrsim 0.5$,
  in sharp contrast to the high-temperature case,
  where smaller $g_x$ is generally preferred, c.f, Eq.~(\ref{eq: HTreSc}).
  This trend can be understood from two main considerations.

  First, as discussed in the previous section, the low-temperature barrier is
  generated by the thermal cubic term $\sim T \phi^{3}$ near $\phi = 0$,
  whose coefficient scales as $E \propto g_x^{3}$.\footnote{
The effective thermal potential in Eq.~(\ref{eq: EffPo}) is derived in the high-temperature limit,
where the $J_B$ functions admit a high-temperature expansion.
In the low-temperature regime, Eq.~(\ref{eq: EffPo}) remains applicable in the field range $\phi \lesssim T_p/g_x$.
In particular, as $\phi\to 0$ the field value is again much smaller than the temperature,
and thus the high-temperature expansion remains valid in this small-field region,
which is the most relevant for generating a thermal barrier.
Nevertheless, Eq.~(\ref{eq: EffPo}) and the high-temperature expansions in Eqs.~(\ref{eq:JBB}) and~(\ref{eq:JBF})
should not be applied indiscriminately over the full temperature range in analyses of supercooled phase transitions.}
  A larger $g_x$ therefore enhances the cubic term and makes it easier to develop a sufficiently strong barrier for the supercooled phase transition.

  Second, the tree-level tachyonic mass term must be lifted to an effective positive value by thermal mass corrections
  proportional to $g_x^{2}$ in order to create and stabilize a local minimum at $\phi=0$.
  A larger $g_x$ thus keeps the symmetric point $\phi=0$ metastable down to lower temperatures, delaying the loss of metastability,
  which is essential for realizing a supercooled transition.\footnote{
  A conformal $U(1)$ Higgs sector typically contains no tree-level $\phi^{2}$ term.
  The local minimum at the origin is instead stabilized by the finite-temperature thermal mass.
  In this setup, a potential barrier can arise even without relying on the thermal cubic contribution $\sim T\,\phi^{3}$,
  and may be generated predominantly by the Coleman-Weinberg potential.
  As a result, the allowed range of the gauge coupling $g_x$ can be more flexible
  than in the conventional gauged $U(1)$ case with a tachyonic mass term.
  Conformal $U(1)$ models also naturally admit $T_{n,p}\ll v$ and $\alpha \gtrsim 1$,
  corresponding to strongly supercooled phase transitions that produce enhanced gravitational wave signals
  with peak frequencies shifted to lower bands.
  Hence conformal hidden sectors are often discussed as natural pathways to nanohertz band signals.}
\end{itemize}

\section{Gravitational waves from first-order phase transition}\label{Sec:GW}

In a $U(1)$ gauge theory, the finite-temperature effective potential can
induce a first-order phase transition and consequently generate a
stochastic gravitational wave background.
In this section, we review the standard computation of the gravitational wave signal,
starting from the nucleation rate inferred from the bounce solution for a given set of model parameters.

\subsection{The tunneling action and transition rate}

As the temperature falls below the critical temperature $T_c$, the effective
potential may develop two local minima separated by a barrier, rendering the
minimum in the symmetric phase metastable.
The transition from the false vacuum to the true vacuum then proceeds via bubble nucleation.
The resulting bubbles expand and collide, sourcing a stochastic gravitational wave background.
At finite temperature, the bubble nucleation rate,
also referred to as the transition rate or the false vacuum decay rate, is given by~\cite{Coleman:1977py,Callan:1977pt,Gould:2021ccf}
\begin{equation}
\Gamma(T)=\mathcal{A}(T)\,{\rm e}^{-S_{3}(T)/T}\,,\label{eq:Gamma-1}
\end{equation}
where $S_{3}(T)\equiv S_{3}(\phi_{b},T)$ is the three-dimensional Euclidean action evaluated
on the $O(3)$ symmetric bounce configuration.
The action is defined as\footnote{
At finite temperature $T$, the Euclidean time direction is compactified with period $\beta \equiv 1/T$.
The Euclidean action thus takes the form
\begin{equation}
S_E(\phi) = \int_0^\beta {\rm d}\tau \int {\rm d}^3x \left[\tfrac{1}{2} (\partial_\tau \phi)^2 + \tfrac{1}{2} (\nabla \phi)^2 +V_{\rm eff}(\phi, T, \xi)\right]\,.
\end{equation}
In the conventional gauge-dependent treatment one often works in Landau gauge, $\xi=0$.
By contrast, in the gauge-independent approach discussed in Section~\ref{sec:GIEA},
we take $V_{\rm eff}$ at leading order, for which it is independent of the gauge-fixing parameter $\xi$,
and thus the resulting bounce solution is manifestly gauge independent.

For thermally assisted bubble nucleation, the dominant saddle point configuration is approximately time-independent ($\partial_\tau \phi \simeq 0$).
In this case the action reduces to
\begin{equation}
S_E \simeq \beta\, S_3(T) = S_3(T) / T\,, \label{eq: SET}
\end{equation}
where $S_3(T)$ is shown in Eq.~(\ref{eq: S3}).}
\begin{equation}
	S_{3}(\phi,T)=\int_{0}^{\infty}{\rm d}^{3}x\,\left[\frac{1}{2}(\nabla\phi)^{2}+\widetilde{V}_{{\rm eff}}(\phi,T)\right]\,, \label{eq: S3}
\end{equation}
with $\widetilde{V}_{{\rm eff}}(\phi,T)\equiv V_{{\rm eff}}(\phi,T)-V_{{\rm eff}}(\phi_{{\rm FV}},T)$
the free energy density shifted relative to the false vacuum.
Varying $S_3$ yields the three-dimensional bounce equation.
With the $O(3)$ symmetry, the bounce depends only on the radial coordinate, $\phi=\phi(r)$ with $r=\sqrt{x^2+y^2+z^2}$,
\begin{equation}
	\frac{{\rm d}^{2}\phi}{{\rm d}r^{2}}+\frac{2}{r}\frac{{\rm d}\phi}{{\rm d}r}=\frac{\partial}{\partial\phi}\widetilde{V}_{{\rm eff}}(\phi,T)\,,
\label{eq: S3EoM}
\end{equation}
supplemented by the boundary conditions
\begin{equation}
\phi\,(r\to \infty) = 0\,,\qquad
\left. \frac{{\rm d}\phi}{{\rm d}r}\right|_{r=0} =0\,.
\end{equation}
In this work the solution is obtained using \textbf{CosmoTransitions}~\cite{Wainwright:2011kj}.

The prefactor $\mathcal{A}(T)$ is given by~\cite{Linde:1981zj}
\begin{equation}
\mathcal{A}(T)=T\biggl[\frac{S_{3}(\phi_{b},T)}{2\pi T}\biggr]^{3/2}\Biggl\{\frac{{\rm det}^{\prime}\bigl[-\nabla^{2}+V^{\prime\prime}(\phi_{b},T)\bigr]}{{\rm det}\bigl[-\nabla^{2}+V^{\prime\prime}(\phi_{{\rm FV}},T)\bigr]}\Biggr\}^{-1/2}\,,\label{eq:AT}
\end{equation}
where $\rm det^{\prime}$ denotes the functional determinant with zero eigenvalues omitted.
For dimensional consistency, $\mathcal{A}(T)$ is commonly approximated as~\cite{Linde:1981zj}
\begin{equation}
\mathcal{A}(T)\simeq T^{4}\biggl[\frac{S_{3}(\phi_{b},T)}{2\pi T}\biggr]^{3/2}\,.
\end{equation}

The false vacuum fraction $P_{f}$ is introduced to track the progress of the phase transition, written as~\cite{Guth:1979bh,Guth:1981uk}
\begin{equation}
P_{f}(t)=\exp\bigl[-\mathcal{V}_{t}^{{\rm ext}}(t)\bigr]\,,
\end{equation}
where $\mathcal{V}_{t}^{{\rm ext}}(T)$ denotes the fractional extended
volume occupied by true vacuum bubbles at time $t$, given by
\begin{equation}
\mathcal{V}_{t}^{{\rm ext}}(t)=\frac{4\pi}{3}\int_{t_{c}}^{t}{\rm d}t^{\prime}\,\Gamma(t^{\prime})\frac{a^{3}(t^{\prime})}{a^{3}(t)}R^{3}(t^{\prime},t)\,,\label{eq:Vext0}
\end{equation}
with
\begin{equation}
R(t^{\prime},t)=\int_{t^{\prime}}^{t}{\rm d}t^{\prime\prime}\,v_{w}\frac{a(t)}{a(t^{\prime\prime})}\,,
\end{equation}
in which an initial bubble radius is often neglected and the bubble
wall velocity $v_{w}$ is assumed to remain constant throughout the
expansion. Thus Eq.~(\ref{eq:Vext0}) becomes
\begin{equation}
\mathcal{V}_{t}^{{\rm ext}}(t)=\frac{4\pi}{3}v_{w}^{3}\int_{t_{c}}^{t}{\rm d}t^{\prime}\,\Gamma(t^{\prime})\biggl[\frac{a(t^{\prime})}{a(t)}\biggr]^{3}\biggl[\int_{t^{\prime}}^{t}{\rm d}t^{\prime\prime}\frac{a(t)}{a(t^{\prime\prime})}\biggr]^{3}\,.\label{eq: Vext}
\end{equation}

\subsection{Characteristic temperatures}\label{sec:CT}

To analyze the transition dynamics and predict the resulting gravitational waves,
we introduce several benchmark temperatures that characterize key stages of the phase transition.

\paragraph*{Critical temperature $T_{c}$}

The critical temperature $T_{c}$ is defined as the temperature at
which the effective potential of the true and false vacua become degenerate,
$V_{{\rm eff}}(\phi_{{\rm TV}},T_{c})=V_{{\rm eff}}(\phi_{{\rm FV}},T_{c})$,
with a potential barrier separating the two minima.

\paragraph*{Nucleation temperature $T_{n}$}

The nucleation temperature $T_n$ is the temperature at which bubbles of the
broken phase begin to form in the false vacuum at an appreciable rate,
signaling the onset of efficient nucleation.
It is defined when one bubble is nucleated within
one Hubble volume per Hubble time
\begin{equation}
\int_{t_{c}}^{t_{n}}\frac{\Gamma}{H^{3}}\,{\rm d}t\simeq1\,,\label{eq:Tn_def}
\end{equation}
where $\Gamma$ is the bubble nucleation rate and $H$ is the Hubble
parameter including the vacuum energy contribution, given by
\begin{equation}
H^{2}=\frac{\rho_{{\rm rad}}+\rho_{{\rm vac}}}{3M_{{\rm Pl}}^{2}}=\frac{1}{3M_{{\rm Pl}}^{2}}\Bigl[\frac{\pi^{2}}{30}g_{\ast}T^{4}+\Delta V_{{\rm eff}}(T)\Bigr]\,,
\end{equation}
where $\Delta V_{{\rm eff}}=V_{{\rm eff}}(\phi_{{\rm FV}},T)-V_{{\rm eff}}(\phi_{{\rm TV}},T)$,
$g_{\ast}$ is the effective number of relativistic degrees of freedom
including both the SM and hidden sectors, and $M_{{\rm Pl}}=2.4\times10^{18}$~GeV denotes the reduced Planck mass.

Given the temperature dependence of both $\Gamma(T)$ and $H(T)$,
it is convenient to express the nucleation condition as a function
of temperature~\cite{Athron:2022mmm}. Assuming an adiabatically expanding Universe, the
conservation of entropy per comoving volume implies
\begin{equation}
\frac{{\rm d}}{{\rm d}t}\bigl[s(t)a^{3}(t)\bigr]=0\;\;\Rightarrow\;\;\frac{{\rm d}s}{{\rm d}t}=-3H(t)s(t)\,,\label{eq:dsdt}
\end{equation}
where $s$ is the entropy density of the plasma and $a$ is the scale
factor. The entropy density is written as
\begin{equation}
s(T)=\frac{\partial p}{\partial T}=-\frac{\partial V}{\partial T}\,,
\end{equation}
which can be combined with Eq.~(\ref{eq:dsdt}) and
obtain
\begin{equation}
\frac{{\rm d}T}{{\rm d}t}\frac{{\rm d}s}{{\rm d}T}=-\frac{{\rm d}T}{{\rm d}t}\frac{\partial^{2}V}{\partial T^{2}}=3H(T)\frac{\partial V}{\partial T}\,,
\end{equation}
yielding
\begin{equation}
\frac{{\rm d}T}{{\rm d}t}=-3H(T)\frac{\partial_{T}V}{\partial_{TT}V}\,,\label{eq:dTdt1}
\end{equation}
where the derivatives $\partial_{T}V=\frac{\partial V}{\partial T}$
and $\partial_{TT}V=\frac{\partial^{2}V}{\partial T^{2}}$ are evaluated
at the false vacuum. In the literature, the MIT bag equation of state~\cite{Chodos:1974je} is often employed, in which
the false vacuum potential is parameterized as
\begin{equation}
V(\phi_{{\rm FV}},T)=aT^{4}+b\,,
\end{equation}
with $a,b$ being temperature-independent coefficients. This bag model
simplifies Eq.~(\ref{eq:dTdt1}) to the familiar form
\begin{equation}
\frac{{\rm d}T}{{\rm d}t}\stackrel{{\rm bag}}{=}-H(T)T\,,\label{eq:dTdt_bag}
\end{equation}
and the nucleation condition defined by Eq.~(\ref{eq:Tn_def})
can then be rewritten as
\begin{equation}
\int_{T_{n}}^{T_{c}}\frac{\Gamma(T)}{H(T)^{4}}\frac{1}{T}\,{\rm d}T\simeq1\,.\label{eq: Tn_def2}
\end{equation}
When the phase transition occurs near the electroweak scale, the nucleation
condition in Eq.~(\ref{eq: Tn_def2}) can be well
approximated by $S_{3}(T_{n})/T_{n}\simeq140$.

For a high-temperature phase transition, the nucleation temperature $T_n$
typically lies only slightly below $T_c$ and marks the onset of the transition.
This is not the case for a supercooled phase transition, where
the onset is instead characterized by the percolation temperature $T_p$.

\paragraph*{Percolation temperature $T_{p}$}

The percolation temperature $T_p$ is the temperature at which
bubbles have nucleated and grown enough that they percolate through space.

For a fast phase transition, the nucleation temperature $T_{n}$ is sufficient
to characterize both the onset and completion of the transition. However,
a slow or supercooled transition is long lasting, during which the
Universe can remain trapped in the false vacuum for an extended period
after it becomes metastable, delaying bubble collision to the temperature much lower than $T_{n}$.
In such case, $T_{p}$ provides a more accurate description of the transition dynamics and is supposed as the reference temperature for gravitational wave calculations, particularly for the strong supercooling in which $T_{p}\ll T_{n}$~\cite{Kobakhidze:2017mru}.

The percolation temperature $T_{p}$ is defined by roughly $70\%$ of the Universe remains in the false vacuum, $P_{f}(T_{p})\approx0.7$~\cite{Leitao:2012tx, Leitao:2015fmj}. In Eq.~(\ref{eq: Vext}),
the ratio of scale factors at different times is given by
\begin{equation}
\frac{a(t_{1})}{a(t_{2})}=\exp\Bigl[\int_{t_{2}}^{t_{1}}{\rm d}t^{\prime}H(t^{\prime})\Bigr]\,.
\end{equation}
Assuming an adiabatic expansion and adopting the MIT bag equation
of state, this ratio can be recast as a function of temperature
\begin{equation}
\frac{a(T_{1})}{a(T_{2})}=\exp\Bigl(\int_{T_{1}}^{T_{2}}{\rm d}T^{\prime}\frac{1}{T^{\prime}}\Bigr)=\frac{T_{2}}{T_{1}}\,,
\end{equation}
leading to
\begin{equation}
P_{f}(T)=\exp\biggl\{-\frac{4\pi}{3}v_{w}^{3}\int_{T}^{T_{c}}{\rm d}T^{\prime}\frac{\Gamma(T^{\prime})}{T^{\prime4}H(T^{\prime})}\Bigl[\int_{T}^{T^{\prime}}\frac{{\rm d}T^{\prime\prime}}{H(T^{\prime\prime})}\Bigr]^{3}\biggr\}\,.
\end{equation}

To ensure the phase transition completes successfully, two conditions
must be checked~\cite{Athron:2022mmm,Athron:2023mer}: (1) the transition has
an \textbf{end temperature $T_e$,} defined by $P_{f}(T_e)=\varepsilon\lesssim0.01$;
(2) the physical volume of the false vacuum $\mathcal{V}_{{\rm phys}}(t)=a^{3}(t)P_{f}(t)$
decreases with time,
\begin{equation}
\frac{{\rm d}\mathcal{V}_{{\rm phys}}}{{\rm d}t}=\mathcal{V}_{{\rm phys}}\Bigl[\frac{{\rm d}}{{\rm d}t}\ln P_{f}(t)+3H(t)\Bigr]\leq0\,,
\end{equation}
giving rise to
\begin{equation}
T\frac{{\rm d}\mathcal{V}_{t}^{{\rm ext}}}{{\rm d}T}+3\leq0\,,
\end{equation}
and this condition should be verified at both $T_{p}$ and $T_e$.

\paragraph*{Reheating temperature $T_{\rm reh}$}

A first-order phase transition releases latent heat and can reheat the plasma
from the percolation temperature $T_p$ to a reheating temperature $T_{\rm reh}$.
Assuming instantaneous reheating, $T_{{\rm reh}}$ is estimated by
\begin{equation}
T_{{\rm reh}}\simeq\bigl[1+ \alpha(T_{p})\bigr]^{1/4}T_{p}\,,
\end{equation}
where $\alpha(T)$ is the transition strength at temperature $T$.
For moderate supercooling ($\alpha\lesssim 1$), one has $T_{\rm reh}\approx T_p$;
whereas for strong supercooling ($\alpha\gg 1$),
the reheating temperature can be significantly higher than $T_p$, or even higher than $T_c$~\cite{Athron:2023mer}.
Our scan indicates that the minimal gauged $U(1)$ dark sector
typically does not realize a very strong supercooled phase transition,
the corresponding reheating temperature generally does not have $T_{\rm reh} > T_c$.
The present gravitational wave amplitude and frequency are obtained by
redshifting their values from the reheating temperature $T_{\rm reh}$
rather than from $T_p$ down to today.\\

To summarize this subsection,
as the Universe cools, the dark $U(1)_x$ sector undergoing a supercooled phase transition
may pass through the following characteristic temperatures,
\begin{equation}
T_c \gg T_n > T_p \gtrsim T_e \gtrsim T_0 \gtrsim 0\,.
\end{equation}
Recall that $T_0$ is defined as the temperature at which $\phi=0$ ceases to be a local minimum.
In a typical supercooled phase transition, $T_p$, $T_e$, and $T_0$ are close to one another.

\subsection{Hydrodynamic parameters}\label{sec:Hydro}

The gravitational wave signal originating from a first-order phase
transition depends sensitively on the characteristic length scale
$L_{\ast}$, which represents the typical scale carrying most of the
released energy. A common estimate identifies $L_{\ast}$ with the
mean bubble separation $R_{\ast}$, which is
\begin{equation}
R_{\ast}=\frac{(8\pi)^{1/3}}{\beta}v_{w}\,, \label{eq: beta}
\end{equation}
where $v_{w}$ is the bubble wall velocity and $\beta$ is the inverse time
scale, defined as
\begin{equation}
\beta=-\frac{{\rm d}}{{\rm d}t}\Bigl(\frac{S_{3}}{T}\Bigr)\biggr|_{t=t_{p}}=H(T)T\frac{{\rm d}}{{\rm d}T}\Bigl(\frac{S_{3}}{T}\Bigr)\biggr|_{T=T_{p}}\,,
\end{equation}
which corresponds to an exponential nucleation rate and is usually
expressed in the following form
\begin{equation}
\frac{\beta}{H_{p}}=T\frac{{\rm d}}{{\rm d}T}\Bigl(\frac{S_{3}}{T}\Bigr)\biggr|_{T=T_{p}}\,.
\end{equation}
For strong supercooled phase transitions, $\beta$ may become tiny
or even negative, indicating the commonly exponential approximation
breaks down. In this case, the mean bubble separation $R_{\ast}$
should instead be directly determined from the bubble number density
\begin{equation}
R_{\ast}(T)\equiv\bigl[n(T)\bigr]^{-1/3}=\biggl[T\int_{T}^{T_{c}}{\rm d}T^{\prime}\frac{\Gamma(T^{\prime})P_{f}(T^{\prime})}{T^{\prime4}H(T^{\prime})}\biggr]^{-1/3}\,.
\end{equation}

The phase transition strength parameter $\alpha$,
which quantifies the energy released during the transition
relative to the radiation energy density of the Universe at the time the gravitational waves are generated, is defined as~\cite{Giese:2020rtr,Giese:2020znk}
\begin{equation}
\alpha\equiv\frac{\bar{\theta}_{f}(T_{p})-\bar{\theta}_{t}(T_{p})}{3w_{f}(T_{p})}\,,\qquad\bar{\theta}\equiv \rho-\frac{p}{c_{s,t}^{2}}\,,\label{eq:alpha}
\end{equation}
where $w_{f}$ denotes the enthalpy density in the false vacuum,
$\bar{\theta}$ represents the \textit{pseudotrace} that is a generalization of the
vacuum energy, and $c_{s}^{2}$ is the speed of sound in the plasma.
Subscripts $f$ and $t$ refer to quantities in the false and true
vacua respectively. The hydrodynamics of the plasma is described by
the energy density $\rho$, the pressure $p$, and the enthalpy density $w$.

The plasma is usually modeled as a perfect relativistic fluid,
with the pressure in the false vacuum determined by the negative finite-temperature free energy density.
For a dark sector first-order phase transition, the pressure receives two contributions:
\begin{itemize}
  \item The dark Higgs sector contribution is encoded in the effective potential $V_{\rm eff}(\phi,T)$,
  which includes all particles that couple to the dark Higgs and therefore contribute to the $\phi$-dependent thermodynamics.
  \item The contribution from additional radiation species that do not couple directly to the dark Higgs
  and therefore do not contribute directly to the $\phi$-dependent part of $V_{\rm eff}$,
  but still belong to the dark thermal bath.
  These species should be included only if they remain relativistic and in kinetic equilibrium with the dark sector at the time of the phase transition.
  In a realistic $U(1)_x$ dark sector, this typically refers to particles that interact efficiently with the dark photon (but not with the dark Higgs),
  provided that the dark photon itself remains in equilibrium with the dark Higgs.
\end{itemize}
The relevant thermodynamic quantities are thus given by~\cite{Giese:2020znk,Feng:2025wvc}
\begin{align}
p & \equiv-V_{{\rm eff}}(\phi_{c},T)+\frac{\pi^{2}}{90}g_{{\rm eff}}(T)T^{4}\,,\label{eq: HDpressure}\\
\rho & \equiv T\frac{\partial p}{\partial T}-p\,,\\
w & \equiv T\frac{\partial p}{\partial T}=p+\rho\,,\\
c_{s}^{2} & \equiv\frac{\delta p}{\delta \rho}\simeq\frac{\partial p/\partial T}{\partial \rho/\partial T}\,,
\end{align}
where the second term in $p$ accounts for the field-independent
contribution from \textit{all relativistic particle species in thermal equilibrium
with the dark thermal bath at temperature $T$}.\footnote{
For a $U(1)_{B-L}$ phase transition, all SM fermions carry $B-L$ charge.
Although the SM fermions are vectorlike under $B-L$ and therefore do not couple directly to the $U(1)_{B-L}$ Higgs,
they interact efficiently with the $U(1)_{B-L}$ gauge boson and thus remain in thermal equilibrium with the $U(1)_{B-L}$ plasma.
Consequently, SM fermions contribute to the second term in Eq.~(\ref{eq: HDpressure})
provided that they are still relativistic at the time of the $U(1)_{B-L}$ phase transition.
This makes the $U(1)_{B-L}$ case qualitatively different from a dark $U(1)_x$ sector phase transition,
in which SM fermions carry no $U(1)_x$ charge.}

During the phase transition, the released energy accelerates bubbles, while the interaction between bubble walls and the surrounding plasma impedes the acceleration. When
the released energy is sufficiently large to overcome this friction,
the wall continues to accelerate without reaching a terminal velocity,
entering a \textit{runaway} regime. The corresponding criterion parameter
is defined as~\cite{Espinosa:2010hh}
\begin{equation}
\alpha_{\infty}\equiv\frac{1}{18}\frac{\sum_{i}g_{i}c_{i}\Delta m_{i}^{2}T_{p}^{2}}{w_{f}(T_{p})}\,,\label{eq:alpha_inf}
\end{equation}
where $c_{i}=1\,(1/2)$ for bosons (fermions), $g_{i}$ denotes the
internal degrees of freedom of particle $i$ and $\Delta m_{i}^{2}=m_{i,t}^{2}-m_{i,f}^{2}$
with the sum running over all species that gain a mass during the
transition. If $\alpha>\alpha_{\infty}$, the bubble wall runs away,
yielding $v_{w}\simeq1$. In the non-runaway regime ($\alpha<\alpha_{\infty}$),
the wall eventually reaches a model-dependent terminal velocity.
We compute the wall velocity $v_{w}(\alpha,c_{s,f}^{2},c_{s,t}^{2},\Psi)$,
where $\Psi \equiv w_t/w_f$ denotes the ratio of the enthalpy densities in the true and false phases,
following~\cite{Ai:2023see,Ai:2024btx}.

Kinetic energy fraction is defined as $K\equiv\rho_{{\rm fl}}/\rho_{p}$
which quantifies the fraction of available vacuum energy converted
into bulk motion of the plasma, where $\rho_{p}$ is the total energy
density at the percolation temperature $T_{p}$, and $\rho_{{\rm fl}}$ is the plasma kinetic energy density,
obtained by integrating over the fluid profile.
The kinetic energy fraction can be further parameterized as
\begin{equation}
K\simeq\frac{\alpha}{1+\alpha}\,\kappa(\alpha,c_{s,f}^{2},c_{s,t}^{2},v_{w})\,,
\end{equation}
where $\kappa$ represents the efficiency factor, determined by the
transition strength, sound speeds in two phases and wall velocity,
with detailed calculations provided in~\cite{Giese:2020rtr,Giese:2020znk}.

\subsection{Gravitational wave power spectrum}

The gravitational wave power spectrum is defined as
\begin{equation}
\Omega_{{\rm GW}}(f)\equiv\frac{1}{\rho_{c}}\frac{{\rm d}\rho_{{\rm GW}}}{{\rm d}\log f}\,,
\end{equation}
where $f$ denotes the frequency and $\rho_{c}$ is the critical density.

The total gravitational wave signal typically receives three primary contributions:
(1) bubble collisions of the scalar field; (2) sound waves in the
bulk plasma; (3) magneto-hydrodynamic turbulence in the plasma~\cite{Kosowsky:1991ua,Hindmarsh:2013xza,Caprini:2006jb}. Accordingly,
\begin{equation}
h^{2}\Omega_{{\rm GW}}(f)=h^{2}\Omega_{\phi}(f)+h^{2}\Omega_{{\rm sw}}(f)+h^{2}\Omega_{{\rm tb}}(f)\,,
\end{equation}
with each component parametrized as
\begin{equation}
h^{2}\Omega(f)=h^{2}\Omega^{{\rm peak}}\mathcal{S}(f)\,,
\end{equation}
where $\Omega^{{\rm peak}}$ is the peak amplitude and $\mathcal{S}$ the spectral-shape function.

Phase transitions in a dark sector can be characterized by two strength parameters,
$\alpha_{{\rm tot}}$ and $\alpha_{h}$, defined as~\cite{Breitbach:2018ddu,Ertas:2021xeh,Bringmann:2023opz,Li:2025nja},
\begin{equation}
\alpha_{{\rm tot}}=\frac{\Delta\bar{\theta}(T_{h,p})}{3\bigl[w_{f}^{v}(T_{v,p})+w_{f}^{h}(T_{h,p})\bigr]}\,,\qquad\alpha_{h}=\frac{\Delta\bar{\theta}(T_{h,p})}{3w_{f}^{h}(T_{h,p})}\,,
\end{equation}
where $\Delta\bar{\theta}=\bar{\theta}_{f}-\bar{\theta}_{t}$, c.f.,
Eq.~(\ref{eq:alpha}). Here, $w_{f}^{v}$ and $w_{f}^{h}$
denote the false vacuum enthalpy densities of the visible and hidden
sector, respectively, and $T_{v,p}$ and $T_{h,p}$ are their corresponding
percolation temperatures.
In this work, we assume that the dark sector shares the same temperature as
the visible sector, $T_h = T_v$, due to efficient interactions between the two sectors.
The total parameter $\alpha_{{\rm tot}}$ determines the amplitude
of gravitational waves, and $\alpha_{h}$ controls
the efficiency factor $\kappa$ which describes the fraction of vacuum
energy transformed into the bulk kinetic energy of the plasma.

Under this framework, the runaway condition introduced in Eq.~(\ref{eq:alpha_inf})
generalizes to the hidden sector as
\begin{equation}
\alpha_{h,\infty}\equiv\frac{1}{18}\frac{\sum_{i}g_{i}c_{i}\Delta m_{i}^{2}T_{h,p}^{2}}{w_{f}^{h}(T_{h,p})}\,,
\end{equation}
where $c_{i}=1\,(1/2)$ for bosons (fermions), $g_{i}$ is the internal
degrees of freedom of particle $i$, and $\Delta m_{i}^{2}=m_{i,t}^{2}-m_{i,f}^{2}$.
In the runaway regime ($\alpha_{h,\infty}<\alpha_{h}$), efficiency
factors are given by
\begin{equation}
\kappa_{{\rm col}}=1-\frac{\alpha_{h,\infty}}{\alpha_{h}}\,,\quad\kappa_{{\rm sw}}=\frac{\alpha_{h,\infty}}{\alpha_{h}}\,\kappa(\alpha_{h,\infty},c_{s,f}^{2},c_{s,t}^{2},v_{w})\,,\quad\kappa_{{\rm tb}}=\epsilon\kappa_{{\rm sw}}\,,
\end{equation}
where $\epsilon$ denotes the fraction of bulk motion that is turbulent.
In the non-runaway regime ($\alpha_{h,\infty}>\alpha_{h}$), the factors are
\begin{equation}
\kappa_{{\rm col}}=0\,,\quad\kappa_{{\rm sw}}=\kappa(\alpha_{h},c_{s,f}^{2},c_{s,t}^{2},v_{w})\,,\quad\kappa_{{\rm tb}}=\epsilon\kappa_{{\rm sw}}\,,
\end{equation}
In this work we adopt $\epsilon=0.1$ according to numerical simulations~\cite{Hindmarsh:2015qta}.

The present gravitational wave amplitude and frequency are obtained by
redshifting their values at production evaluated at the temperature $T_\ast$, given by
\begin{equation}
h^{2}\Omega_{0}(f_{0})=\Bigl(\frac{a_{\ast}}{a_{0}}\Bigr)^{4}\Bigl(\frac{H_{\ast}}{H_{0}}\Bigr)^{2}\Omega_{1}(f_{\ast})=1.67\times10^{-5}\biggl(\frac{100}{g_{{\rm eff}}(T_{\ast})}\biggr)^{1/3}\Omega_{\ast}(f_{\ast})\,,
\end{equation}
and
\begin{equation}
f_{0}=\frac{a_{\ast}}{a_{0}}f_{\ast}=1.65\times10^{-5}{\rm Hz}\,\biggl(\frac{g_{{\rm eff}}(T_{\ast})}{100}\biggr)^{1/6}\biggl(\frac{T_{\ast}}{100\,{\rm GeV}}\biggr)\,,
\end{equation}
where the subscript $0$ denotes present quantities and $\ast$ denotes those evaluated at the time of the gravitational wave production.

As noted in the previous section, a supercooled phase transition releases
latent heat and can reheat the plasma from the percolation temperature $T_p$
to a reheating temperature $T_{\rm reh}$. The present gravitational wave
amplitude and frequency are therefore obtained by evaluating the redshift
factor at $T_\ast = T_{\rm reh}$ rather than at $T_p$.

The gravitational wave power spectra observed today for the three contributions are thus given by
\begin{enumerate}
\item bubble collision
\begin{align}
h^{2}\Omega_{\phi}(f) & =1.67\times10^{-5}\frac{0.11v_{w}^{3}}{0.42+v_{w}^{2}}\Bigl(\frac{100}{g_{{\rm eff}}(T_{{\rm reh}})}\Bigr)^{1/3}\Bigl(\frac{\kappa_{{\rm col}}\alpha_{{\rm tot}}}{1+\alpha_{{\rm tot}}}\Bigr)^{2}\Bigl(\frac{\beta}{H_{\ast}}\Bigr)^{-2}\mathcal{S}_{\phi}(f)\,,\\
\mathcal{S}_{\phi}(f) & =\frac{3.8(f/f_{\phi})^{2.8}}{1+2.8(f/f_{\phi})^{3.8}}\,,\\
f_{\phi} & =1.6\times10^{-7}\,\Bigl(\frac{g_{{\rm eff}}(T_{{\rm reh}})}{100}\Bigr)^{1/6}\Bigl(\frac{T_{{\rm reh}}}{1\,{\rm GeV}}\Bigr)\Bigl(\frac{\beta}{H_{\ast}}\Bigr)\Bigl(\frac{0.62}{1.8-0.1v_{w}+v_{w}^{2}}\Bigr)\,{\rm Hz}\,.
\end{align}
\item sound wave
\begin{align}
h^{2}\Omega_{{\rm sw}}(f) & =2.65\times10^{-6}v_{w}\Bigl(\frac{100}{g_{{\rm eff}}(T_{{\rm reh}})}\Bigr)^{1/3}\Bigl(\frac{\kappa_{{\rm sw}}\alpha_{{\rm tot}}}{1+\alpha_{{\rm tot}}}\Bigr)^{2}\Bigl(\frac{\beta}{H_{\ast}}\Bigr)^{-1}\mathcal{S}_{{\rm sw}}(f)\,,\\
\mathcal{S}_{{\rm sw}}(f) & =\Bigl(\frac{f}{f_{{\rm sw}}}\Bigr)^{3}\biggl[\frac{7}{4+3\bigl(f/f_{{\rm sw}}\bigr)^{2}}\biggr]^{7/2}\,,\\
f_{{\rm sw}} & =1.9\times10^{-7}\,\frac{1}{v_{w}}\Bigl(\frac{g_{{\rm eff}}(T_{{\rm reh}})}{100}\Bigr)^{1/6}\Bigl(\frac{T_{{\rm reh}}}{1\,{\rm GeV}}\Bigr)\Bigl(\frac{\beta}{H_{\ast}}\Bigr)\,{\rm Hz}\,.
\end{align}
\item turbulence
\begin{align}
h^{2}\Omega_{{\rm tb}}(f) & =3.35\times10^{-4}v_{w}\Bigl(\frac{100}{g_{{\rm eff}}(T_{{\rm reh}})}\Bigr)^{1/3}\Bigl(\frac{\kappa_{{\rm tb}}\alpha_{{\rm tot}}}{1+\alpha_{{\rm tot}}}\Bigr)^{3/2}\mathcal{S}_{{\rm tb}}(f)\,,\\
\mathcal{S}_{{\rm tb}}(f) & =\biggl(\frac{f}{f_{{\rm tb}}}\biggr)^{3}\biggl[\frac{1}{1+\bigl(f/f_{{\rm tb}}\bigr)}\biggr]^{11/3}\Bigl(1+8\pi\frac{f}{H_{\ast}^{\prime}}\Bigr)^{-1}\,,\\
f_{{\rm tb}} & =\frac{2.7}{1.9}f_{{\rm sw}}\,,
\end{align}
\end{enumerate}
where $f_{\phi}, f_{{\rm sw}}, f_{{\rm tb}}$ are peak frequencies,
$H_{\ast}$ denotes the Hubble rate at gravitational wave production,
where we take $T_\ast = T_{{\rm reh}}$,
and $H_{\ast}^{\prime}$ is the redshifted Hubble rate, given by
\begin{equation}
H_{\ast}^{\prime}=\Bigl(\frac{a}{a_{0}}\Bigr)H_{\ast}\simeq1.6\times10^{-5}\,{\rm Hz}\,\Bigl(\frac{g_{{\rm eff}}(T_{{\rm reh}})}{100}\Bigr)^{1/6}\Bigl(\frac{T_{{\rm reh}}}{100\,{\rm GeV}}\Bigr)\,.
\end{equation}

\paragraph*{Strong signals from supercooling}

Using the hydrodynamic parameters defined in Section~\ref{sec:Hydro}
together with the gravitational wave power spectrum formulas,
we summarize the main factors by which
supercooled phase transitions can generically produce stronger signals:
\begin{itemize}
  \item From Eq.~(\ref{eq:alpha}), the transition strength parameter $\alpha$ scales inversely
  with the radiation energy density, $\rho_{\rm rad}(T_p)\propto T_p^{4}$.
  Supercooled transitions typically have a much lower percolation temperature $T_p$,
  and thus lead to a large $\alpha$.
  In this regime the fraction of energy available to source gravitational waves,
  $\kappa\alpha/(1+\alpha)$, is enhanced and approaches $\kappa$.
  As a result, the released vacuum energy can constitute a sizable fraction of the total energy density,
  leading to substantially stronger gravitational wave signals.
  \item The parameter $\beta$ characterizes the inverse time scale over which the nucleation rate grows,
  and hence $\beta/H$ controls the typical duration of the transition.
  In supercooled scenarios, nucleation is delayed and then turns on rapidly,
  thus the transition proceeds with larger characteristic bubble sizes; c.f., Eq.~(\ref{eq: beta}).
  Since gravitational waves are sourced by anisotropic stresses
  on length scales of order $R_\ast\sim v_w/\beta$,
  a smaller $\beta/H$ typically enhances the signal amplitude.
  \item In the strong supercooling limit,
  the driving pressure can be large while plasma friction becomes relatively less important.
  Bubble walls may therefore become highly relativistic and can more readily approach runaway behavior.
  This increases the fraction of the released energy stored in the bubble walls or in the scalar field kinetic energy,
  thereby enhancing the collision- and field-sourced contributions to the gravitational wave spectrum.
\end{itemize}

Finally, as can be seen from gravitational wave power spectrum formulas,
the peak frequency parametrically scales as
\begin{equation}
f_{\rm peak} \, \propto\, T_\ast \times \frac{\beta}{H_\ast} \times \big(\text{redshift factors}\big)\,.
\end{equation}
Supercooling typically implies a smaller characteristic temperature $T_\ast$ and often a smaller $\beta/H_\ast$,
both of which push $f_{\rm peak}$ to lower values.
Consequently, supercooled transitions are a well-motivated and frequently studied mechanism
for generating signals in the nanohertz band.

\subsection{Gauge-dependent versus gauge-independent results}

In this subsection, we perform a Monte Carlo scan of the parameter space for
gravitational waves sourced by the first-order phase transition,
using both a gauge-dependent treatment in Landau gauge ($\xi=0$)
and the gauge-independent framework discussed in Section~\ref{Sec:Gauge}.

\subsubsection{Gauge-dependent gravitational wave results}\label{sec:GDGW}

\begin{figure}[t!]
	\centering
	\subfigure[]{
		\includegraphics[scale=0.46,trim=10 10 10 10,clip]{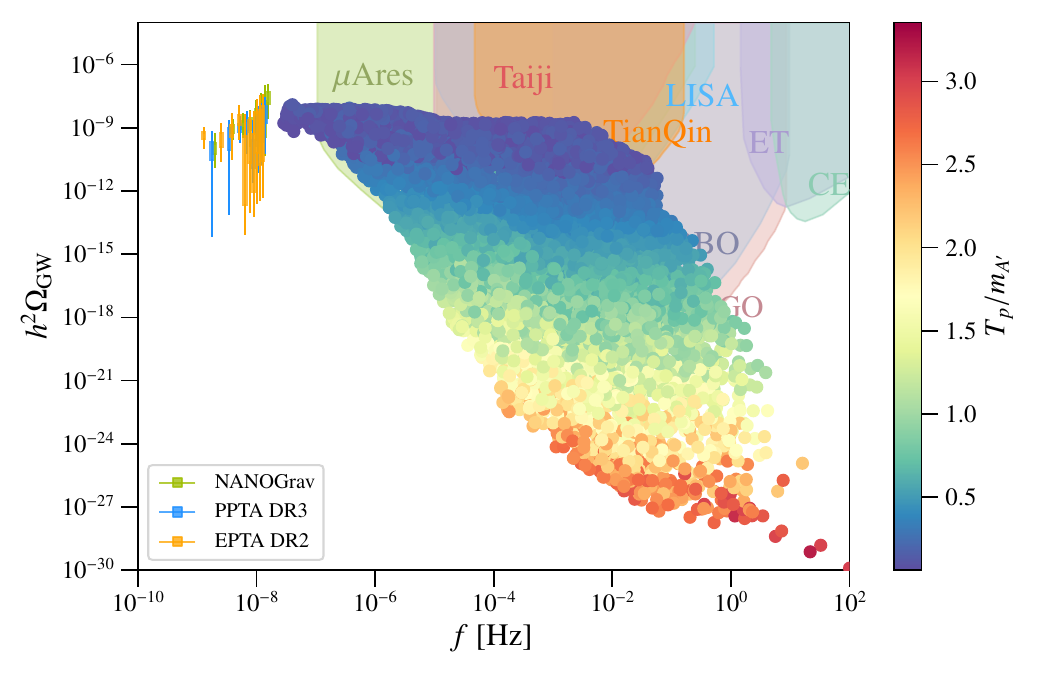} }
	\subfigure[]{
		\includegraphics[scale=0.46,trim=10 10 10 10,clip]{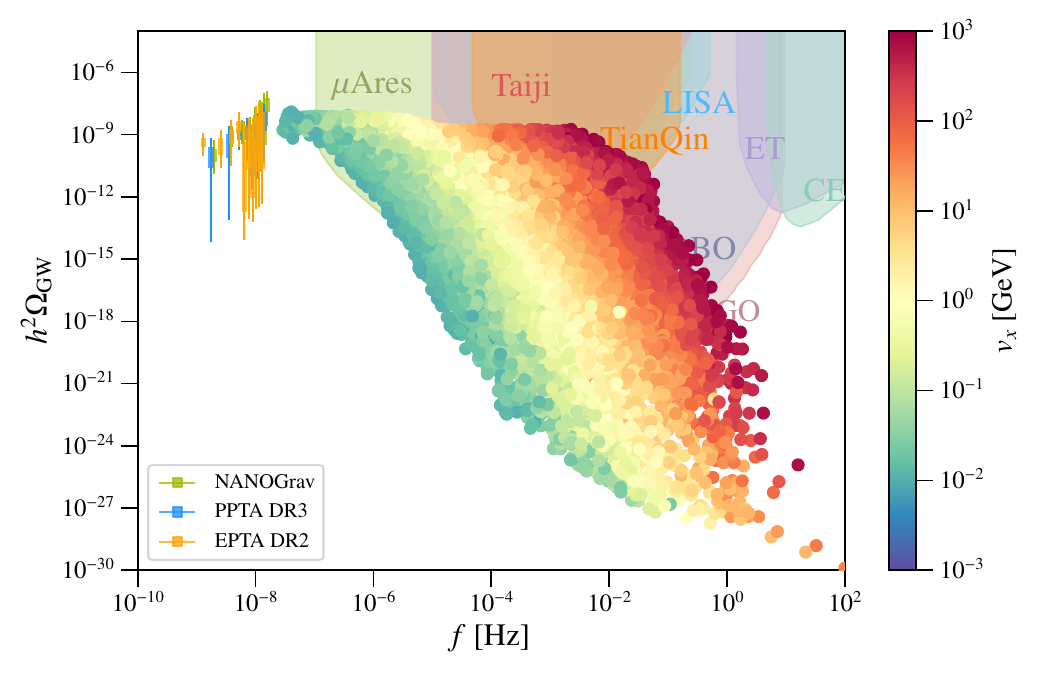} }
    \caption{\label{Fig:GD} Gauge-dependent gravitational wave signals from a Monte Carlo scan over the parameter ranges in Eq.~(\ref{eq:MCrange_GD}) using the finite-temperature effective potential in Landau gauge ($\xi=0$).
    Scatter points show the peak frequency and peak amplitude of the predicted stochastic background from the first-order phase transition.
    Panel (a) is color-coded by the ratio $T_p/m_{A^\prime}$,
    and panel (b) by the $U(1)_x$ symmetry breaking scale $v_x\in[10^{-3},10^{3}]~\mathrm{GeV}$.
    The colored regions denote the power-law integrated sensitivities of current and future detectors, including Taiji~\cite{Ruan:2018tsw}, TianQin~\cite{TianQin:2015yph}, LISA~\cite{LISA:2017pwj}, $\mu$Ares~\cite{Sesana:2019vho}, BBO~\cite{Grojean:2006bp}, U$\_$DECIGO~\cite{Kuroyanagi:2014qaa}, ET~\cite{Punturo:2010zz}, and CE~\cite{LIGOScientific:2016wof}. Box plots show the PTA signals from NANOGrav~\cite{NANOGrav:2023gor}, PPTA~\cite{Reardon:2023gzh}, and EPTA~\cite{EPTA:2023fyk}, with dataset and plotting style following~\cite{Athron:2020sbe}. }
\end{figure}

To begin, we perform a scan using the gauge-dependent finite-temperature effective potential in Landau gauge ($\xi=0$),
as presented in Section~\ref{sec:GDEP}.
We sample $3\times 10^{4}$ parameter points in the ranges
\begin{equation}
	g_{x}\in[10^{-2},1],\;\lambda_{x}\in[10^{-6},10^{-1}],\;v_{x}\in[10^{-3},10^{3}]\,{\rm GeV}\,.\label{eq:MCrange_GD}
\end{equation}
and find $6962$ points for which the phase transition completes successfully.
The corresponding peak signals are shown in Fig.~\ref{Fig:GD}.
The scatter points are color-coded by $T_{p}/m_{A^{\prime}}$ and by $v_{x}$.

Although results obtained from a gauge-dependent potential do \emph{not} constitute definitive gravitational wave predictions for the dark $U(1)$ sector,
the two panels nevertheless reveal clear trends that are instructive for understanding the underlying physics:
\begin{itemize}
  \item The temperature at which the transition occurs is closely correlated with the signal strength:
low-temperature transitions typically yield larger amplitudes,
whereas transitions occurring at higher temperatures tend to produce much weaker signals.
  \item Larger symmetry breaking scales $v_x$ shift the gravitational wave signal toward higher frequencies,
while smaller $v_x$ shift it toward lower frequencies.
\end{itemize}

\begin{figure}[h!]
	\centering
	\includegraphics[scale=0.68,trim=0 20 0 20,clip]{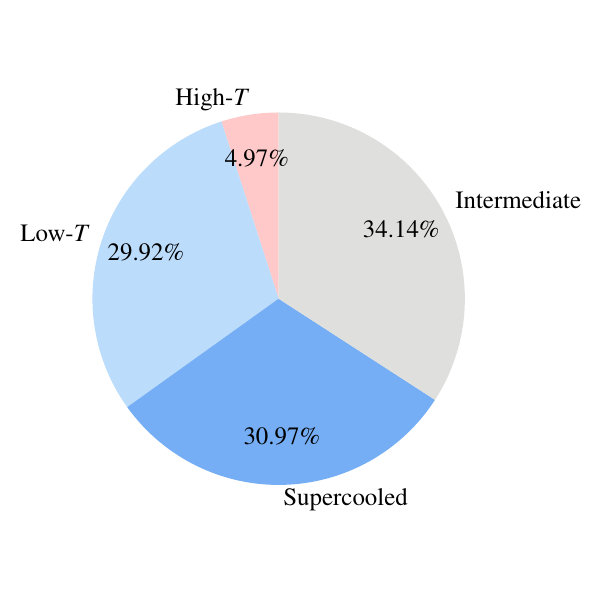}
    \caption{Pie chart showing the composition of phase transition regimes in the gauge-dependent scan (Landau gauge, $\xi=0$)
    over the parameter ranges in Eq.~(\ref{eq:MCrange_GD}).
    The classification is based on the ratio $T_p/m_{A^{\prime}}$ as defined in Table~\ref{TableFrac},
    which also lists the corresponding event counts and fractions.}
	\label{Fig:pie}
\end{figure}

To quantify the range of thermal regimes encountered in the gauge-dependent scan,
we classify the successful transition points into four categories according to the ratio $T_{p}/m_{A^{\prime}}$,
with the supercooled and (approximately) high-temperature regimes being the ones that can be treated within the gauge-independent framework:
\begin{enumerate}
  \item Supercooled phase transition: $T_{p}/m_{A^{\prime}}\lesssim0.2$;
  \item (Moderately) low-temperature phase transition: $0.2 < T_{p}/m_{A^{\prime}} < 0.5$;
  \item Intermediate regime:  $0.5 < T_{p}/m_{A^{\prime}}<2$;
  \item (Approximately) high-temperature phase transition: $T_{p}/m_{A^{\prime}}\gtrsim2$.
\end{enumerate}
The number of points and the corresponding fractions in each category are summarized in Table~\ref{TableFrac},
and the overall composition is illustrated by the pie chart in Fig.~\ref{Fig:pie}.

\begin{table}[h!]
	\centering
		\begin{tabular}{|c|c|c|}
		\hline
		Case & number & fraction\tabularnewline
		\hline
		Supercooled: $T_{p}/m_{A^{\prime}}\lesssim0.2$ & 2156 & 30.97 \%\tabularnewline
		\hline
		(Approx) low-$T$: $0.2 < T_{p}/m_{A^{\prime}} < 0.5$ & 2083 & 29.92 \%\tabularnewline
		\hline
		Intermediate: $0.5<T_{p}/m_{A^{\prime}}<2$ & 2377 & 34.14 \%\tabularnewline
		\hline
		(Approx) high-$T$: $T_{p}/m_{A^{\prime}}\gtrsim2$ & 346 & 4.97 \%\tabularnewline
		\hline
		Total & 6962 & 100 \%\tabularnewline
		\hline
	\end{tabular}
    \caption{Statistics of successful first-order phase transition points obtained in the gauge-dependent calculation (Landau gauge, $\xi=0$).
    We sample $3\times 10^{4}$ parameter points in the ranges of Eq.~(\ref{eq:MCrange_GD}) and find $6962$ points for which the transition percolates successfully. The successful points are classified by the ratio $T_{p}/m_{A^{\prime}}$.
    For each category we report the number of points and the corresponding fraction of the total sample.}
	\label{TableFrac}
\end{table}

From the statistics, we find that supercooled transitions account for roughly $30\%$ of all successful phase transition points.
They typically yield the strongest gravitational wave signals among the regimes considered, as shown in Fig.~\ref{Fig:GD},
and many lie within the projected sensitivity of near-future experiments.
\emph{Moderately} low-temperature transitions with $0.2 < T_{p}/m_{A^{\prime}} < 0.5$
constitute another $\sim 30\%$ of the successful points.
Taken together, the supercooled and moderately low-temperature regimes
are naturally and approximately, respectively,  suited to the low-temperature gauge-independent framework.

We identify an \emph{approximately} high-temperature regime using the criterion $T_{p}/m_{A^{\prime}}\gtrsim 2$,
which already constitutes a small subset of the successful points (below $5\%$).
Truly high-temperature transitions for which the high-temperature expansion is parametrically justified,
$T_{p}/m_{A^{\prime}} \gtrsim 10$, are rare in our scan.
These points generally produce very weak signals and are therefore difficult to detect,
making them phenomenologically less compelling.

The remaining points lie in an intermediate regime, $0.5 < T_{p}/m_{A^{\prime}} < 2$,
and make up about $35\%$ of the successful sample.
Their predicted signals are intermediate between the supercooled and high-temperature cases.
This regime is not applicable to the gauge-independent treatments:
the high-temperature expansion is not valid,
while the low-temperature thermal contributions cannot be regarded as small corrections.
Moreover, the associated gravitational wave signals are typically below the reach of near-future detectors,
rendering this region also less phenomenologically relevant at present.

It should be noted that part of the literature adopts Eq.~(\ref{eq: EffPo}) as the starting point for gravitational wave analyses
and reports many viable points in the intermediate- and low-temperature regimes. Such treatments are not appropriate.
Eq.~(\ref{eq: EffPo}) is obtained by employing the high-temperature expansion of $J_B, J_F$ functions,
which ceases to be valid once the high-temperature condition is no longer satisfied.
Consequently, the resulting gravitational wave spectra in regimes
where the high-temperature expansion is not valid are unreliable, regardless of gauge-dependent issues.

\subsubsection{Gauge-independent gravitational wave predictions}

The gauge-independent effective action in the high- and low-temperature limits is discussed in detail in Section~\ref{Sec:Gauge},
and we scan these two regimes separately.
The ranges of input parameters are chosen as follows:
\begin{align}
	\text{high-}T:\quad & g_{x}\in[10^{-3},0.7],\;\lambda_{x}\in[10^{-9},10^{-2}],\;v_{x}\in[10^{-3},10^{2}]\,{\rm GeV}\,,\\
	\text{low-}T:\quad & g_{x}\in[10^{-2},0.7],\;\lambda_{x}\in[10^{-6},10^{-2}],\;v_{x}\in[10^{-3},10^{2}]\,{\rm GeV}\,.\label{eq:MCrange}
\end{align}

\begin{figure}[t!]
	\centering
	\subfigure[]{
		\includegraphics[scale=0.46,trim=10 10 10 10,clip]{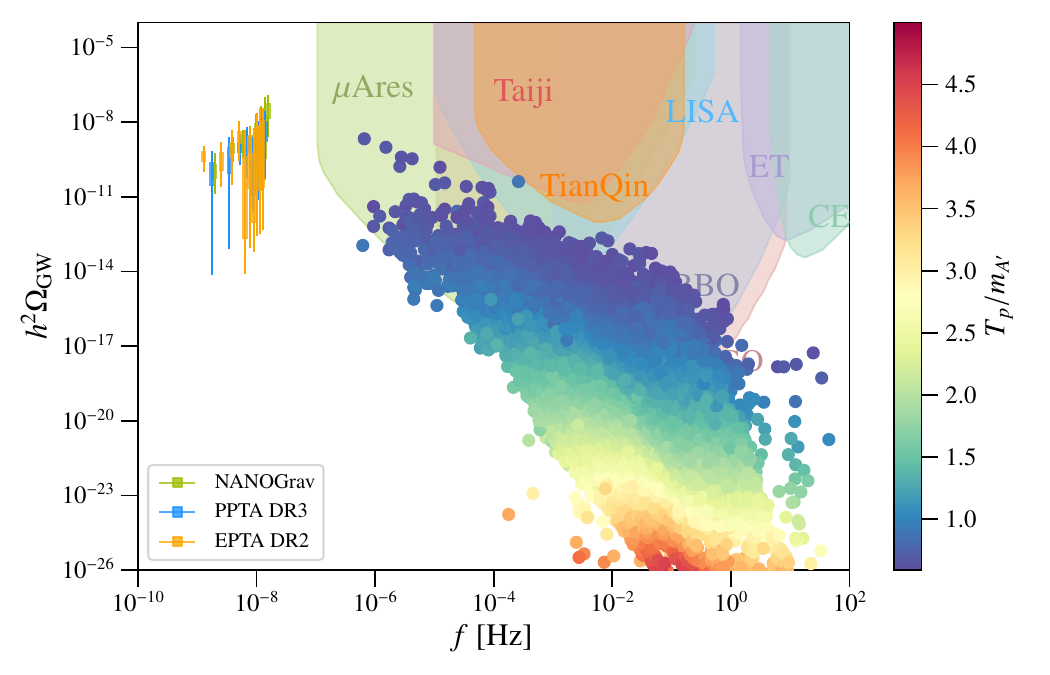} }
	\subfigure[]{
		\includegraphics[scale=0.46,trim=10 10 10 10,clip]{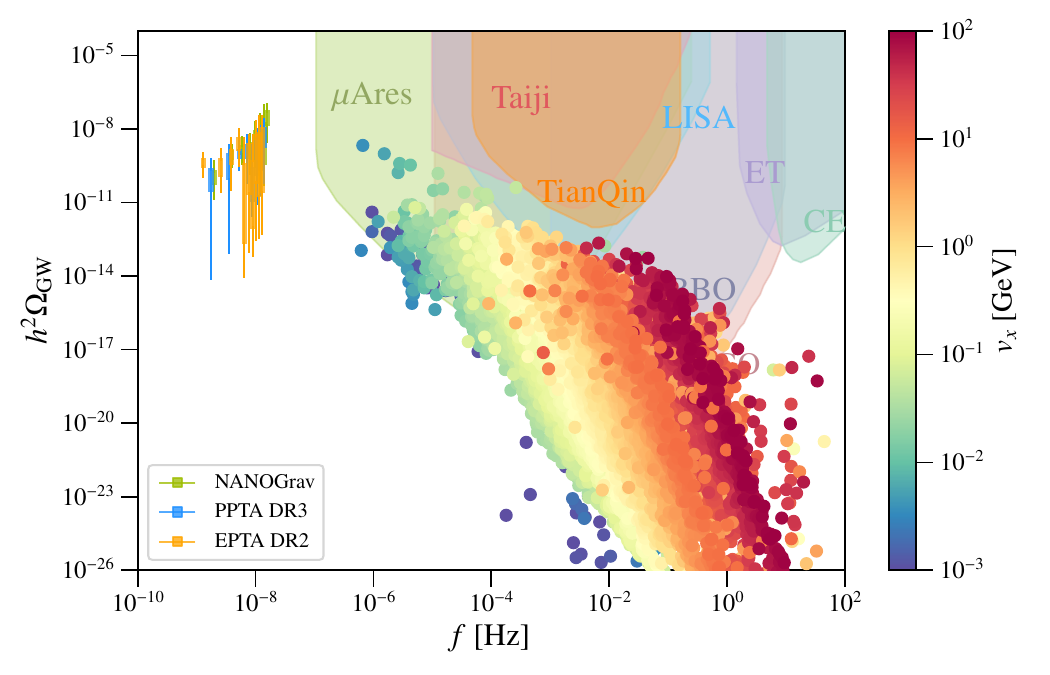} }
    \caption{\label{Fig:HT} Gauge-independent gravitational wave signals in the \textbf{high-temperature regime.}
    Each scatter point shows the peak frequency and peak amplitude of the stochastic background sourced by a first-order phase transition.
    In panel (a), points are color-coded by $T_p/m_{A^\prime}$.
    In panel (b), points are color-coded by the $U(1)_x$ symmetry breaking scale $v_x$, scanned over $v_x\in[10^{-3},10^{2}]~\mathrm{GeV}$.
    All points with $T_p/m_{A^\prime}<2$ violate the high-temperature assumption and \textbf{should be discarded.}}
\end{figure}

\begin{figure}[t!]
	\centering
	\subfigure[]{
		\includegraphics[scale=0.46,trim=10 10 10 10,clip]{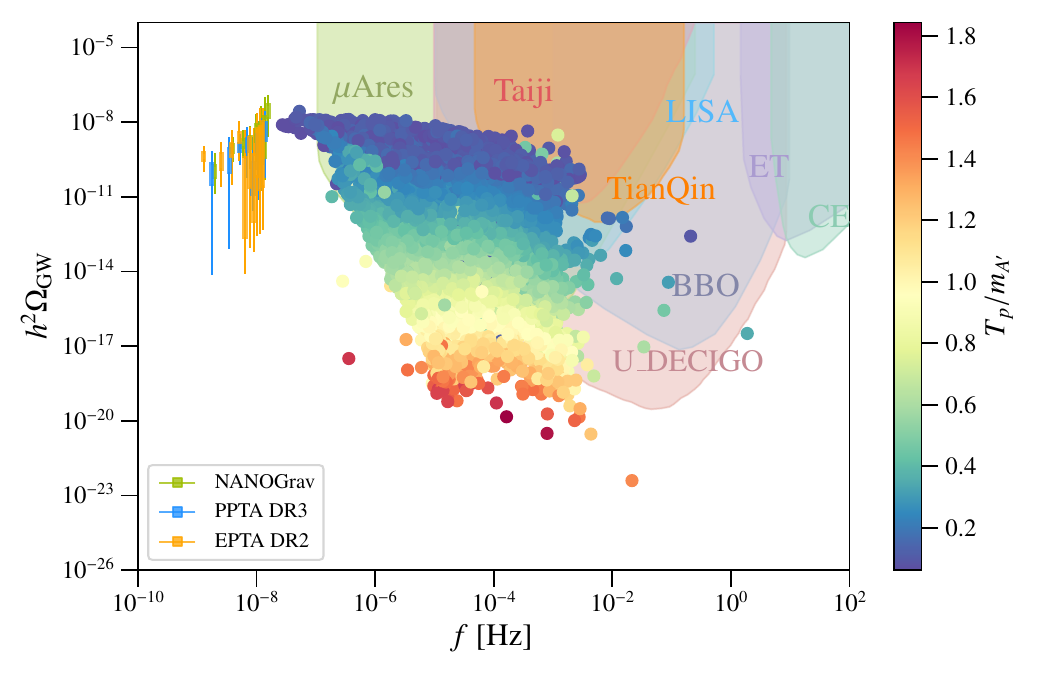} }
	\subfigure[]{
		\includegraphics[scale=0.46,trim=10 10 10 10,clip]{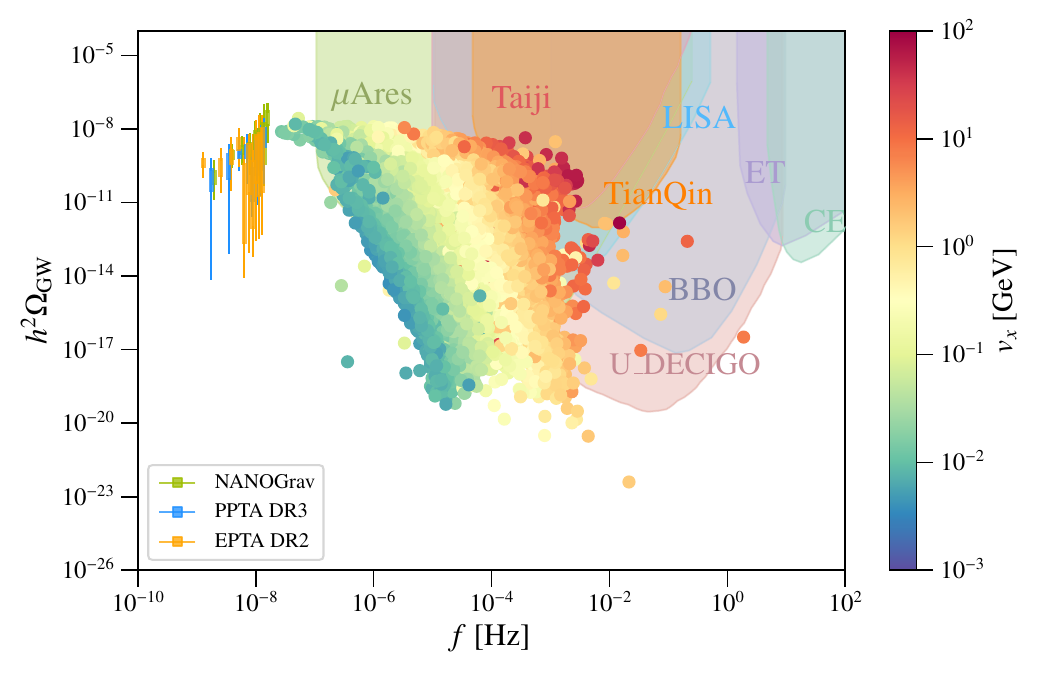} }
    	\caption{\label{Fig:LT} Gauge-independent gravitational wave signals in the \textbf{low-temperature regime.}
    Each scatter point shows the peak frequency and peak amplitude of the stochastic background sourced by a first-order phase transition.
    In panel (a), points are color-coded by $T_p/m_{A^\prime}$.
    In panel (b), points are color-coded by the $U(1)_x$ symmetry breaking scale $v_x$, scanned over $v_x\in[10^{-3},10^{2}]~\mathrm{GeV}$.
    All points with $T_p/m_{A^\prime}>0.5$ violate the low-temperature assumption and \textbf{should be discarded.}}
\end{figure}

For each regime, we sample $3\times 10^{4}$ points and compute the resulting stochastic gravitational wave background.
The results are displayed in the $\{f,\,\Omega_{\rm GW}h^{2}\}$ plane,
where each scatter point corresponds to the peak frequency and peak amplitude for a given parameter set.
The high-temperature scan is shown in Fig.~\ref{Fig:HT}, and the low-temperature scan in Fig.~\ref{Fig:LT}.
To illustrate the dependence on key parameters, we color-code the points by the ratio $T_{p}/m_{A^{\prime}}$
in the left panels and by $v_x$ in the right panels of Figs.~\ref{Fig:HT} and~\ref{Fig:LT}.
We recover the same qualitative trend observed in the gauge-dependent scan:
larger $v_x$ typically shifts the signal toward higher peak frequencies.

It is important to note, however, that the raw outputs of these scans include parameter points that lie
outside the temperature-limit assumptions underlying the corresponding gauge-independent treatments.
In particular, the high-temperature scan contains points with $T_{p}/m_{A^{\prime}}<2$,
which do not satisfy the (approximately) high-temperature criterion and must be discarded.
Likewise, the low-temperature scan contains points with $T_{p}/m_{A^{\prime}}>0.5$,
which fall outside the (approximately) low-temperature domain and must be discarded.

\begin{figure}[t!]
	\centering
	\includegraphics[scale=0.7,trim=10 10 10 5,clip]{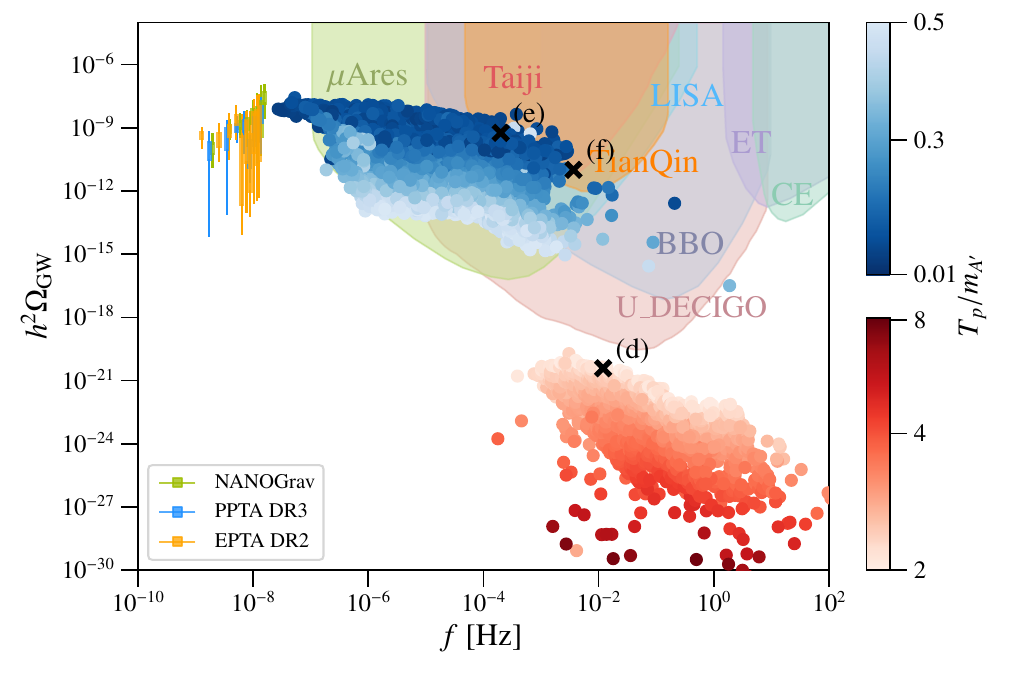}
    \caption{\label{Fig:combine} \textbf{Concrete gauge-independent gravitational wave predictions for the minimal gauged $U(1)_x$ dark sector,}
    obtained by combining the physically consistent points selected from Fig.~\ref{Fig:HT} ($T_p/m_{A^\prime} \gtrsim 2$) and Fig.~\ref{Fig:LT} ($T_p/m_{A^\prime} \lesssim 0.5$).
    The gap corresponds to the intermediate regime $0.5 < T_p/m_{A^\prime} < 2$, where neither the high- nor low-temperature gauge-independent treatment is applicable.
    Low-temperature points are shown in blue, with darker shades corresponding to smaller $T_p/m_{A^\prime}$ (stronger supercooling).
    High-temperature points are shown in red, with darker shades corresponding to larger $T_p/m_{A^\prime}$.
    Black crosses mark three benchmark models for Case~2, for which the dark matter evolution is computed explicitly (see Table~\ref{TableCase2}),
    providing benchmark targets for multi-messenger searches of the corresponding $U(1)$ model.
    In particular, benchmark models (e) and (f) lie within the projected sensitivity of Taiji, TianQin, and LISA.}
\end{figure}

To ensure consistency with the assumptions adopted in the preceding analysis,
we retain only points with $T_p/m_{A^{\prime}}\gtrsim 2$ as belonging to the (approximately) high-temperature regime,
and points with $T_p/m_{A^{\prime}}\lesssim 0.5$ as belonging to the (approximately) low-temperature regime.
Within the latter, points with $T_p/m_{A'}\lesssim 0.2$ correspond to genuinely supercooled transitions.
The surviving points from the two scans are combined in Fig.~\ref{Fig:combine},
which provides \textbf{concrete, gauge-independent predictions for the minimal gauged $U(1)_x$ dark sector.}

The empty gap in Fig.~\ref{Fig:combine} corresponds to the intermediate regime $0.5 < T_p/m_{A^\prime} < 2$,
where the high-temperature expansion is not valid and the low-temperature thermal contributions cannot be treated as small corrections.
Consequently, neither the high- nor low-temperature gauge-independent framework applies,
and concrete predictions in this intermediate regime are not currently under theoretical control.
Moreover, the corresponding gravitational wave signals are typically weaker than those from supercooled transitions
and are generally below the projected reach of near-future experiments,
rendering this regime less phenomenologically relevant at present.

As can be seen from Fig.~\ref{Fig:combine},
in the minimal $U(1)$ dark sector,
stochastic gravitational wave signals within the reach of current and planned detectors arise predominantly from supercooled phase transitions.
In particular, $U(1)_x$ vev in the range $v_x\in[10,100]~\mathrm{MeV}$
can yield peak frequencies approaching the nanohertz band,
offering a plausible interpretation of the PTA signal region.
While for $v_x\in[1,100]~\mathrm{GeV}$ the predicted peak frequencies
fall into the millihertz range targeted by proposed space-based interferometers such as Taiji, TianQin, and LISA,
providing a promising discovery window.

\begin{figure}[h]
	\centering
	\includegraphics[scale=0.7,trim=10 10 10 10,clip]{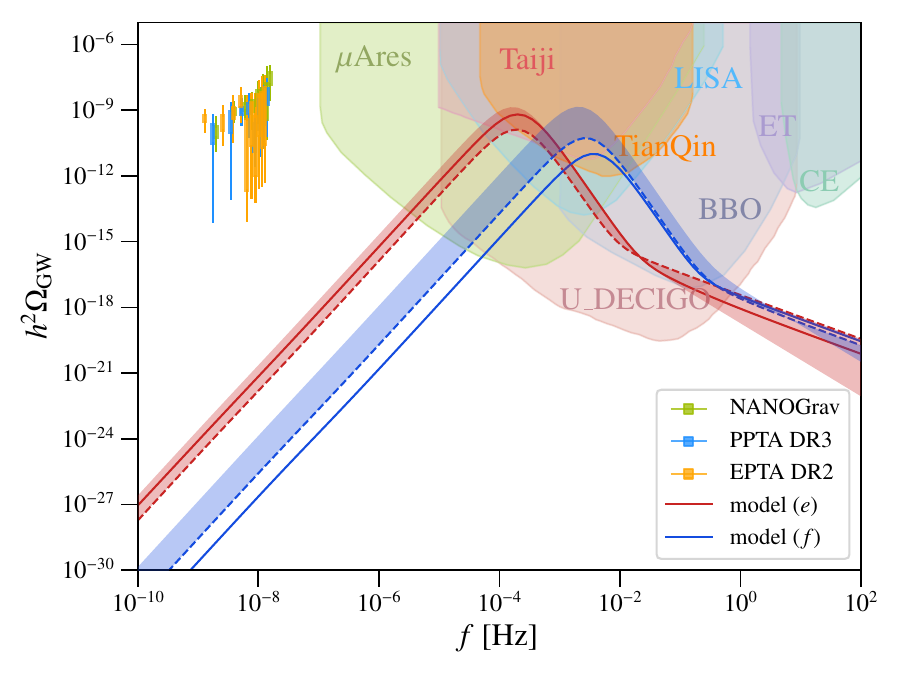}
    \caption{Comparison of the gauge-independent treatment with the gauge-dependent computation.
    Shown are the gravitational wave spectra $\Omega_{\rm GW}h^{2}$ as functions of frequency
    for benchmark models (e) and (f) in the low-temperature regime.
    The shaded bands indicate the gauge-dependent results obtained by varying the gauge parameter over $\xi\in[0,5]$.
    Solid curves denote the gauge-independent predictions, while dashed curves show the gauge-dependent results in Landau gauge ($\xi=0$).}
	\label{Fig:compareL}
\end{figure}

To illustrate the impact of gauge dependence on the predicted gravitational wave signals,
we explicitly compare our gauge-independent treatment with the gauge-dependent calculation in Fig.~\ref{Fig:compareL}.
As representative examples, we consider benchmark models (e) and (f), which undergo supercooled phase transitions,
and present the resulting gravitational wave spectra $\Omega_{\rm GW}h^{2}$ as functions of frequency.
The gauge-independent predictions are shown as solid curves.
The gauge-dependent results are obtained by varying the gauge parameter over $\xi\in[0,5]$
and are displayed as shaded bands; the dashed curves indicate the Landau gauge case with $\xi=0$.
One observes substantial deviations between the gauge-independent predictions and the gauge-dependent results.

\section{Evolution of dark sector particles}\label{Sec:Ph}

A successful dark sector model should also furnish viable dark matter candidates.
With the steadily improving sensitivity of dark matter detection experiments and collider searches,
many conventional freeze-out scenarios, especially in the traditional WIMP mass range $\mathcal{O}(10 - 100~\mathrm{GeV})$,
are increasingly constrained. In large regions of parameter space, the annihilation of dark matter into SM states,
\begin{equation}
XX~({\rm or}~X\overline{X}) \xrightarrow{\ \text{annihilation}\ } \text{SM particles}\,,
\end{equation}
is no longer efficient enough to yield the observed relic abundance without conflicting with existing bounds~\cite{Arcadi:2024ukq,Feng:2024blk}.

Within the minimal $U(1)_x$ setup, two well-motivated possibilities arise, as discussed in Section~\ref{Sec:U1ex}.
First, the dark photon can serve as dark matter provided it is stable,
which typically requires vanishing or negligibly small kinetic mixing with the hypercharge gauge field.
In this limit its relic abundance is set by Higgs portal interactions.
Second, a vectorlike dark fermion can be the dark matter candidate,
communicating with the SM predominantly through the dark photon portal,
since it does not couple directly to the dark Higgs in the minimal field content,
a non-negligible kinetic mixing is then required to maintain efficient interactions.
Alternatively, one may introduce two vectorlike fermions $\chi_1$ and $\chi_2$
with an allowed Yukawa interaction $\sim H_x\,\overline{\chi_1}\chi_2$
for an appropriate assignment of $U(1)_x$ charges.

In this paper, we consider both classes of dark matter candidates in the secluded regime,
where they annihilate into light dark sector mediators that subsequently decay into SM states.

\subsection{Boltzmann equations for dark particles}

\subsubsection{Dark photon dark matter from the Higgs portal}

The dominant annihilation channel for dark photon dark matter proceeds via
\begin{equation}
A^\prime A^\prime  \xrightarrow{\ \text{annihilation}\ } h_x h_x \xrightarrow{\quad\text{decay}\quad} \text{SM\ particles}\,.
\end{equation}
The coupled Boltzmann equations governing $Y_{h_x}$ and $Y_{A^\prime}$ are given by
\begin{align}
	\frac{\mathrm{d}Y_{h_{x}}}{\mathrm{d}T}= & \;-\frac{s}{T\overline{H}}\sum_{i\in\mathrm{SM}}\Bigl\{2\bigl[(Y_{h_{x}}^{\mathrm{eq}})^{2}-Y_{h_{x}}^{2}\bigr]\left\langle \sigma v\right\rangle _{h_{x}h_{x}\to\{i\overline{i},hh,ZZ,W^{\pm}W^{\mp}\}}\nonumber \\
	& +(Y_{h}Y_{h_{x}}^{\mathrm{eq}}-Y_{h}Y_{h_{x}})\left\langle \sigma v\right\rangle _{hh_{x}\to\{i\overline{i},hh,ZZ,W^{\pm}W^{\mp}\}}\nonumber \\
	& +2Y_{A^{\prime}}^{2}\left\langle \sigma v\right\rangle _{A^{\prime}A^{\prime}\to h_{x}h_{x}}-2Y_{h_{x}}^{2}\left\langle \sigma v\right\rangle _{h_{x}h_{x}\to A^{\prime}A^{\prime}}\nonumber \\
	& +Y_{A^{\prime}}^{2}\left\langle \sigma v\right\rangle _{A^{\prime}A^{\prime}\to hh_{x}}-Y_{h}Y_{h_{x}}\left\langle \sigma v\right\rangle _{hh_{x}\to A^{\prime}A^{\prime}}\nonumber \\
	& +2\,\theta(m_{h}-2m_{h_{x}})\bigl(-Y_{h_{x}}^{2}\left\langle \sigma v\right\rangle _{h_{x}h_{x}\to h}+\tfrac{1}{s}Y_{h}\langle\Gamma\rangle_{h\to h_{x}h_{x}}\bigr)\nonumber \\
	& +\theta(m_{h_{x}}-2m_{i})\,\tfrac{1}{s}(Y_{h_{x}}^{{\rm eq}}-Y_{h_{x}})\langle\Gamma\rangle_{h_{x}\to i\overline{i}}\nonumber \\
	& +\theta(m_{h_{x}}-2m_{A^{\prime}})\bigl(Y_{A^{\prime}}^{2}\left\langle \sigma v\right\rangle _{A^{\prime}A^{\prime}\to h_{x}}-\tfrac{1}{s}Y_{h_{x}}\langle\Gamma\rangle_{h_{x}\to A^{\prime}A^{\prime}}\bigr)\Bigr\}\,,\label{eq: Case1_Yhx}\\
	\frac{\mathrm{d}Y_{A^{\prime}}}{\mathrm{d}T}= & \;-2\times\frac{s}{T\overline{H}}\sum_{i\in\mathrm{SM}}\Bigl\{\bigl[(Y_{A^{\prime}}^{\mathrm{eq}})^{2}-Y_{A^{\prime}}^{2}\bigr]\left\langle \sigma v\right\rangle _{A^{\prime}A^{\prime}\to\{i\overline{i},hh,ZZ,W^{\pm}W^{\mp}\}}\nonumber \\
	& -Y_{A^{\prime}}^{2}\left\langle \sigma v\right\rangle _{A^{\prime}A^{\prime}\to h_{x}h_{x}}+Y_{h_{x}}^{2}\left\langle \sigma v\right\rangle _{h_{x}h_{x}\to A^{\prime}A^{\prime}}\nonumber \\
	& -Y_{A^{\prime}}^{2}\left\langle \sigma v\right\rangle _{A^{\prime}A^{\prime}\to hh_{x}}+Y_{h}Y_{h_{x}}\left\langle \sigma v\right\rangle _{hh_{x}\to A^{\prime}A^{\prime}}\nonumber \\
	& +\theta(m_{h}-2m_{A^{\prime}})\bigl(-Y_{A^{\prime}}^{2}\left\langle \sigma v\right\rangle _{A^{\prime}A^{\prime}\to h}+\tfrac{1}{s}Y_{h}\langle\Gamma\rangle_{h\to A^{\prime}A^{\prime}}\bigr)\nonumber \\
	& +\theta(m_{h_{x}}-2m_{A^{\prime}})\bigl(-Y_{A^{\prime}}^{2}\left\langle \sigma v\right\rangle _{A^{\prime}A^{\prime}\to h_{x}}+\tfrac{1}{s}Y_{h_{x}}\langle\Gamma\rangle_{h_{x}\to A^{\prime}A^{\prime}}\bigr)\Bigr\}\,,
	\label{eq: Case1_YA}
\end{align}
where
\begin{equation}
	\overline{H}=\frac{H}{1+\frac{1}{3}\frac{T}{h_{{\rm eff}}}\frac{{\rm d}h_{{\rm eff}}}{{\rm d}T}}=\sqrt{\frac{\pi^{2}g_{{\rm eff}}}{90}}\frac{T^{2}/M_{{\rm Pl}}}{1+\frac{1}{3}\frac{T}{h_{{\rm eff}}}\frac{{\rm d}h_{{\rm eff}}}{{\rm d}T}}\,.
\end{equation}

\subsubsection{Dark fermion dark matter from the $U(1)$ portal}

The benchmark models we consider are chosen such that
the dark Higgs $h_x$ remains in thermal equilibrium and decays into SM fermions before BBN,
thus not affecting the number density evolution of the dark matter or the dark photon.
We therefore focus on the coupled evolution of $\chi$ and $A^\prime$.
The dominant annihilation channel for dark fermion dark matter is
\begin{equation}
\chi\bar{\chi}\xrightarrow{\quad Z,\ A^\prime\quad} {\rm SM\ particles}\,.
\end{equation}
The coupled Boltzmann equations for $Y_\chi$ and $Y_{A^\prime}$ are given by
\begin{align}
	\frac{\mathrm{d}Y_{\chi}}{\mathrm{d}T}= & \;-\frac{s}{T\overline{H}}\sum_{i\in{\rm SM}}\Bigl\{\bigl[(Y_{\chi}^{\mathrm{eq}})^{2}-Y_{\chi}^{2}\bigr]\langle\sigma v\rangle_{\chi\overline{\chi}\rightarrow\{i\overline{i},ZZ\}}\nonumber \\
	& -Y_{\chi}^{2}\langle\sigma v\rangle_{\chi\overline{\chi}\to A^{\prime}Z}+Y_{A^{\prime}}Y_{Z}\langle\sigma v\rangle_{A^{\prime}Z\to\chi\overline{\chi}}\nonumber \\
	& -Y_{\chi}^{2}\langle\sigma v\rangle_{\chi\overline{\chi}\to A^{\prime}A^{\prime}}+Y_{A^{\prime}}^{2}\langle\sigma v\rangle_{A^{\prime}A^{\prime}\to\chi\overline{\chi}}\nonumber \\
	& +\theta(m_{A^{\prime}}-2m_{\chi})\bigl(-Y_{\chi}^{2}\langle\sigma v\rangle_{\overline{\chi}\chi\rightarrow A^{\prime}}+\tfrac{1}{s}Y_{A^{\prime}}\langle\Gamma\rangle_{A^{\prime}\to\chi\overline{\chi}}\bigr)\Big\}\,, \label{eq: Case2_YD}\\
	\frac{\mathrm{d}Y_{A^{\prime}}}{\mathrm{d}T}= & \;-\frac{s}{T\overline{H}}\sum_{i\in{\rm SM}}\Bigl\{(Y_{A^{\prime}}^{{\rm eq}}-Y_{A^{\prime}})\left[Y_{Z}\langle\sigma v\rangle_{A^{\prime}Z\to i\overline{i}}+Y_{\gamma}\langle\sigma v\rangle_{A^{\prime}\gamma\to i\overline{i}}\right.\nonumber \\
	& \left.+2Y_{i}\bigl(\langle\sigma v\rangle_{i A^{\prime}\to iZ}+\langle\sigma v\rangle_{iA^{\prime}\to i\gamma}\bigr)\right]\nonumber \\
	& +Y_{\chi}^{2}\langle\sigma v\rangle_{\chi\overline{\chi}\to A^{\prime}A^{\prime}}-Y_{A^{\prime}}^{2}\langle\sigma v\rangle_{A^{\prime}A^{\prime}\to\chi\overline{\chi}}\nonumber \\
	& +\theta(m_{A^{\prime}}-2m_{\chi})\bigl(Y_{\chi}^{2}\langle\sigma v\rangle_{\overline{\chi}\chi\rightarrow A^{\prime}}-\tfrac{1}{s}Y_{A^{\prime}}\langle\Gamma\rangle_{A^{\prime}\to\chi\overline{\chi}}\bigr)\nonumber \\
	& +\theta(m_{A^{\prime}}-2m_{i})\bigl(Y_{i}^{2}\langle\sigma v\rangle_{i\overline{i}\rightarrow A^{\prime}}-\tfrac{1}{s}Y_{A^{\prime}}\langle\Gamma\rangle_{A^{\prime}\to i\overline{i}}\bigr)\Big\}\,.
	\label{eq: Case2_YA}
\end{align}

\subsection{Benchmark models}\label{sec:BM}

\paragraph*{Case-1: dark photon dark matter}

Current bounds on the mixing parameter $|\sin\theta\,|$, c.f., Eq.~(\ref{eq: HMang}) arise from
rare meson decays, beam-dump experiments, collider searches, and astrophysical observations,
see~\cite{Beacham:2019nyx,Arcadi:2021mag,NA62:2025upx} for recent summaries.
Colliders also constrain invisible Higgs decays.
The ATLAS Collaboration reports ${\rm Br}(h\to{\rm inv})<0.107$ at $95\%$ CL~\cite{ATLAS:2023tkt},
while CMS finds ${\rm Br}(h\to{\rm inv})<0.15$ at $95\%$ CL~\cite{CMS:2023sdw}.
The invisible branching fraction is
\begin{equation}
	{\rm Br}(h\to{\rm inv})=\frac{\Gamma_{h\to{\rm inv}}}{\Gamma_{h\to{\rm SM}}+\Gamma_{h\to{\rm inv}}}\,,\qquad\Gamma_{h\to{\rm inv}}=\Gamma_{h\to h_{x}h_{x}}+\Gamma_{h\to A^{\prime}A^{\prime}}\,,
\end{equation}
where for $m_h\simeq125~{\rm GeV}$ the SM Higgs width is $\Gamma_{h\to{\rm SM}}=4.1~{\rm MeV}$~\cite{LHCHiggsCrossSectionWorkingGroup:2016ypw}.
Since the dark Higgs $h_x$ ultimately decays into SM fermions, its lifetime must be shorter than $\mathcal{O}(1~{\rm s})$ to avoid spoiling big-bang nucleosynthesis.

For this scenario we present three benchmark models in Table~\ref{TableCase1},
with $v_x=$ 100, 200, 450~GeV and a mass hierarchy $m_{A^{\prime}}>m_{h_x}$.
This hierarchy is required to realize a first-order phase transition and hence to generate a stochastic gravitational wave signal.
All benchmarks are consistent with current experimental bounds.

For each benchmark we compute both the number density evolution of the relevant dark sector species and the gravitational wave signal from the phase transition.
The Boltzmann equations governing $Y_{h_x}$ and $Y_{A^{\prime}}$ are given in Eqs.~(\ref{eq: Case1_Yhx})--(\ref{eq: Case1_YA}).
As the Universe cools, the dark sector undergoes a first-order phase transition during which $U(1)_x$
is spontaneously broken and the dark photon $A^{\prime}$ acquires a mass.
After the transition, $A^{\prime}$ annihilates into lighter states, including the dark Higgs $h_x$ and SM particles,
and subsequently freezes out,  yielding the observed dark matter relic abundance.
The dark Higgs $h_x$ is unstable and decays into SM fermions before big bang nucleosynthesis (BBN).

Although these benchmarks satisfy the dark matter relic density constraint,
they typically require a tiny dark gauge coupling $g_x$ to avoid excessive secluded depletion of the dark photon.
As a result, the phase transition is relatively weak and the corresponding gravitational wave signal is suppressed.
Moreover, for these benchmarks we find that $T_p/m_{A^{\prime}}$ lies outside the validity ranges of our gauge-independent treatments in both the high- and low-temperature limits.

\begin{table}[h!]
	\centering
	\resizebox{\textwidth}{10.3mm}{
		\begin{tabular}{|c|c|c|c|c|c|c|c|c|c|c|}
			\hline
			Model & $v_{x}$ & $g_{x}$ & $\lambda_{x}$ & $\lambda_{{\rm mix}}$ & $m_{h_{x}}$ & $m_{A^{\prime}}$ & $\Omega_{A^{\prime}}h^{2}$ & $\tau_{h_{x}}$ & $T_{p}$ & $T_{p}/m_{A^{\prime}}$\tabularnewline
			\hline
			(a) & $100$ & $0.0221$ & $1.84\times10^{-7}$ & $1\times10^{-5}$ & $0.061$ & $2.21$ & $0.12$ & $2.61\times10^{-3}$ & $3.90$ & $1.77$\tabularnewline
			\hline
			(b) & $200$ & $0.0372$ & $1.00\times10^{-6}$ & $1\times10^{-5}$ & $0.28$ & $7.44$ & $0.12$ & $5.19\times10^{-7}$ & $10.85$ & $1.46$\tabularnewline
			\hline
			(c) & $450$ & $0.0525$ & $2.95\times10^{-6}$ & $1\times10^{-6}$ & $1.09$ & $23.63$ & $0.12$ & $9.64\times10^{-7}$ & $29.69$ & $1.26$\tabularnewline
			\hline
	\end{tabular}}
    \caption{Benchmark models for Case~1, in which the dark photon is the dark matter candidate
    and the dark sector communicates with the SM through the Higgs portal.
    All masses are in GeV.
    For each benchmark we compute both the dark sector number density evolution and the first-order phase transition dynamics.
    The Higgs mixing parameter $\lambda_{\rm mix}$ satisfies current experimental bounds,
    and the $h_x$ lifetimes (in seconds) are ensured to be shorter than $1~\mathrm{s}$.
    We also list the percolation temperature $T_p$ and the ratio $T_p/m_{A^{\prime}}$ for each benchmark.}
	\label{TableCase1}
\end{table}

\paragraph*{Case-2: dark fermion dark matter}

In Case~2, the dark matter candidate is a vectorlike dark fermion $\chi$ with no direct Yukawa coupling to the dark Higgs $h_x$.
As a result, $\chi$ affects the finite-temperature effective potential $V_{\rm eff}(\phi_c,T)$ only indirectly through its contribution to the longitudinal dark photon thermal mass $\Pi_{A^\prime}$, c.f., Eq.~(\ref{eq: PiZp}).
Consequently, in the gauge-independent calculation the coefficient appearing in $\Pi_{A^\prime}$ is modified to $2/3$.

The kinetic-mixing parameter $\delta$ in this work denotes the mixing between $U(1)_x$ and the hypercharge gauge field $U(1)_Y$, c.f., Eq.~(\ref{eq:KM}).
Experimental searches constrain the kinetic mixing between the dark photon and the electromagnetic photon, parameterized by $\boldsymbol{\delta}$.
The two parameters are related by~\cite{Feng:2023ubl,Feng:2024nkh}
\begin{equation}
	\delta \approx \frac{\sqrt{g_{2}^{2}+g_{Y}^{2}}}{g_{2}}\boldsymbol{\delta}\,.\label{eq:delta}
\end{equation}

\begin{figure}[h!]
	\centering
	\includegraphics[scale=0.65,trim=10 10 10 5,clip]{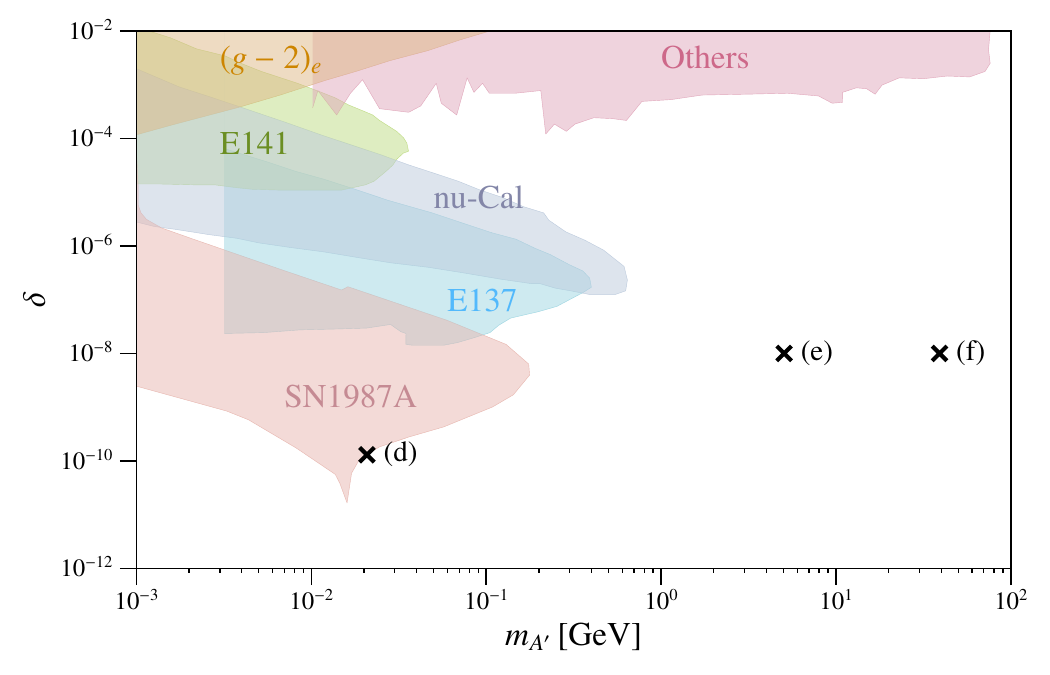}
	\caption{
    Experimental constraints on the kinetic mixing parameter $\delta$, defined in Eqs.~(\ref{eq:KM}) and~(\ref{eq:delta}),
    for dark photon masses $m_{A^{\prime}}\in[1{\rm MeV},\,100{\rm GeV}]$.
    The three Case~2 benchmark models (Table~\ref{TableCase2}) are shown as black crosses and are consistent with all current experimental bounds.
    Constraints on the kinetic mixing for dark photons with $m_{A^{\prime}}\gtrsim$~1~MeV arise from
    supernova (SN1987A)~\cite{Chang:2016ntp,Caputo:2025avc}, $(g-2)_{e}$~\cite{Pospelov:2008zw} and from various beam dump experiments: E141~\cite{Riordan:1987aw}, E137~\cite{Bjorken:1988as,Batell:2014mga,Marsicano:2018krp}, $\nu$-Cal~\cite{Blumlein:2011mv,Blumlein:2013cua}.
    The pink region in the upper portion of the plot, labeled as ``Others'',
    collects constraints from collider and fixed-target experiments, including:
    CMS~\cite{CMS:2019kiy}, BaBar~\cite{BaBar:2014zli}, LHCb~\cite{LHCb:2019vmc}, KLOE~\cite{KLOE-2:2011hhj,KLOE-2:2012lii,KLOE-2:2014qxg,KLOE-2:2016ydq} and A1~\cite{Merkel:2014avp}.}
	\label{Fig:KMbound}
\end{figure}

For Case~2, we choose three benchmark models, listed in Table~\ref{TableCase2}, with $v_x=$ 1, 10, 100~GeV.
For each benchmark we compute both the dark sector number density evolution and the associated gravitational wave signal.
All benchmarks satisfy current experimental constraints on the kinetic mixing parameter $\delta$ and are shown in Fig.~\ref{Fig:KMbound}.
The Boltzmann equations governing the evolution of $Y_\chi$ and $Y_{A^\prime}$ are given in Eqs.~(\ref{eq: Case2_YD})--(\ref{eq: Case2_YA}).

In model (d), the dark fermion is heavier than the dark photon, $m_\chi>m_{A^\prime}$.
As the Universe cools, $\chi\bar{\chi}$ annihilates efficiently into dark photons and yields the observed relic abundance.
The dark photons subsequently decay into $e^+e^-$ pairs before BBN.
For models~(e) and (f), we instead choose $m_\chi$ to lie slightly below $m_{A^\prime}$.
This near-degeneracy allows for a larger $g_x$ while still reproducing the correct $\Omega_\chi h^{2}$ via the process $A^{\prime}A^{\prime}\to\chi\overline{\chi}$.
We indicate these three Case~2 benchmarks in Figs.~\ref{Fig:combine} and~\ref{Fig:KMbound}.
Benchmark models (e) and (f) lie within the projected sensitivity of Taiji, TianQin, and LISA,
and thus provide promising multi-messenger targets for a genuinely minimal gauged $U(1)$ dark sector.

\begin{table}[h!]
	\centering
	\resizebox{\textwidth}{10.5mm}{
		\begin{tabular}{|c|c|c|c|c|c|c|c|c|c|c|}
			\hline
			Model & $v_{x}$ & $g_{x}$ & $\lambda_{x}$ & $\delta$ & $m_{\chi}$ & $m_{A^{\prime}}$ & $\Omega_{\chi}h^{2}$ & $\tau_{A^{\prime}}$ & $T_{p}$ & $T_{p}/m_{A^{\prime}}$\tabularnewline
			\hline
			(d) & $1$ & $0.0208$ & $1.93\times10^{-7}$ & $1.3\times10^{-10}$ & $0.35$ & $0.0208$ & $0.12$ & $0.35$ & $0.0423$ & $2.03$\tabularnewline
			\hline
			(e) & $10$ & 0.505 & $1.17\times10^{-4}$ & $1\times10^{-8}$ & $4.50$ & 5.05 & 0.12 & $1.05\times10^{-7}$ & $0.54$ & $0.107$\tabularnewline
			\hline
			(f) & $100$ & 0.391 & $3.13\times10^{-5}$ & $1\times10^{-8}$ & $38.50$ & 39.10 & 0.12 & $1.14\times10^{-8}$ & $3.34$ & $0.085$\tabularnewline
			\hline
	\end{tabular}}
    \caption{Benchmark models for Case~2, in which the dark matter candidate is a vectorlike fermion $\chi$ that communicates with the SM through $U(1)$ portal via kinetic mixing.
    All masses are in GeV and lifetimes in seconds.
    For each benchmark we compute both the dark sector number density evolution and the first-order phase transition dynamics.
    All benchmarks satisfy the relevant constraints on the kinetic mixing parameter $\delta$ (see Fig.~\ref{Fig:KMbound}).
    The corresponding peak gravitational wave signals are shown in Fig.~\ref{Fig:combine}.
    We also list the percolation temperature $T_p$ and the ratio $T_p/m_{A^{\prime}}$ for each benchmark.}
	\label{TableCase2}
\end{table}

\subsection{Phenomenological implications}\label{sec:DMimp}

The continuing non-observation of weakly interacting massive particles (WIMP),
especially in the $\mathcal{O}(10-100~{\rm GeV})$ region,
has substantially challenged a large class of traditional dark matter scenarios.
Current limits from direct and indirect detection, together with collider searches,
have significantly narrowed the viable parameter space for many WIMP candidates.
As a result, the measured relic abundance has increasingly become a stringent constraint
on model building rather than an automatic target that can be achieved across broad regions of parameter space.

At the level of thermal freeze-out, the tension is particularly transparent:
with present experimental sensitivities, many WIMP-like candidates
that remain compatible with direct detection bounds cannot maintain
sufficient annihilation into SM final states,
which leads to an overproduction of dark matter.
In hidden sector constructions, this difficulty can be even more pronounced,
since the interactions connecting the dark sector to the SM
are typically mediated by portal couplings that must be small.
Consequently, annihilation directly into SM particles can be inefficient.
A common alternative is secluded annihilation,
in which dark matter annihilates dominantly into lighter dark mediators
that subsequently decay into SM states,
thereby achieving an efficient depletion of the relic abundance
while alleviating dark matter experimental constraints.

In the minimal $U(1)_x$ framework considered here,
these general considerations manifest themselves clearly in the benchmark models.
As illustrated in Fig.~\ref{Fig:combine}, parameter points with dark fermion dark matter can simultaneously
achieve sufficient annihilation to match the observed relic density
and produce a potentially detectable stochastic gravitational wave signal from the first-order phase transition.
By contrast, dark photon dark matter realized through secluded freeze-out
cannot accommodate both requirements:
regions of parameter space that yield an observable gravitational wave signal
often require a sizable gauge coupling,
while such a coupling tends to deplete the dark photon relic abundance too efficiently,
making it difficult to retain the correct relic density.
This complementarity is not surprising.
Detectable gravitational wave signals from a first-order phase transition
generally favor a sizable gauge coupling,
which strengthens the dynamics of the transition and enhances the signal.
However, the same sizable coupling often increases the efficiency of dark photon annihilation,
reducing the relic density below the observed value via freeze-out.

Dark fermion dark matter can annihilate into dark photons,
which then decay into SM particles via portal interactions.
The presence of an additional mass parameter $m_\chi$ offers additional flexibility.
In particular, choosing $m_\chi$ close to the dark photon mass
can suppress the annihilation efficiency and prevent excessive depletion,
enabling simultaneous compatibility with the relic abundance and gravitational wave detectability.

It should be noted that in the minimal $U(1)_x$ setup,
the dark photon and dark Higgs masses are $m_{A^\prime} = g_x v_x$ and $m_{h_x}=\sqrt{2\lambda}\, v_x$, respectively.
Given the scaling relation between the Higgs quartic coupling and the gauge coupling discussed in detail in Section~\ref{Sec:Gauge}
in both high- and low-temperature regimes, the dark photon is generically heavier than the dark Higgs.
Accordingly, in all benchmark models considered here we impose $m_{A'}>m_{h_x}$,
a hierarchy that is required for a dark sector phase transition in this minimal setup.

Finally, we emphasize that production via freeze-in can remain viable for
both classes of dark matter candidates even when the gauge coupling is sizable.
For dark fermion dark matter, previous studies, see, e.g.,~\cite{Aboubrahim:2020lnr,Feng:2023ubl},\footnote{
Although these references consider a simplified $U(1)_x$ setup consisting only of a vectorlike fermion and a dark photon
whose mass arises from the St\"uckelberg mechanism,
they already illustrate the broad viability and flexibility of freeze-in production in $U(1)$ portal scenarios.}
demonstrate that large regions of parameter space can satisfy existing experimental constraints
while reproducing the observed relic abundance.
For dark photon dark matter, even if the hidden sector possesses a sizable gauge coupling,
the Higgs portal mixing can be adjusted to generate an appropriate abundance.
More broadly, freeze-in production can offer greater freedom to maintain a sizable gauge coupling,
and hence potentially large gravitational wave signals,
while still achieving the correct relic density.
A systematic exploration of this possibility,
in particular for dark photon dark matter in the minimal gauged $U(1)_x$ setup,
is left open for future work.

\section{Conclusion and outlook}\label{Sec:Con}

In this work, we investigated gauge-independent predictions for stochastic gravitational waves
sourced by first-order phase transitions in a minimal gauged $U(1)_x$ dark sector.
This addresses a central obstacle to making reliable predictions in gauge theories:
the gauge dependence of the finite-temperature effective potential and the associated tunneling action.
To obtain model-intrinsic and reproducible results,
we adopted a gauge-independent formulation of the effective action based on the Nielsen identity,
combined with a controlled derivative expansion and power counting.
This framework yields a gauge-independent tunneling action,
enabling a well-defined treatment of nucleation dynamics and, consequently, robust gravitational wave predictions.
The resulting signals span a wide range of peak frequencies relevant to pulsar timing arrays and planned space-based interferometers,
and they exhibit systematic trends across parameter space.

We performed Monte Carlo explorations of the parameter space using both the conventional gauge-dependent approach
and, separately, the gauge-independent framework in the high- and low-temperature limits.
The gauge-dependent scan is useful for building intuition and reveals clear qualitative trends:
the $U(1)_x$ symmetry breaking scale $v_x$ largely controls the characteristic frequency,
while the phase transition temperature regime correlates strongly with the signal strength,
with supercooled transitions typically producing larger amplitudes.
At the same time, an explicit comparison shows that gauge-dependent spectra can depart significantly
from the gauge-independent predictions when the gauge-fixing parameter is varied,
underscoring that the gauge choice can materially impact phenomenological interpretations.

Using the gauge-independent effective action, we scanned the high- and low-temperature limits separately
and retained only those parameter points that are self-consistent with the corresponding temperature-limit assumptions.
The surviving points shown in Fig.~\ref{Fig:combine} hence provide concrete, gauge-independent predictions
for the minimal gauged $U(1)_x$ dark sector,
directly mapping viable microscopic parameters to detector-facing peak frequencies and amplitudes.
We find that, in the minimal $U(1)_x$ dark sector,
detectable signals in current and planned experiments arise predominantly from supercooled phase transitions.
In particular, a $U(1)_x$ symmetry breaking scale $v_x\in[10,100]~\mathrm{MeV}$ can yield peak frequencies approaching the nanohertz band,
overlapping the PTA signal region;
whereas $v_x\in[1,100]~\mathrm{GeV}$ leads to peak frequencies in the millihertz range targeted by proposed space-based interferometers such as Taiji, TianQin, and LISA.

A notable outcome of our scan is that supercooled transitions make up a sizable fraction of viable points and, when realized, tend to populate the most sensitive regions of the experimental landscape. By contrast, parametrically high-temperature transitions are comparatively rare and typically produce weaker signals. Between these limits lies an intermediate regime in which neither the high- nor low-temperature gauge-independent treatment is applicable.
Bridging this gap with a fully gauge-independent framework remains an important open direction.

Finally, we connected the phase transition phenomenology to viable dark matter candidates within the same minimal field content.
We considered both dark photon dark matter via the Higgs portal,
and dark fermion dark matter via the dark photon portal, in secluded settings,
and provided benchmark targets that enable genuine multi-messenger tests.
The benchmarks illustrate a clear complementarity:
parameter regions capable of producing potentially detectable gravitational waves typically favor sizable gauge dynamics,
which can be in tension with reproducing the observed relic abundance for dark photon dark matter via secluded freeze-out.
By contrast, dark fermion dark matter retains greater flexibility and can more readily accommodate both relic density consistency and gravitational wave visibility.
More broadly, freeze-in production channels may further expand the viable parameter space
while maintaining sizable gauge couplings, and merit systematic exploration in future work.

Across all temperature regimes, we find a robust feature of the parameter region that supports a first-order phase transition:
the dark photon must be heavier than the dark Higgs.
If nature were to realize the opposite hierarchy, a $U(1)_x$ breaking phase transition
in the minimal dark sector and thus the associated gravitational wave signal will not occur.


Taken together, this work delivers a gauge-independent, end-to-end pipeline that connects a minimal hidden sector Lagrangian
to nucleation dynamics, stochastic gravitational wave spectra, and cosmologically viable dark matter benchmarks.
Our results make clear that gauge independence is not merely a technical refinement,
but a necessary condition for obtaining physically meaningful gravitational wave predictions in gauge theories.
Within this framework, even the minimal gauged $U(1)_x$ dark sector yields predictive, detector-relevant targets,
most notably in the supercooled regime,
that are complementary to collider probes and conventional dark matter searches.

Looking ahead, the setup studied here admits several natural extensions, including additional hidden states,
gauge-independent conformal $U(1)$ realizations, and explorations of phase transitions at larger symmetry breaking scales.
More broadly, the lesson is general:
whenever gauge fields and spontaneous symmetry breaking play a dynamical role,
gravitational wave predictions should be formulated in gauge-independent terms
to enable reliable comparisons with experimental data and across different theoretical studies.\\

\noindent\textbf{Acknowledgments:}

This work is
supported in part by the National Natural Science Foundation of China under Grant No. 11935009,
and Tianjin University Self-Innovation Fund Extreme Basic Research Project Grant No. 2025XJ21-0007.


\begin{thebibliography}{99}
\bibitem{Anderson:1991zb}
G.~W.~Anderson and L.~J.~Hall,
Phys. Rev. D \textbf{45}, 2685-2698 (1992)
doi:10.1103/PhysRevD.45.2685

\bibitem{Carrington:1991hz}
M.~E.~Carrington,
Phys. Rev. D \textbf{45}, 2933-2944 (1992)
doi:10.1103/PhysRevD.45.2933

\bibitem{Arnold:1992rz}
P.~B.~Arnold and O.~Espinosa,
Phys. Rev. D \textbf{47}, 3546 (1993)
[erratum: Phys. Rev. D \textbf{50}, 6662 (1994)]
doi:10.1103/PhysRevD.47.3546
[arXiv:hep-ph/9212235 [hep-ph]].

\bibitem{Kajantie:1996mn}
K.~Kajantie, M.~Laine, K.~Rummukainen and M.~E.~Shaposhnikov,
Phys. Rev. Lett. \textbf{77}, 2887-2890 (1996)
doi:10.1103/PhysRevLett.77.2887
[arXiv:hep-ph/9605288 [hep-ph]].

\bibitem{Grojean:2004xa}
C.~Grojean, G.~Servant and J.~D.~Wells,
Phys. Rev. D \textbf{71}, 036001 (2005)
doi:10.1103/PhysRevD.71.036001
[arXiv:hep-ph/0407019 [hep-ph]].

\bibitem{Morrissey:2012db}
D.~E.~Morrissey and M.~J.~Ramsey-Musolf,
New J. Phys. \textbf{14}, 125003 (2012)
doi:10.1088/1367-2630/14/12/125003
[arXiv:1206.2942 [hep-ph]].

\bibitem{Athron:2023xlk}
P.~Athron, C.~Bal{\'a}zs, A.~Fowlie, L.~Morris and L.~Wu,
Prog. Part. Nucl. Phys. \textbf{135}, 104094 (2024)
doi:10.1016/j.ppnp.2023.104094
[arXiv:2305.02357 [hep-ph]].

\bibitem{Schwaller:2015tja}
P.~Schwaller,
Phys. Rev. Lett. \textbf{115}, no.18, 181101 (2015)
doi:10.1103/PhysRevLett.115.181101
[arXiv:1504.07263 [hep-ph]].

\bibitem{Jaeckel:2016jlh}
J.~Jaeckel, V.~V.~Khoze and M.~Spannowsky,
Phys. Rev. D \textbf{94}, no.10, 103519 (2016)
doi:10.1103/PhysRevD.94.103519
[arXiv:1602.03901 [hep-ph]].

\bibitem{Jinno:2016knw}
R.~Jinno and M.~Takimoto,
Phys. Rev. D \textbf{95}, no.1, 015020 (2017)
doi:10.1103/PhysRevD.95.015020
[arXiv:1604.05035 [hep-ph]].

\bibitem{Addazi:2016fbj}
A.~Addazi,
Mod. Phys. Lett. A \textbf{32}, no.08, 1750049 (2017)
doi:10.1142/S0217732317500493
[arXiv:1607.08057 [hep-ph]].

\bibitem{Chao:2017vrq}
W.~Chao, H.~K.~Guo and J.~Shu,
JCAP \textbf{09}, 009 (2017)
doi:10.1088/1475-7516/2017/09/009
[arXiv:1702.02698 [hep-ph]].

\bibitem{Addazi:2017gpt}
A.~Addazi and A.~Marciano,
Chin. Phys. C \textbf{42}, no.2, 023107 (2018)
doi:10.1088/1674-1137/42/2/023107
[arXiv:1703.03248 [hep-ph]].

\bibitem{Marzola:2017jzl}
L.~Marzola, A.~Racioppi and V.~Vaskonen,
Eur. Phys. J. C \textbf{77}, no.7, 484 (2017)
doi:10.1140/epjc/s10052-017-4996-1
[arXiv:1704.01034 [hep-ph]].

\bibitem{Baldes:2018emh}
I.~Baldes and C.~Garcia-Cely,
JHEP \textbf{05}, 190 (2019)
doi:10.1007/JHEP05(2019)190
[arXiv:1809.01198 [hep-ph]].

\bibitem{Marzo:2018nov}
C.~Marzo, L.~Marzola and V.~Vaskonen,
Eur. Phys. J. C \textbf{79}, no.7, 601 (2019)
doi:10.1140/epjc/s10052-019-7076-x
[arXiv:1811.11169 [hep-ph]].

\bibitem{Breitbach:2018ddu}
M.~Breitbach, J.~Kopp, E.~Madge, T.~Opferkuch and P.~Schwaller,
JCAP \textbf{07}, 007 (2019)
doi:10.1088/1475-7516/2019/07/007
[arXiv:1811.11175 [hep-ph]].

\bibitem{Fairbairn:2019xog}
M.~Fairbairn, E.~Hardy and A.~Wickens,
JHEP \textbf{07}, 044 (2019)
doi:10.1007/JHEP07(2019)044
[arXiv:1901.11038 [hep-ph]].

\bibitem{Nakai:2020oit}
Y.~Nakai, M.~Suzuki, F.~Takahashi and M.~Yamada,
Phys. Lett. B \textbf{816}, 136238 (2021)
doi:10.1016/j.physletb.2021.136238
[arXiv:2009.09754 [astro-ph.CO]].

\bibitem{Addazi:2020zcj}
A.~Addazi, Y.~F.~Cai, Q.~Gan, A.~Marciano and K.~Zeng,
Sci. China Phys. Mech. Astron. \textbf{64}, no.9, 290411 (2021)
doi:10.1007/s11433-021-1724-6
[arXiv:2009.10327 [hep-ph]].

\bibitem{Kierkla:2022odc}
M.~Kierkla, A.~Karam and B.~Swiezewska,
JHEP \textbf{03}, 007 (2023)
doi:10.1007/JHEP03(2023)007
[arXiv:2210.07075 [astro-ph.CO]].

\bibitem{Fujikura:2023lkn}
K.~Fujikura, S.~Girmohanta, Y.~Nakai and M.~Suzuki,
Phys. Lett. B \textbf{846}, 138203 (2023)
doi:10.1016/j.physletb.2023.138203
[arXiv:2306.17086 [hep-ph]].

\bibitem{Ertas:2021xeh}
F.~Ertas, F.~Kahlhoefer and C.~Tasillo,
JCAP \textbf{02}, no.02, 014 (2022)
doi:10.1088/1475-7516/2022/02/014
[arXiv:2109.06208 [astro-ph.CO]].

\bibitem{Wang:2022akn}
W.~Wang, W.~L.~Xu and J.~M.~Yang,
Eur. Phys. J. Plus \textbf{138}, no.9, 781 (2023)
doi:10.1140/epjp/s13360-023-04412-4
[arXiv:2209.11408 [hep-ph]].

\bibitem{Chen:2023rrl}
Z.~Chen, K.~Ye and M.~Zhang,
Phys. Rev. D \textbf{107}, no.9, 095027 (2023)
doi:10.1103/PhysRevD.107.095027
[arXiv:2303.11820 [hep-ph]].

\bibitem{Bringmann:2023opz}
T.~Bringmann, P.~F.~Depta, T.~Konstandin, K.~Schmidt-Hoberg and C.~Tasillo,
JCAP \textbf{11}, 053 (2023)
doi:10.1088/1475-7516/2023/11/053
[arXiv:2306.09411 [astro-ph.CO]].

\bibitem{Li:2023bxy}
S.~P.~Li and K.~P.~Xie,
Phys. Rev. D \textbf{108}, no.5, 055018 (2023)
doi:10.1103/PhysRevD.108.055018
[arXiv:2307.01086 [hep-ph]].

\bibitem{Kanemura:2023jiw}
S.~Kanemura and S.~P.~Li,
JCAP \textbf{03}, 005 (2024)
doi:10.1088/1475-7516/2024/03/005
[arXiv:2308.16390 [hep-ph]].

\bibitem{Bringmann:2023iuz}
T.~Bringmann, T.~E.~Gonzalo, F.~Kahlhoefer, J.~Matuszak and C.~Tasillo,
JCAP \textbf{05}, 065 (2024)
doi:10.1088/1475-7516/2024/05/065
[arXiv:2311.06346 [astro-ph.CO]].

\bibitem{Banik:2024zwj}
A.~Banik, Y.~Cui, Y.~D.~Tsai and Y.~Tsai,
[arXiv:2412.16282 [hep-ph]].

\bibitem{Feng:2024pab}
W.~Z.~Feng, J.~Li and P.~Nath,
Phys. Rev. D \textbf{110}, no.1, 015020 (2024)
doi:10.1103/PhysRevD.110.015020
[arXiv:2403.09558 [hep-ph]].

\bibitem{Balan:2025uke}
S.~Balan, T.~Bringmann, F.~Kahlhoefer, J.~Matuszak and C.~Tasillo,
JCAP \textbf{08}, 062 (2025)
doi:10.1088/1475-7516/2025/08/062
[arXiv:2502.19478 [hep-ph]].

\bibitem{Baules:2025pww}
V.~Baules and N.~Okada,
[arXiv:2508.13527 [hep-ph]].

\bibitem{Goncalves:2024lrk}
J.~Gon{\c{c}}alves, D.~Marfatia, A.~P.~Morais and R.~Pasechnik,
JHEP \textbf{02}, 110 (2025)
doi:10.1007/JHEP02(2025)110
[arXiv:2412.02645 [hep-ph]].

\bibitem{Costa:2025csj}
F.~Costa, J.~Hoefken Zink, M.~Lucente, S.~Pascoli and S.~Rosauro-Alcaraz,
Phys. Lett. B \textbf{868}, 139634 (2025)
doi:10.1016/j.physletb.2025.139634
[arXiv:2501.15649 [hep-ph]].

\bibitem{Li:2025nja}
J.~Li and P.~Nath,
Phys. Rev. D \textbf{111}, no.12, 123007 (2025)
doi:10.1103/79cb-rssl
[arXiv:2501.14986 [hep-ph]].

\bibitem{Wang:2026wwp}
Z.~Wang,
[arXiv:2601.04340 [astro-ph.CO]].

\bibitem{Feng:2025wvc}
W.~Z.~Feng, J.~Li, P.~Nath and Z.~H.~Ye,
Phys. Rev. D \textbf{113}, no.6, 063504 (2026)
doi:10.1103/kvq2-glq5
[arXiv:2510.13770 [hep-ph]].

\bibitem{Houtz:2025ogg}
R.~Houtz, M.~Ulloa and M.~West,
[arXiv:2511.23467 [hep-ph]].

\bibitem{Mahapatra:2026fyv}
S.~Mahapatra, P.~K.~Paul and N.~Sahu,
[arXiv:2601.12319 [hep-ph]].

\bibitem{Metaxas:1995ab}
D.~Metaxas and E.~J.~Weinberg,
Phys. Rev. D \textbf{53}, 836-843 (1996)
doi:10.1103/PhysRevD.53.836
[arXiv:hep-ph/9507381 [hep-ph]].

\bibitem{Patel:2011th}
H.~H.~Patel and M.~J.~Ramsey-Musolf,
JHEP \textbf{07}, 029 (2011)
doi:10.1007/JHEP07(2011)029
[arXiv:1101.4665 [hep-ph]].

\bibitem{Garny:2012cg}
M.~Garny and T.~Konstandin,
JHEP \textbf{07}, 189 (2012)
doi:10.1007/JHEP07(2012)189
[arXiv:1205.3392 [hep-ph]].

\bibitem{Andreassen:2014eha}
A.~Andreassen, W.~Frost and M.~D.~Schwartz,
Phys. Rev. D \textbf{91}, no.1, 016009 (2015)
doi:10.1103/PhysRevD.91.016009
[arXiv:1408.0287 [hep-ph]].

\bibitem{Andreassen:2014gha}
A.~Andreassen, W.~Frost and M.~D.~Schwartz,
Phys. Rev. Lett. \textbf{113}, no.24, 241801 (2014)
doi:10.1103/PhysRevLett.113.241801
[arXiv:1408.0292 [hep-ph]].

\bibitem{Arunasalam:2021zrs}
S.~Arunasalam and M.~J.~Ramsey-Musolf,
JHEP \textbf{08}, 138 (2022)
doi:10.1007/JHEP08(2022)138
[arXiv:2105.07588 [hep-ph]].

\bibitem{Lofgren:2021ogg}
J.~L{\"o}fgren, M.~J.~Ramsey-Musolf, P.~Schicho and T.~V.~I.~Tenkanen,
Phys. Rev. Lett. \textbf{130}, no.25, 251801 (2023)
doi:10.1103/PhysRevLett.130.251801
[arXiv:2112.05472 [hep-ph]].

\bibitem{Hirvonen:2021zej}
J.~Hirvonen, J.~L{\"o}fgren, M.~J.~Ramsey-Musolf, P.~Schicho and T.~V.~I.~Tenkanen,
JHEP \textbf{07}, 135 (2022)
doi:10.1007/JHEP07(2022)135
[arXiv:2112.08912 [hep-ph]].

\bibitem{Zhu:2025pht}
Y.~Zhu, J.~Liu, R.~Qin and L.~Bian,
Phys. Rev. D \textbf{112}, no.1, 015018 (2025)
doi:10.1103/f4gr-hycg
[arXiv:2503.19566 [hep-ph]].

\bibitem{Liu:2025ipj}
J.~Liu, R.~Qin and L.~Bian,
[arXiv:2512.05565 [hep-ph]].

\bibitem{Liu:2026ask}
J.~Liu, R.~Qin and L.~Bian,
[arXiv:2601.05793 [hep-ph]].

\bibitem{Leitao:2015fmj}
L.~Leitao and A.~Megevand,
JCAP \textbf{05}, 037 (2016)
doi:10.1088/1475-7516/2016/05/037
[arXiv:1512.08962 [astro-ph.CO]].

\bibitem{Megevand:2016lpr}
A.~Megevand and S.~Ramirez,
Nucl. Phys. B \textbf{919}, 74-109 (2017)
doi:10.1016/j.nuclphysb.2017.03.009
[arXiv:1611.05853 [astro-ph.CO]].

\bibitem{Kobakhidze:2017mru}
A.~Kobakhidze, C.~Lagger, A.~Manning and J.~Yue,
Eur. Phys. J. C \textbf{77}, no.8, 570 (2017)
doi:10.1140/epjc/s10052-017-5132-y
[arXiv:1703.06552 [hep-ph]].

\bibitem{Iso:2017uuu}
S.~Iso, P.~D.~Serpico and K.~Shimada,
Phys. Rev. Lett. \textbf{119}, no.14, 141301 (2017)
doi:10.1103/PhysRevLett.119.141301
[arXiv:1704.04955 [hep-ph]].

\bibitem{Ellis:2019oqb}
J.~Ellis, M.~Lewicki, J.~M.~No and V.~Vaskonen,
JCAP \textbf{06}, 024 (2019)
doi:10.1088/1475-7516/2019/06/024
[arXiv:1903.09642 [hep-ph]].

\bibitem{Wang:2020jrd}
X.~Wang, F.~P.~Huang and X.~Zhang,
JCAP \textbf{05}, 045 (2020)
doi:10.1088/1475-7516/2020/05/045
[arXiv:2003.08892 [hep-ph]].

\bibitem{Ellis:2020nnr}
J.~Ellis, M.~Lewicki and V.~Vaskonen,
JCAP \textbf{11}, 020 (2020)
doi:10.1088/1475-7516/2020/11/020
[arXiv:2007.15586 [astro-ph.CO]].

\bibitem{Kawana:2022fum}
K.~Kawana,
Phys. Rev. D \textbf{105}, no.10, 103515 (2022)
doi:10.1103/PhysRevD.105.103515
[arXiv:2201.00560 [hep-ph]].

\bibitem{Freese:2022qrl}
K.~Freese and M.~W.~Winkler,
Phys. Rev. D \textbf{106}, no.10, 103523 (2022)
doi:10.1103/PhysRevD.106.103523
[arXiv:2208.03330 [astro-ph.CO]].

\bibitem{Lewicki:2022pdb}
M.~Lewicki and V.~Vaskonen,
Eur. Phys. J. C \textbf{83}, no.2, 109 (2023)
doi:10.1140/epjc/s10052-023-11241-3
[arXiv:2208.11697 [astro-ph.CO]].

\bibitem{Sagunski:2023ynd}
L.~Sagunski, P.~Schicho and D.~Schmitt,
Phys. Rev. D \textbf{107}, no.12, 123512 (2023)
doi:10.1103/PhysRevD.107.123512
[arXiv:2303.02450 [hep-ph]].

\bibitem{Madge:2023dxc}
E.~Madge, E.~Morgante, C.~Puchades-Ib{\'a}{\~n}ez, N.~Ramberg, W.~Ratzinger, S.~Schenk and P.~Schwaller,
JHEP \textbf{10}, 171 (2023)
doi:10.1007/JHEP10(2023)171
[arXiv:2306.14856 [hep-ph]].

\bibitem{Athron:2023mer}
P.~Athron, A.~Fowlie, C.~T.~Lu, L.~Morris, L.~Wu, Y.~Wu and Z.~Xu,
Phys. Rev. Lett. \textbf{132}, no.22, 221001 (2024)
doi:10.1103/PhysRevLett.132.221001
[arXiv:2306.17239 [hep-ph]].

\bibitem{Ghosh:2023aum}
T.~Ghosh, A.~Ghoshal, H.~K.~Guo, F.~Hajkarim, S.~F.~King, K.~Sinha, X.~Wang and G.~White,
JCAP \textbf{05}, 100 (2024)
doi:10.1088/1475-7516/2024/05/100
[arXiv:2307.02259 [astro-ph.HE]].

\bibitem{Athron:2023rfq}
P.~Athron, L.~Morris and Z.~Xu,
JCAP \textbf{05}, 075 (2024)
doi:10.1088/1475-7516/2024/05/075
[arXiv:2309.05474 [hep-ph]].

\bibitem{Athron:2025pog}
P.~Athron, S.~Datta and Z.~Y.~Zhang,
[arXiv:2511.10288 [hep-ph]].

\bibitem{Pascoli:2026tuu}
S.~Pascoli, S.~Rosauro-Alcaraz and M.~Zandi,
[arXiv:2602.02829 [hep-ph]].

\bibitem{Feldman:2007wj}
D.~Feldman, Z.~Liu and P.~Nath,
Phys. Rev. D \textbf{75}, 115001 (2007)
doi:10.1103/PhysRevD.75.115001
[arXiv:hep-ph/0702123 [hep-ph]].

\bibitem{Feng:2023ubl}
W.~Z.~Feng, Z.~H.~Zhang and K.~Y.~Zhang,
JCAP \textbf{05}, 112 (2024)
doi:10.1088/1475-7516/2024/05/112
[arXiv:2312.03837 [hep-ph]].

\bibitem{Dolan:1973qd}
L.~Dolan and R.~Jackiw,
Phys. Rev. D \textbf{9}, 3320-3341 (1974)
doi:10.1103/PhysRevD.9.3320

\bibitem{Curtin:2016urg}
D.~Curtin, P.~Meade and H.~Ramani,
Eur. Phys. J. C \textbf{78}, no.9, 787 (2018)
doi:10.1140/epjc/s10052-018-6268-0
[arXiv:1612.00466 [hep-ph]].

\bibitem{Laine:2016hma}
M.~Laine and A.~Vuorinen,
Lect. Notes Phys. \textbf{925}, pp.1-281 (2016)
Springer, 2016,
doi:10.1007/978-3-319-31933-9
[arXiv:1701.01554 [hep-ph]].

\bibitem{Ginsparg:1980ef}
P.~H.~Ginsparg,
Nucl. Phys. B \textbf{170}, 388-408 (1980)
doi:10.1016/0550-3213(80)90418-6

\bibitem{Appelquist:1981vg}
T.~Appelquist and R.~D.~Pisarski,
Phys. Rev. D \textbf{23}, 2305 (1981)
doi:10.1103/PhysRevD.23.2305

\bibitem{Nadkarni:1982kb}
S.~Nadkarni,
Phys. Rev. D \textbf{27}, 917 (1983)
doi:10.1103/PhysRevD.27.917

\bibitem{Farakos:1994kx}
K.~Farakos, K.~Kajantie, K.~Rummukainen and M.~E.~Shaposhnikov,
Nucl. Phys. B \textbf{425}, 67-109 (1994)
doi:10.1016/0550-3213(94)90173-2
[arXiv:hep-ph/9404201 [hep-ph]].

\bibitem{Kajantie:1995dw}
K.~Kajantie, M.~Laine, K.~Rummukainen and M.~E.~Shaposhnikov,
Nucl. Phys. B \textbf{458}, 90-136 (1996)
doi:10.1016/0550-3213(95)00549-8
[arXiv:hep-ph/9508379 [hep-ph]].

\bibitem{Braaten:1995cm}
E.~Braaten and A.~Nieto,
Phys. Rev. D \textbf{51}, 6990-7006 (1995)
doi:10.1103/PhysRevD.51.6990
[arXiv:hep-ph/9501375 [hep-ph]].

\bibitem{Parwani:1991gq}
R.~R.~Parwani,
Phys. Rev. D \textbf{45}, 4695 (1992)
[erratum: Phys. Rev. D \textbf{48}, 5965 (1993)]
doi:10.1103/PhysRevD.45.4695
[arXiv:hep-ph/9204216 [hep-ph]].

\bibitem{Nielsen:1975fs}
N.~K.~Nielsen,
Nucl. Phys. B \textbf{101}, 173-188 (1975)
doi:10.1016/0550-3213(75)90301-6

\bibitem{Coleman:1977py}
S.~R.~Coleman,
Phys. Rev. D \textbf{15}, 2929-2936 (1977)
[erratum: Phys. Rev. D \textbf{16}, 1248 (1977)]
doi:10.1103/PhysRevD.16.1248

\bibitem{Callan:1977pt}
C.~G.~Callan, Jr. and S.~R.~Coleman,
Phys. Rev. D \textbf{16}, 1762-1768 (1977)
doi:10.1103/PhysRevD.16.1762

\bibitem{Gould:2021ccf}
O.~Gould and J.~Hirvonen,
Phys. Rev. D \textbf{104}, no.9, 096015 (2021)
doi:10.1103/PhysRevD.104.096015
[arXiv:2108.04377 [hep-ph]].

\bibitem{Wainwright:2011kj}
C.~L.~Wainwright,
Comput. Phys. Commun. \textbf{183}, 2006-2013 (2012)
doi:10.1016/j.cpc.2012.04.004
[arXiv:1109.4189 [hep-ph]].

\bibitem{Linde:1981zj}
A.~D.~Linde,
Nucl. Phys. B \textbf{216}, 421 (1983)
[erratum: Nucl. Phys. B \textbf{223}, 544 (1983)]
doi:10.1016/0550-3213(83)90072-X

\bibitem{Guth:1979bh}
A.~H.~Guth and S.~H.~H.~Tye,
Phys. Rev. Lett. \textbf{44}, 631 (1980)
[erratum: Phys. Rev. Lett. \textbf{44}, 963 (1980)]
doi:10.1103/PhysRevLett.44.631

\bibitem{Guth:1981uk}
A.~H.~Guth and E.~J.~Weinberg,
Phys. Rev. D \textbf{23}, 876 (1981)
doi:10.1103/PhysRevD.23.876

\bibitem{Athron:2022mmm}
P.~Athron, C.~Bal{\'a}zs and L.~Morris,
JCAP \textbf{03}, 006 (2023)
doi:10.1088/1475-7516/2023/03/006
[arXiv:2212.07559 [hep-ph]].

\bibitem{Chodos:1974je}
A.~Chodos, R.~L.~Jaffe, K.~Johnson, C.~B.~Thorn and V.~F.~Weisskopf,
Phys. Rev. D \textbf{9}, 3471-3495 (1974)
doi:10.1103/PhysRevD.9.3471

\bibitem{Leitao:2012tx}
L.~Leitao, A.~Megevand and A.~D.~Sanchez,
JCAP \textbf{10}, 024 (2012)
doi:10.1088/1475-7516/2012/10/024
[arXiv:1205.3070 [astro-ph.CO]].

\bibitem{Giese:2020rtr}
F.~Giese, T.~Konstandin and J.~van de Vis,
JCAP \textbf{07}, no.07, 057 (2020)
doi:10.1088/1475-7516/2020/07/057
[arXiv:2004.06995 [astro-ph.CO]].

\bibitem{Giese:2020znk}
F.~Giese, T.~Konstandin, K.~Schmitz and J.~van de Vis,
JCAP \textbf{01}, 072 (2021)
doi:10.1088/1475-7516/2021/01/072
[arXiv:2010.09744 [astro-ph.CO]].

\bibitem{Espinosa:2010hh}
J.~R.~Espinosa, T.~Konstandin, J.~M.~No and G.~Servant,
JCAP \textbf{06}, 028 (2010)
doi:10.1088/1475-7516/2010/06/028
[arXiv:1004.4187 [hep-ph]].

\bibitem{Ai:2023see}
W.~Y.~Ai, B.~Laurent and J.~van de Vis,
JCAP \textbf{07}, 002 (2023)
doi:10.1088/1475-7516/2023/07/002
[arXiv:2303.10171 [astro-ph.CO]].

\bibitem{Ai:2024btx}
W.~Y.~Ai, B.~Laurent and J.~van de Vis,
JHEP \textbf{02}, 119 (2025)
doi:10.1007/JHEP02(2025)119
[arXiv:2411.13641 [hep-ph]].

\bibitem{Kosowsky:1991ua}
A.~Kosowsky, M.~S.~Turner and R.~Watkins,
Phys. Rev. D \textbf{45}, 4514-4535 (1992)
doi:10.1103/PhysRevD.45.4514

\bibitem{Hindmarsh:2013xza}
M.~Hindmarsh, S.~J.~Huber, K.~Rummukainen and D.~J.~Weir,
Phys. Rev. Lett. \textbf{112}, 041301 (2014)
doi:10.1103/PhysRevLett.112.041301
[arXiv:1304.2433 [hep-ph]].

\bibitem{Caprini:2006jb}
C.~Caprini and R.~Durrer,
Phys. Rev. D \textbf{74}, 063521 (2006)
doi:10.1103/PhysRevD.74.063521
[arXiv:astro-ph/0603476 [astro-ph]].

\bibitem{Hindmarsh:2015qta}
M.~Hindmarsh, S.~J.~Huber, K.~Rummukainen and D.~J.~Weir,
Phys. Rev. D \textbf{92}, no.12, 123009 (2015)
doi:10.1103/PhysRevD.92.123009
[arXiv:1504.03291 [astro-ph.CO]].

\bibitem{Ruan:2018tsw}
W.~H.~Ruan, Z.~K.~Guo, R.~G.~Cai and Y.~Z.~Zhang,
Int. J. Mod. Phys. A \textbf{35}, no.17, 2050075 (2020)
doi:10.1142/S0217751X2050075X
[arXiv:1807.09495 [gr-qc]].

\bibitem{TianQin:2015yph}
J.~Luo \textit{et al.} [TianQin],
Class. Quant. Grav. \textbf{33}, no.3, 035010 (2016)
doi:10.1088/0264-9381/33/3/035010
[arXiv:1512.02076 [astro-ph.IM]].

\bibitem{LISA:2017pwj}
P.~Amaro-Seoane \textit{et al.} [LISA],
[arXiv:1702.00786 [astro-ph.IM]].

\bibitem{Sesana:2019vho}
A.~Sesana, N.~Korsakova, M.~A.~Sedda, V.~Baibhav, E.~Barausse, S.~Barke, E.~Berti, M.~Bonetti, P.~R.~Capelo and C.~Caprini, \textit{et al.}
Exper. Astron. \textbf{51}, no.3, 1333-1383 (2021)
doi:10.1007/s10686-021-09709-9
[arXiv:1908.11391 [astro-ph.IM]].

\bibitem{Grojean:2006bp}
C.~Grojean and G.~Servant,
Phys. Rev. D \textbf{75}, 043507 (2007)
doi:10.1103/PhysRevD.75.043507
[arXiv:hep-ph/0607107 [hep-ph]].

\bibitem{Kuroyanagi:2014qaa}
S.~Kuroyanagi, S.~Tsujikawa, T.~Chiba and N.~Sugiyama,
Phys. Rev. D \textbf{90}, no.6, 063513 (2014)
doi:10.1103/PhysRevD.90.063513
[arXiv:1406.1369 [astro-ph.CO]].

\bibitem{Punturo:2010zz}
M.~Punturo, M.~Abernathy, F.~Acernese, B.~Allen, N.~Andersson, K.~Arun, F.~Barone, B.~Barr, M.~Barsuglia and M.~Beker, \textit{et al.}
Class. Quant. Grav. \textbf{27}, 194002 (2010)
doi:10.1088/0264-9381/27/19/194002

\bibitem{LIGOScientific:2016wof}
B.~P.~Abbott \textit{et al.} [LIGO Scientific],
Class. Quant. Grav. \textbf{34}, no.4, 044001 (2017)
doi:10.1088/1361-6382/aa51f4
[arXiv:1607.08697 [astro-ph.IM]].

\bibitem{NANOGrav:2023gor}
G.~Agazie \textit{et al.} [NANOGrav],
Astrophys. J. Lett. \textbf{951}, no.1, L8 (2023)
doi:10.3847/2041-8213/acdac6
[arXiv:2306.16213 [astro-ph.HE]].

\bibitem{Reardon:2023gzh}
D.~J.~Reardon, A.~Zic, R.~M.~Shannon, G.~B.~Hobbs, M.~Bailes, V.~Di Marco, A.~Kapur, A.~F.~Rogers, E.~Thrane and J.~Askew, \textit{et al.}
Astrophys. J. Lett. \textbf{951}, no.1, L6 (2023)
doi:10.3847/2041-8213/acdd02
[arXiv:2306.16215 [astro-ph.HE]].

\bibitem{EPTA:2023fyk}
J.~Antoniadis \textit{et al.} [EPTA and InPTA:],
Astron. Astrophys. \textbf{678}, A50 (2023)
doi:10.1051/0004-6361/202346844
[arXiv:2306.16214 [astro-ph.HE]].

\bibitem{Athron:2020sbe}
P.~Athron, C.~Bal{\'a}zs, A.~Fowlie and Y.~Zhang,
Eur. Phys. J. C \textbf{80}, no.6, 567 (2020)
doi:10.1140/epjc/s10052-020-8035-2
[arXiv:2003.02859 [hep-ph]].

\bibitem{Arcadi:2024ukq}
G.~Arcadi, D.~Cabo-Almeida, M.~Dutra, P.~Ghosh, M.~Lindner, Y.~Mambrini, J.~P.~Neto, M.~Pierre, S.~Profumo and F.~S.~Queiroz,
Eur. Phys. J. C \textbf{85}, no.2, 152 (2025)
doi:10.1140/epjc/s10052-024-13672-y
[arXiv:2403.15860 [hep-ph]].

\bibitem{Feng:2024blk}
W.~Z.~Feng and Z.~H.~Zhang,
[arXiv:2409.17217 [hep-ph]].

\bibitem{Beacham:2019nyx}
J.~Beacham, C.~Burrage, D.~Curtin, A.~De Roeck, J.~Evans, J.~L.~Feng, C.~Gatto, S.~Gninenko, A.~Hartin and I.~Irastorza, \textit{et al.}
J. Phys. G \textbf{47}, no.1, 010501 (2020)
doi:10.1088/1361-6471/ab4cd2
[arXiv:1901.09966 [hep-ex]].

\bibitem{Arcadi:2021mag}
G.~Arcadi, A.~Djouadi and M.~Kado,
Eur. Phys. J. C \textbf{81}, no.7, 653 (2021)
doi:10.1140/epjc/s10052-021-09411-2
[arXiv:2101.02507 [hep-ph]].

\bibitem{NA62:2025upx}
E.~Cortina Gil \textit{et al.} [NA62],
JHEP \textbf{11}, 143 (2025)
doi:10.1007/JHEP11(2025)143
[arXiv:2507.17286 [hep-ex]].

\bibitem{ATLAS:2023tkt}
G.~Aad \textit{et al.} [ATLAS],
Phys. Lett. B \textbf{842}, 137963 (2023)
doi:10.1016/j.physletb.2023.137963
[arXiv:2301.10731 [hep-ex]].

\bibitem{CMS:2023sdw}
A.~Tumasyan \textit{et al.} [CMS],
Eur. Phys. J. C \textbf{83}, no.10, 933 (2023)
doi:10.1140/epjc/s10052-023-11952-7
[arXiv:2303.01214 [hep-ex]].

\bibitem{LHCHiggsCrossSectionWorkingGroup:2016ypw}
D.~de Florian \textit{et al.} [LHC Higgs Cross Section Working Group],
CERN Yellow Rep. Monogr. \textbf{2}, 1-869 (2017)
doi:10.23731/CYRM-2017-002
[arXiv:1610.07922 [hep-ph]].

\bibitem{Feng:2024nkh}
W.~Z.~Feng and Z.~H.~Zhang,
Phys. Rev. D \textbf{112}, no.3, 035004 (2025)
doi:10.1103/vshn-gbdp
[arXiv:2405.19431 [hep-ph]].

\bibitem{Chang:2016ntp}
J.~H.~Chang, R.~Essig and S.~D.~McDermott,
JHEP \textbf{01}, 107 (2017)
doi:10.1007/JHEP01(2017)107
[arXiv:1611.03864 [hep-ph]].

\bibitem{Caputo:2025avc}
A.~Caputo, J.~Park and S.~Yun,
[arXiv:2511.15785 [hep-ph]].

\bibitem{Pospelov:2008zw}
M.~Pospelov,
Phys. Rev. D \textbf{80}, 095002 (2009)
doi:10.1103/PhysRevD.80.095002
[arXiv:0811.1030 [hep-ph]].

\bibitem{Riordan:1987aw}
E.~M.~Riordan, M.~W.~Krasny, K.~Lang, P.~De Barbaro, A.~Bodek, S.~Dasu, N.~Varelas, X.~Wang, R.~G.~Arnold and D.~Benton, \textit{et al.}
Phys. Rev. Lett. \textbf{59}, 755 (1987)
doi:10.1103/PhysRevLett.59.755

\bibitem{Bjorken:1988as}
J.~D.~Bjorken, S.~Ecklund, W.~R.~Nelson, A.~Abashian, C.~Church, B.~Lu, L.~W.~Mo, T.~A.~Nunamaker and P.~Rassmann,
Phys. Rev. D \textbf{38}, 3375 (1988)
doi:10.1103/PhysRevD.38.3375

\bibitem{Batell:2014mga}
B.~Batell, R.~Essig and Z.~Surujon,
Phys. Rev. Lett. \textbf{113}, no.17, 171802 (2014)
doi:10.1103/PhysRevLett.113.171802
[arXiv:1406.2698 [hep-ph]].

\bibitem{Marsicano:2018krp}
L.~Marsicano, M.~Battaglieri, M.~Bondi', C.~D.~R.~Carvajal, A.~Celentano, M.~De Napoli, R.~De Vita, E.~Nardi, M.~Raggi and P.~Valente,
Phys. Rev. D \textbf{98}, no.1, 015031 (2018)
doi:10.1103/PhysRevD.98.015031
[arXiv:1802.03794 [hep-ex]].

\bibitem{Blumlein:2011mv}
J.~Blumlein and J.~Brunner,
Phys. Lett. B \textbf{701}, 155-159 (2011)
doi:10.1016/j.physletb.2011.05.046
[arXiv:1104.2747 [hep-ex]].

\bibitem{Blumlein:2013cua}
J.~Bl{\"u}mlein and J.~Brunner,
Phys. Lett. B \textbf{731}, 320-326 (2014)
doi:10.1016/j.physletb.2014.02.029
[arXiv:1311.3870 [hep-ph]].

\bibitem{CMS:2019kiy}
 [CMS],
CMS-PAS-EXO-19-018.

\bibitem{BaBar:2014zli}
J.~P.~Lees \textit{et al.} [BaBar],
Phys. Rev. Lett. \textbf{113}, no.20, 201801 (2014)
doi:10.1103/PhysRevLett.113.201801
[arXiv:1406.2980 [hep-ex]].

\bibitem{LHCb:2019vmc}
R.~Aaij \textit{et al.} [LHCb],
Phys. Rev. Lett. \textbf{124}, no.4, 041801 (2020)
doi:10.1103/PhysRevLett.124.041801
[arXiv:1910.06926 [hep-ex]].

\bibitem{KLOE-2:2011hhj}
F.~Archilli \textit{et al.} [KLOE-2],
Phys. Lett. B \textbf{706}, 251-255 (2012)
doi:10.1016/j.physletb.2011.11.033
[arXiv:1110.0411 [hep-ex]].

\bibitem{KLOE-2:2012lii}
D.~Babusci \textit{et al.} [KLOE-2],
Phys. Lett. B \textbf{720}, 111-115 (2013)
doi:10.1016/j.physletb.2013.01.067
[arXiv:1210.3927 [hep-ex]].

\bibitem{KLOE-2:2014qxg}
D.~Babusci \textit{et al.} [KLOE-2],
Phys. Lett. B \textbf{736}, 459-464 (2014)
doi:10.1016/j.physletb.2014.08.005
[arXiv:1404.7772 [hep-ex]].

\bibitem{KLOE-2:2016ydq}
A.~Anastasi \textit{et al.} [KLOE-2],
Phys. Lett. B \textbf{757}, 356-361 (2016)
doi:10.1016/j.physletb.2016.04.019
[arXiv:1603.06086 [hep-ex]].

\bibitem{Merkel:2014avp}
H.~Merkel, P.~Achenbach, C.~Ayerbe Gayoso, T.~Beranek, J.~Bericic, J.~C.~Bernauer, R.~B{\"o}hm, D.~Bosnar, L.~Correa and L.~Debenjak, \textit{et al.}
Phys. Rev. Lett. \textbf{112}, no.22, 221802 (2014)
doi:10.1103/PhysRevLett.112.221802
[arXiv:1404.5502 [hep-ex]].

\bibitem{Aboubrahim:2020lnr}
A.~Aboubrahim, W.~Z.~Feng, P.~Nath and Z.~Y.~Wang,
Phys. Rev. D \textbf{103}, no.7, 075014 (2021)
doi:10.1103/PhysRevD.103.075014
[arXiv:2008.00529 [hep-ph]].

\end{thebibliography}
\end{document}